\documentclass[zpreprint,zbstdefault]{./LaTeX/zeus/zeus_paper}
\usepackage[english]{babel}
\newcommand{\ZcoosysA}{%
The ZEUS coordinate system is a right-handed Cartesian system, with the $Z$
axis pointing in the proton beam direction, referred to as the ``forward
direction'', and the $X$ axis pointing left towards the center of HERA.
The coordinate origin is at the nominal interaction point.\xspace}

\newcommand{\ZcoosysfnA}{\footnote{\ZcoosysA}}

\chardef\usc=95
\chardef\til=126
\catcode`\@=11 
\DeclareRobustCommand\xdotspace{\futurelet\@let@token\@xdotspace}
\def\@xdotspace{%
  \ifx\@let@token.\else
  \ifx\@let@token\bgroup.\else
  \ifx\@let@token\egroup.\else
  \ifx\@let@token\/.\else
  \ifx\@let@token\ .\else
  \ifx\@let@token~.\else
  \ifx\@let@token!.\else
  \ifx\@let@token,.\else
  \ifx\@let@token:.\else
  \ifx\@let@token;.\else
  \ifx\@let@token?.\else
  \ifx\@let@token/.\else
  \ifx\@let@token'.\else
  \ifx\@let@token).\else
  \ifx\@let@token-.\else
  \ifx\@let@token\@xobeysp.\else
  \ifx\@let@token\space.\else
  \ifx\@let@token\@sptoken.\else
   .\space
   \fi\fi\fi\fi\fi\fi\fi\fi\fi\fi\fi\fi\fi\fi\fi\fi\fi\fi}
\catcode`\@=12 

\newcommand{\stru}[2]{%
   \relax\ifmmode\hbox{\vrule height#1 depth#2 width0pt}%
   \else\vrule height#1 depth#2 width0pt\fi}

\newcommand{\Ronum}[1]{\uppercase\expandafter{\romannumeral#1}}
\newcommand{\ronum}[1]{\expandafter{\romannumeral#1}}
\DeclareRobustCommand{\LaTeXZ}{%
  \LaTeX\kern-.05em4\kern-.1em
  {\raisebox{-0.2ex}{$\scriptstyle\text{ZEUS}$}}\xspace}

\newcommand{\eq}[1]{Eq.~(\ref{eq-#1})}

\newcommand{\fig}[1]{Fig.~\ref{fig-#1}}
\newcommand{\Fig}[1]{Figure~\ref{fig-#1}}
\newcommand{\figand}[2]{Figs.~\ref{fig-#1} and~\ref{fig-#2}}

\newcommand{\tab}[1]{Table~\ref{tab-#1}}

\newcommand{\taband}[2]{Tables~\ref{tab-#1} and~\ref{tab-#2}}

\newcommand{\sect}[1]{Section~\ref{sec-#1}}

\DeclareMathAlphabet{\mathbf}{OT1}{cmr}{bx}{sl}
\newcommand{\eVdist}{\kern-0.06667em}

\newcommand{\Gev}{{\text{Ge}\eVdist\text{V\/}}}

\newcommand{\mev}{{\,\text{Me}\eVdist\text{V\/}}}
\newcommand{\gev}{{\,\text{Ge}\eVdist\text{V\/}}}

\newcommand{\pb}{\,\text{pb}}

\newcommand{\mm}{\,\text{mm}}
\newcommand{\cm}{\,\text{cm}}

\newcommand{\ns}{\,\text{ns}}

\newcommand{\rad}{\,\text{rad}}

\newcommand{\mrad}{\,\text{mrad}}

\newcommand{\Tesla}{\,\text{T}}

\newcommand{\slashfrac}[2]{%
  \raisebox{0.5ex}{\ensuremath #1}\kern-0.12em/\kern-0.08em
  \raisebox{-.8ex}{\ensuremath #2}}

\newcommand{\sqr}[3]{%
    {\vcenter{\hrule height.#3ex\hbox{\vrule width.#2ex height#1ex
     \kern#1ex\vrule width.#3ex}\hrule height.#2ex}}}

\catcode`\@=11 
\newcommand{\parenbar}{\mathpalette\p@renb@r}
\def\p@renb@r#1#2{\vbox{%
  \ifx#1\scriptscriptstyle \dimen@.7em\dimen@ii.2em\else
  \ifx#1\scriptstyle \dimen@.8em\dimen@ii.25em\else
  \dimen@1em\dimen@ii.4em\fi\fi \offinterlineskip
  \ialign{\hfill##\hfill\cr
    \vbox{\hrule width\dimen@ii}\cr
    \noalign{\vskip-.3ex}%
    \hbox to\dimen@{$\mathchar300\hfil\mathchar301$}\cr
    \noalign{\vskip-.3ex}%
    $#1#2$\cr}}}
\catcode`\@=12 

\newcommand{\gh}{{\gamma_h}}

\newcommand{\DA}{{\rm DA}}

\newcommand{\IP}{{\rm I$\kern-0.01667em$P}\xspace}
\newcommand{\JB}{{\rm JB}}

\mathchardef\qsm=63
\mathchardef\pls=43
\mathchardef\mns=512
\mathchardef\plm=518
\mathchardef\eql=61
\mathchardef\smallleft=300
\mathchardef\smallright=301
\mathchardef\les=316
\mathchardef\gre=318
\mathchardef\leq=532
\mathchardef\grq=533

\catcode`\@=11 
\newcounter{pict@width}
\newcounter{pict@height}
\newlength{\pict@scale}
\newcommand{\psfigadd}[4]{%
\setcounter{pict@width}{1*\ratio{#2+\pict@scale/2}{\pict@scale}}
\setcounter{pict@height}{1*\ratio{#3+\pict@scale/2}{\pict@scale}}
\setlength{\unitlength}{\pict@scale}
\hbox to #2{\hspace{-\fill}\begin{picture}(\thepict@width,\thepict@height)
\put(0,0){\psfig{figure=#1,width=#2,height=#3,clip=}}
\SetScale{0.283466457}
\SetWidth{1.763889}
{#4}
\end{picture}}
}
\newcounter{pict@widthfst}
\newcounter{pict@widthscd}
\newcounter{pict@widthtot}
\newcommand{\psfigaddtwo}[7]{%
\setcounter{pict@widthfst}{1*\ratio{#2+\pict@scale/2}{\pict@scale}}
\setcounter{pict@widthscd}{1*\ratio{#2+#4+\pict@scale/2}{\pict@scale}}
\setcounter{pict@widthtot}{1*\ratio{#2+#4+#6+\pict@scale/2}{\pict@scale}}
\setcounter{pict@height}{1*\ratio{#3+\pict@scale/2}{\pict@scale}}
\setlength{\unitlength}{\pict@scale}
\hbox{\hspace{-\fill}\begin{picture}(\thepict@widthtot,\thepict@height)
\put(0,0){\psfig{figure=#1,width=#2,height=#3,clip=}}
\put(\thepict@widthscd,0){\psfig{figure=#5,width=#6,height=#3,clip=}}
\SetScale{0.283466457}
\SetWidth{1.763889}
{#7}
\end{picture}}
}
\newcommand{\psfigror}[4]{%
\setcounter{pict@width}{1*\ratio{#2+\pict@scale/2}{\pict@scale}}
\setcounter{pict@height}{1*\ratio{#3+\pict@scale/2}{\pict@scale}}
\setlength{\unitlength}{\pict@scale}
\hbox{\begin{picture}(\thepict@width,\thepict@height)
\put(0,\thepict@height){\psfig{figure=#1,width=#3,height=#2,clip=,angle=270}}
\SetScale{0.283466457}
\SetWidth{1.763889}
{#4}
\end{picture}}
}
\newcommand{\psfigrol}[4]{%
\setcounter{pict@width}{1*\ratio{#2+\pict@scale/2}{\pict@scale}}
\setcounter{pict@height}{1*\ratio{#3+\pict@scale/2}{\pict@scale}}
\setlength{\unitlength}{\pict@scale}
\hbox{\begin{picture}(\thepict@width,\thepict@height)
\put(0,0){\psfig{figure=#1,width=#3,height=#2,clip=,angle=90}}
\SetScale{0.283466457}
\SetWidth{1.763889}
{#4}
\end{picture}}
}
\catcode`\@=12 
\newlength\listtextwidth

\catcode`\@=11 
\newlength{\@tabfninsert}
\newlength{\@tabfnwidth}
\newcommand{\tabfootnote}[2]{%
  \setlength{\@tabfninsert}{0.8em}
  \setlength{\@tabfnwidth}{\textwidth}
  \addtolength{\@tabfnwidth}{-\@tabfninsert}
  \addtolength{\@tabfnwidth}{-0.4em}
  \noindent\makebox[\@tabfninsert][r]{\footnotesize$^{#1}$\hfil}\hfill%
  \parbox[t]{\@tabfnwidth}{\footnotesize #2\hfill}}
\catcode`\@=12 

\def\@evenhead{\underline{\protect\parbox{\textwidth}{\bf\boldmath\protect\rule[-0.2cm]{0pt}{0.2cm}{\Large\thepage}\hfill\leftmark\hfill~}}}%
\def\@oddhead{\underline{\protect\parbox{\textwidth}{\bf\boldmath\protect\rule[-0.2cm]{0pt}{0.2cm}~\hfill\rightmark\hfill{\Large\thepage}}}}%

\newcommand{\etamax}{{\eta_{\rm{max}}}}

\newcommand{\trk}{{\rm trk}}

\newcommand{\Pom}{{\text{I\kern-0.12em P}}}

\newcommand{\qq}       {\mbox{$Q^{2}$}}

\newcommand{\sigqq}    {\mbox{$d\sigma/d\qq$}}
\newcommand{\gevv}     {\mbox{${\rm Ge\kern -0.1em V}^{2}$}}
\newcommand{\Et}       {\mbox{$E_{T}$}}

\newcommand{\Pt}       {\mbox{$P_{T}$}}
\newcommand{\qqda}     {\mbox{$Q^{2}\dab$}}

\newcommand{\dab}      {_{\rm \scriptscriptstyle DA}}
\newcommand{\jbb}      {_{\rm \scriptscriptstyle JB}}

\newcommand{\sigx}     {\mbox{$d\sigma/dx$}}
\newcommand{\sigy}     {\mbox{$d\sigma/dy$}}

\newcommand{\qqc}      {\mbox{$Q^{2}_{c}$}}
\newcommand{\lw}[1]{\smash{\lower1.8ex\hbox{#1}}}

\def\RefDJango{{\cite{%
proc:hera:1991:1419,*spi:www:djangoh11%
}}\xspace}
\def\RefJetSet{{\cite{%
cpc:39:347,*cpc:43:367,*cpc:82:74%
}}\xspace}
\def\RefRecMethod{{\cite{%
proc:hera:1991:23,*proc:hera:1991:43%
}}\xspace}
\def\citeH1Refs{{\cite{%
np:b470:3,*np:b497:3,*epj:c13:609,*epj:c19:269,*h1:9900%
}}\xspace}
\def\citeCTD{{\cite{%
nim:a279:290,*npps:b32:181,*nim:a338:254%
}}\xspace}
\def\citeCAL{{\cite{%
nim:a309:77,*nim:a309:101,*nim:a321:356,*nim:a336:23%
}}\xspace}

\begin{document}
\title{ \boldmath High-$Q^2$ neutral current cross sections
in $e^+p$ deep inelastic scattering at $\sqrt{s}=318 \gev$ }
                    
\author{ZEUS Collaboration}
\draftversion{3.3}
\prepnum{DESY-03-214}
\date{December 2003}

%
\abstract{
Cross sections for $e^+p$ neutral current deep inelastic scattering have been
measured at a centre-of-mass energy of $\sqrt{s}=318 \gev$ with the ZEUS detector at HERA using an integrated luminosity of $63.2 \pb^{-1}$. The
double-differential cross section, $d^2\sigma / dx dQ^2$, is presented for
$200 \gev^2 < Q^2 < 30\,000 \gev^2$ and for $0.005 < x < 0.65$. The single-differential cross-sections $d\sigma / dQ^2$,
$d\sigma / dx$ and $d\sigma / dy$ are presented for $Q^2 > 200 \gev^2$. The effect of $Z$-boson exchange is seen in $d\sigma / dx$ measured for $Q^2
> 10\,000 \gev^2$.  The data presented here were combined with ZEUS $e^+ p$
neutral current data taken at $\sqrt{s}=300 \gev$ and the structure function
$F_2^{\rm em}$ was extracted. All results agree well with the predictions of the Standard Model.  }

%
\makezeustitle
\def\3{\ss}                                                                                        
\pagenumbering{Roman}                                                                              
                                                   %
\begin{center}                                                                                     
{                      \Large  The ZEUS Collaboration              }                               
\end{center}                                                                                       
  S.~Chekanov,                                                                                     
  M.~Derrick,                                                                                      
  D.~Krakauer,                                                                                     
  J.H.~Loizides$^{   1}$,                                                                          
  S.~Magill,                                                                                       
  S.~Miglioranzi$^{   1}$,                                                                         
  B.~Musgrave,                                                                                     
  J.~Repond,                                                                                       
  R.~Yoshida\\                                                                                     
 {\it Argonne National Laboratory, Argonne, Illinois 60439-4815}, USA~$^{n}$                       
\par \filbreak                                                                                     
  M.C.K.~Mattingly \\                                                                              
 {\it Andrews University, Berrien Springs, Michigan 49104-0380}, USA
% ...
\newpage                                                                                           
$^{\    1}$ also affiliated with University College London, London, UK \\                          
% ...
                                                           %
                                                           %
\newpage   
                                                           %
                                                           %
\begin{tabular}[h]{rp{14cm}}                                                                       
$^{a}$ &  supported by the Natural Sciences and Engineering Research                               
          Council of Canada (NSERC) \\                                                             
$^{b}$ &  supported by the German Federal Ministry for Education and                               
          Research (BMBF), under contract numbers HZ1GUA 2, HZ1GUB 0, HZ1PDA 5, HZ1VFA 5\\         
$^{c}$ &  supported by the MINERVA Gesellschaft f\"ur Forschung GmbH, the                          
          Israel Science Foundation, the U.S.-Israel Binational Science                            
          Foundation and the Benozyio Center                                                       
          for High Energy Physics\\                                                                
$^{d}$ &  supported by the German-Israeli Foundation and the Israel Science                        
          Foundation\\                                                                             
$^{e}$ &  supported by the Italian National Institute for Nuclear Physics (INFN) \\                
$^{f}$ &  supported by the Japanese Ministry of Education, Culture,                                
          Sports, Science and Technology (MEXT) and its grants for                                 
          Scientific Research\\                                                                    
$^{g}$ &  supported by the Korean Ministry of Education and Korea Science                          
          and Engineering Foundation\\                                                             
$^{h}$ &  supported by the Netherlands Foundation for Research on Matter (FOM)\\                   
$^{i}$ &  supported by the Polish State Committee for Scientific Research,                         
          grant no. 620/E-77/SPB/DESY/P-03/DZ 117/2003-2005\\                                      
$^{j}$ &  partially supported by the German Federal Ministry for Education                         
          and Research (BMBF)\\                                                                    
$^{k}$ &  partly supported by the Russian Ministry of Industry, Science                            
          and Technology through its grant for Scientific Research on High                         
          Energy Physics\\                                                                         
$^{l}$ &  supported by the Spanish Ministry of Education and Science                               
          through funds provided by CICYT\\                                                        
$^{m}$ &  supported by the Particle Physics and Astronomy Research Council, UK\\                   
$^{n}$ &  supported by the US Department of Energy\\                                               
$^{o}$ &  supported by the US National Science Foundation\\                                        
$^{p}$ &  supported by the Polish State Committee for Scientific Research,                         
          grant no. 112/E-356/SPUB/DESY/P-03/DZ 116/2003-2005,2 P03B 13922\\                       
$^{q}$ &  supported by the Polish State Committee for Scientific Research,                         
          grant no. 115/E-343/SPUB-M/DESY/P-03/DZ 121/2001-2002, 2 P03B 07022\\                    
\end{tabular}                                                                                      
                                                           %
                                                           %

\pagenumbering{arabic} 
\pagestyle{plain}
%
\section{Introduction}
\label{sec-Introduction}

Neutral current (NC) deep inelastic scattering (DIS) is described in terms of 
the space-like exchange of a virtual photon and a
virtual $Z$ boson. The photon-exchange contribution dominates when the
four-momentum-transfer squared, $Q^2$, is much less than the square of the
$Z$-boson mass, $M_Z^2$. The effect of $Z$ exchange is comparable in
magnitude to that of photon exchange when $Q^2 \sim M_Z^2$. The parity-violating 
part of the $Z$ exchange contribution increases the cross
section for $e^-p$ NC DIS and decreases that for $e^+p$ NC DIS over what would be
expected for pure single-photon exchange. The comparison of the $e^- p$ NC DIS 
cross section to that for
$e^+ p$ NC DIS therefore provides a direct way to observe the effect of
$Z$-exchange in the scattering of charged leptons on protons.

The ZEUS and H1 collaborations have each measured both the $e^- p$ and the
$e^+ p$ NC DIS cross sections up to a $Q^2$ of $30\,000\gev^2$
\cite{epj:c11:427,epj:c13:609,epj:c21:33,epj:c21:443,epj:c28:175,np:b470:3,np:b497:3,epj:c19:269,epj:c30:1}.  
When HERA ran at a centre-of-mass energy $\sqrt{s} = 300 \gev$, $e^+p$ data sets were collected, 
whereas both $e^+ p$ and $e^- p$ data were
collected in 1998-2000 at $\sqrt{s} = 318 \gev$. The measured $e^\pm p$ NC DIS
cross sections are well described at next-to-leading order (NLO) in quantum
chromodynamics (QCD) by the Standard Model (SM) prediction including both photon- and
$Z$-exchange contributions.

This paper presents the measurement of the NC $e^+p$ DIS cross-section $d^2\sigma / dx dQ^2$ for
$200 \gev^2 < Q^2 < 30\,000 \gev^2$ and $0.005 < x < 0.65$, together with $d\sigma / dQ^2$, 
$d\sigma / dx$ and $d\sigma / dy$ for $Q^2 > 200 \gev^2$,
where $x$ and $y$ are the Bjorken scaling variables. The data were
collected in 1999 and 2000 at $\sqrt{s} = 318 \gev$ and correspond to an integrated
luminosity of $63.2 \pb^{-1}$.  The results are compared to recent ZEUS
measurements of the $e^-p$ NC DIS cross sections \cite{epj:c28:175} and to SM 
predictions. The structure function $F_2^{\rm em}$ was extracted by combining the data
presented here with the ZEUS measurement of $d^2 \sigma / dx dQ^2$ for NC
DIS at $\sqrt{s} = 300$~GeV~\cite{epj:c21:443}, and compared to measurements by the H1
collaboration and by fixed-target experiments.

%
%

\section{Standard Model cross sections}
\label{sec-KineXSect}

For longitudinally unpolarised beams, the NC DIS differential cross section,
$d^2\sigma_{\rm Born}/dx \,dQ^2$, for the reaction $e^+p \rightarrow e^+X$ can
be written at leading order in the electroweak interaction
as~\cite{pl:b201:369,ijmp:a13:3385}:
\begin{equation}
  \frac{d^2\sigma_{\rm Born}(e^\pm p)}{dx \,dQ^2} =  
  \frac{2 \pi \alpha^2}{x Q^4}
  \left[
    Y_+ F_2 \left(x, Q^2 \right) \mp Y_- xF_3 \left(x, Q^2 \right) - 
    y^2 F_L \left(x, Q^2 \right)
  \right] \ ,
  \label{eq-Born}
\end{equation}
where $y=Q^{2}/xs$ (neglecting the masses of the incoming particles) and 
$Y_{\pm} \equiv 1 \pm (1-y)^2$ and $\alpha$ denotes the fine-structure
constant.  At leading order (LO) in QCD, the longitudinal structure
function, $F_L$, is zero and the structure functions $F_2$ and $xF_3$ can be
expressed as products of electroweak couplings and parton density functions
(PDFs) as follows:
\begin{alignat*}{2}
  F_2  &=& &  x \sum\limits_{f}\,
  A_f
  (q_f + \bar{q}_f) \ ,                                    \nonumber \\
  &&&                                                      \nonumber \\[-2.2em]
  &&&                                                                \\[-0.8em]
  xF_3 &=& & x \sum\limits_{f}\,
  B_f
  (q_f - \bar{q}_f) \ ,                                    \nonumber
\end{alignat*}
where $x q_f(x,\qq)$ are the quark and $x {\bar q_f}(x,\qq)$ the
anti-quark PDFs and $f$ runs over the five active quark flavours;
$A_f$ and $B_f$ contain products of electroweak couplings and
ratios of photon and $Z$-boson propagators.
For convenience, the reduced cross section, $\tilde{\sigma}$, can be
defined as 
\begin{equation}
  \tilde{\sigma} =    \frac{x \, Q^4}{2 \pi \alpha^2 Y_+} \
                      \frac{d^2\sigma_{\rm Born}}{dx\, dQ^2} \ . \nonumber
\end{equation}

All cross-section calculations presented in this paper have been
performed using NLO QCD.
These calculations predict that the contribution of $F_L$ to
$d^2\sigma_{\rm Born}/dx dQ^2$ is approximately $1.5\%$, averaged
over the kinematic range considered in this paper. 
However, in the region of small $x$, near $Q^2 = 250 \gev^2$, the
$F_L$ contribution to the cross section can be as large as $17\%$.

%
\section{The ZEUS experiment at HERA}
\label{sec-zeus}

For the data analysed in the present study, HERA accelerated positrons to an
energy of $E_e = 27.5 \gev$ and protons to an energy of $E_p = 920\gev$,
yielding $\sqrt{s} = 318 \gev$. The inter-bunch spacing of the beams was $96 \ns$. 
In normal running, some radiofrequency buckets in both the positron
and the proton ring were left empty to study single-beam backgrounds.

A detailed description of the ZEUS detector can be found
elsewhere~\cite{zeus:1993:bluebook}. A brief outline of the components that are
most relevant for this analysis is given below.

The high-resolution uranium--scintillator calorimeter (CAL)~\citeCAL consists of
three parts: the forward (FCAL), the barrel (BCAL) and the rear (RCAL)
calorimeters. Each part is subdivided into towers and each tower is
longitudinally segmented into one electromagnetic section (EMC) and either one
(in RCAL) or two (in BCAL and FCAL) hadronic sections (HAC). The smallest
subdivision of the calorimeter is called a cell.  The CAL energy resolutions,
measured under test-beam conditions, are $\sigma(E)/E=0.18/\sqrt{E}$ for
positrons and $\sigma(E)/E=0.35/\sqrt{E}$ for hadrons, with $E$ in $\Gev$. The
timing resolution of the CAL is $\sim 1 \ns$ for energy deposits greater than
$4.5 \gev$.

Presampler detectors~\cite{nim:a382:419,*bpres:2000:625} are mounted in front of the
CAL. They consist of scintillator tiles
matching the calorimeter towers and measure signals from particle showers
created by interactions in the material lying between the interaction point and
the calorimeter.

The RCAL is instrumented with a layer of $3 \times 3 \cm^2$
silicon-pad detectors at a depth of 3.3 radiation lengths. This
hadron-electron separator (HES) \cite{nim:a277:176} is used to improve
the positron-angle measurement.

Charged particles are tracked in the central tracking detector (CTD)~\citeCTD,
which operates in a magnetic field of $1.43\Tesla$ provided by a thin
superconducting solenoid. The CTD consists of 72~cylindrical drift-chamber
layers, organised in nine superlayers covering the polar-angle\ZcoosysfnA\
region \mbox{$15^\circ<\theta<164^\circ$}. The transverse-momentum resolution
for full-length tracks is $\sigma(p_T)/p_T=0.0058 \,
p_T\oplus0.0065\oplus0.0014/p_T$, with $p_T$ in \gev.

The luminosity is measured using the Bethe-Heitler reaction $ep
\rightarrow e\gamma p$ \cite{desy-92-066,*zfp:c63:391,*acpp:b32:2025}.
The resulting small-angle photons were measured by the luminosity
monitor, a lead-scintillator calorimeter placed in the HERA tunnel 107
m from the interaction point in the positron beam direction. In addition
a lead-scintillator calorimeter placed 35 m from the interaction point 
was used to measure positrons scattered through small angles.

%
\section{Monte Carlo simulation}
\label{sec-MC}

Monte Carlo (MC) simulations were used to evaluate the efficiency for selecting
events, to determine the accuracy of the kinematic reconstruction, to estimate
the background rate, and to extrapolate the measured cross sections to the full
kinematic range. A sufficient number of events was generated to ensure that
statistical uncertainties from the MC samples were negligible in comparison to
those of the data.

Neutral current DIS events were simulated including radiative effects, using the
{\sc heracles} 4.6.1~\cite{cpc:69:155} program with the {\sc djangoh} 1.1
~\RefDJango interface to the hadronisation programs and using CTEQ5D
\cite{epj:c12:375} PDFs. In {\sc heracles}, $\mathcal{O}(\alpha)$ electroweak
corrections for initial- and final-state radiation, vertex and propagator
corrections and two-boson exchange are included. Values from the Particle Data
Group~\cite{epj:c15:1} were used for the Fermi constant, $G_F$, and the masses of the
$Z$ boson and the top quark. The Higgs-boson mass was set to $100 \gev$. The
colour-dipole model of {\sc ariadne} 4.10~\cite{cpc:71:15} was used to
simulate the $\mathcal{O}(\alpha_S)$ plus leading-logarithmic corrections to the
quark-parton model. The MEPS model of {\sc lepto} 6.5
~\cite{cpc:101:108} was used as a check. Both programs use the Lund string model of
{\sc jetset} 7.4~\RefJetSet for the hadronisation.  Diffractive events,
characterised by a suppression of particle production between the current jet and the
proton remnant, were generated using the
{\sc rapgap}~2.08/06~\cite{cpc:86:147} generator and appropriately mixed with
the non-diffractive NC DIS sample.  The contribution of diffractive events was
obtained by fitting the $\etamax$ distribution\footnote{The quantity $\etamax$ is defined as the pseudorapidity of the CAL energy deposit with the lowest polar angle and an energy above $400 \mev$.} of the data with a linear combination of
non-diffractive and diffractive MC samples whilst preserving the overall
normalisation. The fraction of diffractive events in the MC sample was 6.2\%. 
Photoproduction events,
including both direct and resolved processes, were simulated using
{\sc herwig}~6.1~\cite{cpc:67:465} to study backgrounds. The normalisation of the photoproduction
MC was determined from a sample of events in which the positron was detected in the
positron calorimeter of the luminosity monitor~\cite{thesis:schnurbusch:2002}.

The ZEUS detector response was simulated using a program based on {\sc geant}
3.13~\cite{tech:cern-dd-ee-84-1}. The generated events were passed through the
detector simulation, subjected to the same trigger requirements as the data and
processed by the same reconstruction programs.

The vertex distribution in data is a crucial input to the MC simulation for the
correct evaluation of the event-selection efficiency. Therefore, the $Z$-vertex
distribution used in the MC simulation was determined from a sample of NC DIS
events in which the event-selection efficiency was independent of $Z$.

%
\section{Event characteristics and kinematic reconstruction}
\label{sec-Char}

Neutral current events at high $Q^2$ are characterised by the presence of a high-energy
isolated positron in the final state. The transverse momentum of the scattered positron balances
that of the hadronic final state, resulting in a small net transverse momentum,
\Pt. The measured net transverse momentum and the net transverse energy, $E_T$, are
defined by:
\begin{alignat}{2}
  P_T^2 & = & P_X^2 + P_Y^2 = & \left( \sum\limits_{i} E_i \sin \theta_i \cos
    \phi_i \right)^2+ \left( \sum\limits_{i} E_i \sin \theta_i \sin \phi_i
  \right)^2,
  \label{eq-PT2}\\ 
  E_T & = & \sum\limits_{i} E_i \sin \theta_i, \nonumber
\end{alignat}
where the sums run over all calorimeter energy deposits, $E_i$, with polar and
azimuthal angles $\theta_i$ and $\phi_i$ with respect to the event vertex,
respectively.  The variable $\delta$ is also used in the event selection and is
defined as:
\begin{equation}
  \delta \equiv \sum\limits_{i} (E-p_Z)_{i} = \sum\limits_{i} ( E_i - E_i \cos
  \theta_{i} )
  \label{eq-Delta}
\end{equation}
where the sum runs over all calorimeter energy deposits $E_i$  (uncorrected for detector  
effects in the trigger, but corrected in the offline analysis as discussed below) with
polar angles $\theta_i$.  Conservation of energy and longitudinal momentum,
$p_Z$, requires $\delta = 2E_e= 55 \gev$ if all final-state particles are
detected and perfectly measured. Undetected particles that escape through the
forward beam-hole have a negligible effect on $\delta$. However, particles lost
through the rear beam hole can lead to a substantial reduction in
$\delta$ such as is the case in photoproduction events, in
which the positron emerges at very small scattering angles, or in events in which an
initial-state bremsstrahlung photon is emitted.

For the present study, the CAL energy deposits were separated into those
associated with the scattered positron and all other energy deposits. The sum of
the latter is referred to as the hadronic energy.  The spatial distribution of
the hadronic energy, together with the reconstructed vertex position, were used
to evaluate the hadronic polar angle, $\gamma_h$  (see Section~\ref{sec-GammaHad}), which, 
in the naive quark-parton model, corresponds to the polar angle of the struck quark.

The reconstruction of $x$, $\qq$ and $y$ was performed using the double angle (DA)
method~\RefRecMethod. This method uses the polar angle of the scattered positron
and the hadronic angle, $\gamma_h$, to obtain estimators of the kinematic
variables, $x_\DA$, $y_\DA$ and $Q^2_\DA$. The DA method is insensitive to
uncertainties in the overall energy scale of the calorimeter. However, it is
sensitive to initial-state QED radiation and, in addition, an accurate
simulation of the hadronic final state is necessary. In the event selection, 
$y$ calculated using the electron method ($y_e$) and the Jacquet-Blondel method 
\cite{proc:epfacility:1979:391} ($y_\JB$) were also used. 

The relative resolution in $\qq$ was $\sim 3\%$ over the kinematic range
covered.  The relative resolution in $x$ varied from $15\%$ in the lowest $\qq$
bins (see \sect{Binning}) to $\sim 4\%$ in the highest $\qq$ region. The
relative resolution in $y$ was $\sim 10 \%$ in the lowest $\qq$ bins, decreasing
to $1\%$ for high $y$ values in the highest $\qq$ bins.

%
\section{Positron reconstruction}
\label{sec-RecoEFinder}

\subsection{Positron identification}
To identify and reconstruct the scattered positron, an algorithm was
used that combines calorimeter and CTD information \cite{epj:c11:427}.  The
algorithm starts by identifying CAL clusters that are topologically consistent
with an electromagnetic shower. The clusters were required to have an energy of at least
$10 \gev$ and, if the positron candidate fell within the acceptance of the CTD,
a track was required which, when extrapolated, passed within $10 \cm$ of the 
cluster centre at the shower maximum. Such a track will be referred to as a
``matched'' track.  A positron candidate was considered to lie within the CTD
acceptance if a matched track emerging from the reconstructed event vertex
traversed at least four of the nine superlayers of the CTD. For the nominal
interaction point, i.e.  $Z = 0$, this requirement corresponds to the angular
range $23^\circ < \theta_e < 156^\circ$. Monte Carlo studies~\cite{thesis:amaya:2001} 
showed that the overall efficiency for finding the scattered positron is $\sim 95\%$ 
for a scattered positron energy, $E'_e$, greater than $10 \gev$ and 
$Q^2 < 15\,000 \gev^2$, decreasing to $\sim 85\%$ for $Q^2 > 30\,000 \gev^2$.

\subsection{Positron-energy determination}
\label{sec-EePrime}

The scattered-positron energy was determined from the calorimeter deposit since,
above $10 \gev$, the calorimeter energy resolution is better than the momentum
resolution of the CTD. The measured energy was corrected for the energy lost in
inactive material in front of the CAL. The presampler was used in the RCAL,
while in the B/FCAL a detailed material map was used~\cite{epj:c21:443}. To
render the energy response uniform across the face of the calorimeter, a
correction obtained from fits to the non-uniformity pattern in
data and in the MC simulation~\cite{epj:c11:427} was made. The corrections were
determined separately for the BCAL and the RCAL.  After these corrections, the
non-uniformities were greatly reduced and the data were well reproduced by the
MC simulation~\cite{thesis:moritz:2001}. Too few positrons were scattered into
the FCAL for such a correction to be derived.

After applying the corrections described above, the positron-energy resolution
was 10\% at $E'_e = 10 \gev$, falling to 5\% for $E'_e \gtrsim 20 \gev$.  The scale
uncertainty on the energies of the scattered positrons detected in the BCAL was
$\pm 1\%$.  For positrons detected in the RCAL, the scale uncertainty was $\pm
1.7\%$ at $10 \gev$, falling linearly to $\pm 1 \%$ for positrons with energies
of $15 \gev$ and above~\cite{epj:c21:443}. A scale uncertainty of $\pm 3\%$ was 
assigned to positrons reconstructed in the FCAL~\cite{epj:c11:427}.

\subsection{Determination of the positron polar angle}

Studies~\cite{thesis:amaya:2001} showed that the
angular resolution of tracks is superior to that for calorimeter
clusters. Hence, in the CTD acceptance region, which contains $98.8\%$ of the
events, $\theta_e$ was determined using the matched track. For candidates outside this
region, the position of the calorimeter cluster was used together with the event vertex
to determine the positron angle.

The CAL was aligned with respect to the CTD using the positron tracks
extrapolated to the face of the calorimeter with the aid of a detailed map of
the magnetic field. This allowed the BCAL to be aligned to precisions of $\pm
0.3 \mm$ in the $Z$ direction and $\pm0.6 \mrad$ in the azimuthal angle,
$\phi$~\cite{thesis:kappes:2001}. For the alignment of the RCAL, the position of
the extrapolated track was compared to that determined by the
HES\cite{thesis:kappes:2001}. The precision of the alignment was $\pm 0.3 \mm$
($\pm 0.6 \mm$) in the $X$ ($Z$) direction and $\pm 0.9 \mrad$ in $\phi$. In all
cases, the precision was sufficient to render resulting systematic uncertainties
on the cross sections negligible.

The resolution in $\theta_e$ was obtained by comparing the MC-generated angle to
that obtained after applying the detector simulation, reconstruction and
correction algorithms. The resulting resolution for positrons was $2 \mrad$ for $\theta_e < 23^\circ$, $3 \mrad$ for $23^\circ < \theta_e < 156^\circ$ and $5 \mrad$ for $\theta_e > 156^\circ$.

%
\section{Reconstruction of the hadronic system}
\label{sec-HadRecons}

\subsection{Hadronic-energy determination}
\label{sec-HadEne}

The hadronic-energy deposits were corrected for energy loss in the material
between the interaction point and the calorimeter using the material maps
implemented in the detector-simulation package. After applying all corrections,
the measured resolution for the transverse momentum of the hadronic final state, $P_{T,h}$, was
about $13\%$ ($11\%$) at $P_{T,h} = 20 \gev$ in BCAL (FCAL), decreasing to 8\%
(7.5\%) at $P_{T,h} = 60 \gev$.  The uncertainties in the hadronic energy
scales of the FCAL and the BCAL were $\pm 1\%$, while for the RCAL the
uncertainty was $\pm 2\%$~\cite{hep-ex-0307043}.

\subsection{Determination of the hadronic polar angle $\boldmath \gamma_h$}
\label{sec-GammaHad}
 The angle $\gamma_h$ is given by \cite{proc:hera:1991:23}
\begin{equation*}
  \cos \gamma_h = \frac{P^2_{T,h} - \delta_h^2}{P^2_{T,h} + \delta_h^2},
\end{equation*}
where $P_{T,h}$ and $\delta_h$ were calculated from Eqns. (\ref{eq-PT2}) 
and (\ref{eq-Delta}) using only the hadronic energy. 
Particles scattered by interactions in the
material between the primary vertex and the CAL generate energy
deposits in the CAL that bias the reconstructed value of $\gamma_h$.
To minimise this bias, an algorithm was developed in which CAL
clusters with energies below $3 \gev$ and with polar angles larger
than an angle $\gamma_{\rm max}$ were removed \cite{epj:c11:427}.  The
value of $\gamma_{\rm max}$ was derived from a NC MC sample by
minimising the bias in the reconstructed hadronic variables.

The resolution of $\gamma_h$ is below $15 \mrad$ for $\gamma_h < 0.2
\rad$, increasing to $100 \mrad$ at $\gamma_h \approx 2 \rad$. 
These resolutions dominate the uncertainties on the kinematic variables.

%
\section{Event selection}
\label{Sect:EvtSel}

\subsection{Trigger}
\label{sec-Trigger}

ZEUS operates a three-level trigger system~\cite{zeus:1993:bluebook,smith:1992}. 
At the first-level trigger, only coarse calorimeter and tracking information is available.
Events were selected using criteria based on an energy deposit in the CAL consistent with 
an isolated positron. In addition, events with high $E_{T}$ in coincidence with a CTD track 
were accepted. At the second level, a requirement on $\delta$ was used to select NC DIS 
events and timing information from the calorimeter was used to reject events inconsistent 
with the bunch-crossing time. At the third level, events were fully reconstructed on a
computer farm. The requirements were similar to, but looser than, the
offline cuts described below; a simpler and generally more
efficient (but less pure) positron finder was used.

The main uncertainty in the trigger efficiency comes from the first level.
The data and MC simulation agree to within $\sim 0.5\%$ and the
overall efficiency is close to
$100\%$~\cite{thesis:schnurbusch:2002}.

\subsection{Offline selection}
\label{sec-selec:offline}

The following criteria were applied offline: 
\begin{itemize}
\item positrons, identified as described in \sect{RecoEFinder}, were required to
  satisfy the following criteria:
  \begin{itemize}       
    \item to reduce background, isolated positrons were selected by
      requiring that less than $5 \gev$ be deposited in calorimeter
      cells not associated with the scattered positron, inside an
      $\eta$-$\phi$ cone of radius $R_{\rm cone} = 0.8$ centred on the
      positron. For those positrons with a matched track, the momentum of the
      track, $p_\trk$, was required to be at least $5 \gev$. For positrons outside 
      the forward tracking acceptance of the CTD, the tracking
      requirement in the positron selection was replaced by a cut on the
      transverse momentum of the positron, $p_T^e > 30 \gev$. For positrons outside 
      the backward tracking acceptance of the CTD, no track was required;
    \item a fiducial-volume cut was applied to the positron to guarantee that
      the experimental acceptance was well understood. It excluded the upper
      part of the central RCAL area occluded by the
      cryogenic supply for the solenoid magnet as well as the transition regions
      between the three parts of the CAL~\cite{thesis:goncalo:2003,thesis:liu:2003};
\end{itemize}
\item to ensure that event quantities were accurately determined, a
  reconstructed vertex with $-50 < Z < 50 \cm$ was required, a range
  consistent with the $ep$ interaction region. A small fraction of the
  proton-beam current was contained in satellite bunches, which were
  shifted by $\pm 4.8 \ns$ with respect to the nominal bunch-crossing
  time, resulting in a few of the $ep$ interactions occurring $\pm 72
  \cm$ from the nominal interaction point. This cut rejects $ep$
  events from these regions;
\item to suppress photoproduction events, in which the scattered positron escaped through
  the beam hole in the RCAL, $\delta$ was required to be greater than $38 \gev$.
  This cut also
  rejected events with large initial-state QED radiation.  The requirement $\delta<65
  \gev$ removed ``overlay'' events in which a normal DIS event coincided with
  additional energy deposits in the RCAL from some other reaction. This requirement had 
  a negligible effect on the efficiency for selecting NC DIS events. For positrons
  outside the forward tracking acceptance of the CTD, the lower $\delta$ cut was
  raised to $44 \gev$;
\item to further reduce background from photoproduction events, $y_e$ was required to 
  satisfy $y_e < 0.95$;
\item the net transverse momentum, \Pt, is expected to be close to
  zero for true NC events and was measured with an uncertainty approximately
  proportional to $\sqrt{\Et}$.  To remove cosmic rays and beam-related
  backgrounds, $P_T / \sqrt{E_T}$ was required to be less than $4\sqrt{\Gev}$;
\item to reduce the size of the QED radiative corrections, elastic
  Compton scattering events ($e p \rightarrow e \gamma p$) were
  removed. This was done using an algorithm that searched
  for an additional photon candidate and discarded the event if the
  sum of the energies associated with the positron and photon
  candidates was within $2 \gev$ of the total energy measured in the
  calorimeter. The contribution from deeply virtual 
  Compton scattering was estimated to be negligible;
\item in events with low $\gamma_h$, a large amount of energy is
  deposited near the inner edges of the FCAL or escapes through the forward beam
  pipe. As the MC simulation of the very forward energy flow is problematic
  ~\cite{thesis:goncalo:2003}, events where $\gamma_h$, extrapolated to
  the FCAL surface, lay within a circle of radius $20 \cm$ around the forward
  beam line were removed. For an interaction at the nominal interaction point,
  this FCAL circle cut corresponds to a lower $\gamma_h$ cut of $90 \mrad$. 
  This cut rejects events at very low $y$ that have high $x$;
\item the kinematic range over which the MC generator is valid does not extend
  to very low $y$ at high $x$. To avoid these regions of phase space, in
  addition to the previous cut, $y\jbb(1-x\dab)^2$ was required to be greater
  than $0.004$ \cite{cpc:81:381,*spi:www:django6}.
\end{itemize}

\begin{figure}
  \vspace*{-2.0cm}
  \begin{center}
    \includegraphics[width=0.8\textwidth]{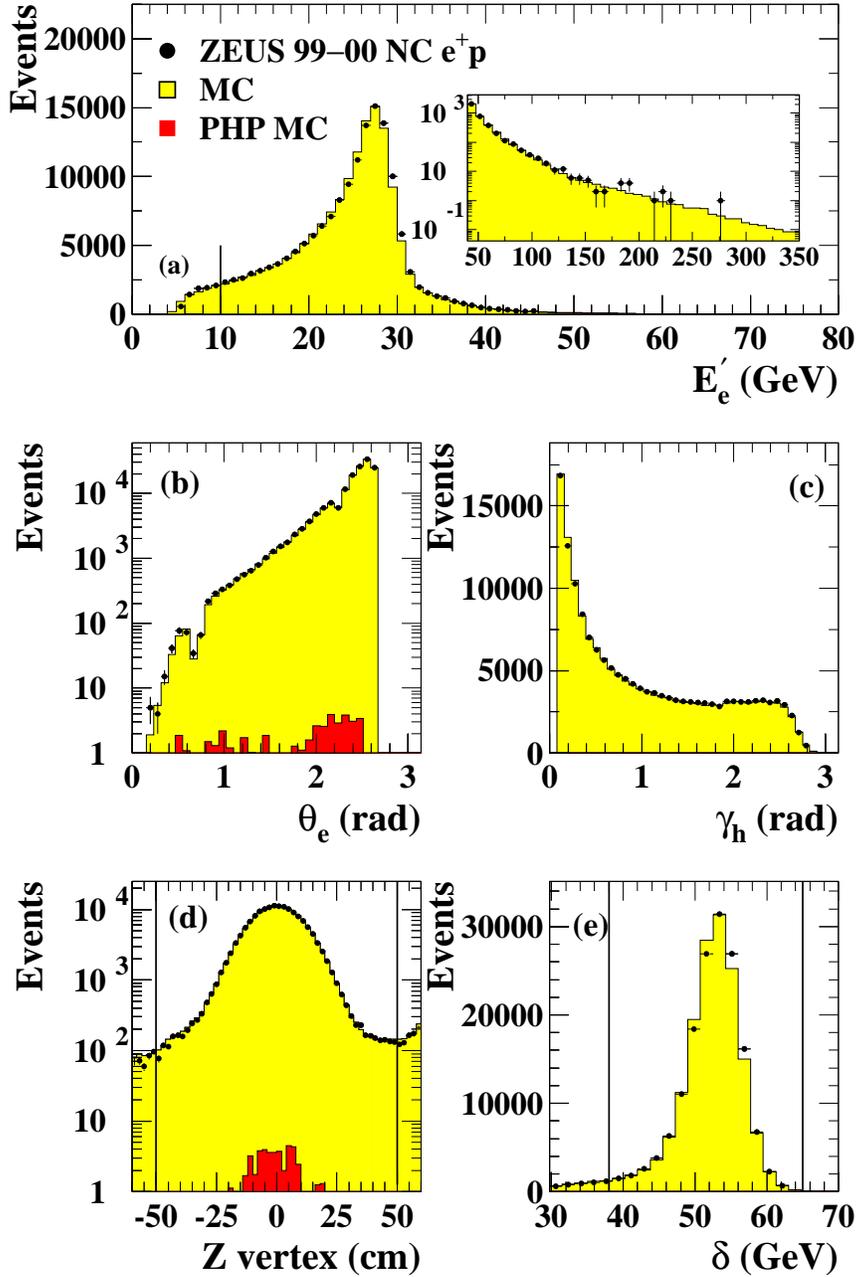}
  \end{center}
  \caption{ Comparison of $e^+p$ data (points) and MC simulation (histograms)
    for: (a) the energy of the scattered positron, $E_{e}^{\prime}$ (the inset
    shows the high-energy part of the distribution);
    (b) the angle of the scattered positron, $\theta_e$;
    (c) the hadronic angle, $\gamma_h$;
    (d) the $Z$ coordinate of the event vertex, and 
    (e) the $\delta$ variable.
    The vertical lines indicate the cut boundaries described in the
    text. The darker histogram visible in the $\theta_e$ and $Z$ vertex 
    figures corresponds to the photoproduction background.
  }
  \label{fig-CtrlPlts}
\end{figure}

A total of $156\,962$ events with $\qqda > 185 \gev^2$ satisfied the above 
criteria. Data distributions are compared to the sum of the signal and
photoproduction MC samples in \fig{CtrlPlts}. 
The signal MC includes a diffractive component, as discussed in Section~\ref{sec-MC}. Good
agreement between data and MC simulation is seen over the full range of most
variables. Imperfections in the MC simulation can be seen in the  disagreement 
between data and MC simulation that occurs in the region of the kinematic peak 
($E_e' \approx E_e$) in the positron energy distribution, 
and correspondingly in the peak region of the $\delta$ distribution.  
The effect of these differences on the cross section measurements are evaluated 
 in the systematic uncertainties assigned to the positron energy scale and to the 
positron energy resolution (see \sect{SysErr}).

The photoproduction background was $< 0.3\%$ over most of the kinematic range
covered, rising to $\sim 1.7\%$ at high $y$. From the study of empty
positron and proton buckets, it was concluded that possible backgrounds
associated with non-$ep$ collisions could be neglected.

%
\section{Results}
\label{Sect:Results}

\subsection{Binning, acceptance and cross-section determination}
\label{sec-Binning}

\begin{figure}
  \begin{center}
    \includegraphics[width=.9\textwidth]{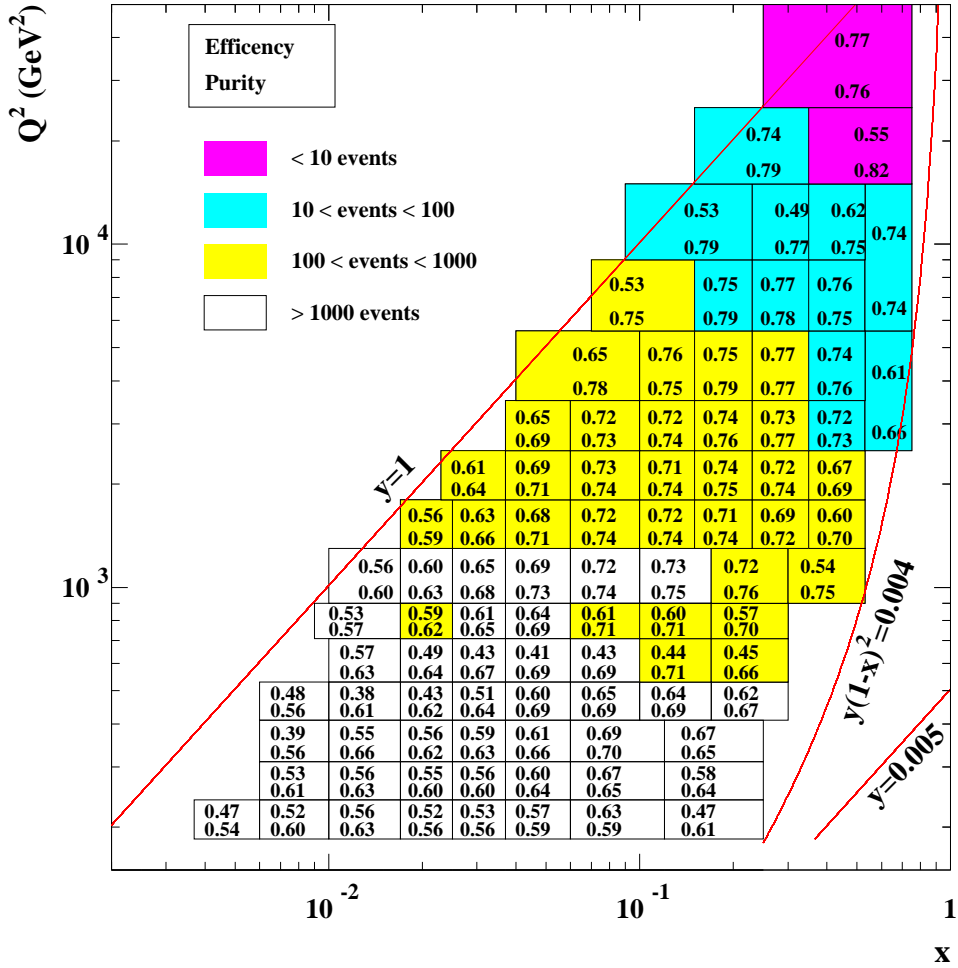}
  \end{center}
  \caption{ Bins used in the extraction of the double-differential cross
    section.  The solid diagonal lines are isolines of $y$ drawn for $y=1$ (the
    kinematic limit) and $y=0.005$. The curved line indicates the cut on
    $y\jbb(1-x\dab)^2$ described in Section~\ref{sec-selec:offline}. An
    indication of the approximate number of events from the final sample that
    lie in each bin is given by the shading level. The efficiency and purity 
    for each bin is shown }
  \label{fig-KinePlan}
\end{figure}

The bin sizes used for the determination of the single- and double-differential
cross sections were chosen commensurate with the resolutions. \Fig{KinePlan}
shows the kinematic region used in extracting the $e^+p$ double-differential
cross section. The number of events per bin decreases from $\sim 7\,000$ in the
lowest-$Q^2$ bins to five in the bin at the highest $Q^2$ and $x$. The
efficiency after all selection cuts (defined as the number of events generated
and reconstructed in a bin after all selection cuts divided by the number of
events that were generated in that bin) varied between 50\% and 80\%. In some
medium-$Q^2$ bins, dominated by events in which the positron is scattered into
the region between the R/BCAL at $\theta_e \sim 2.25 \rad$, the efficiency
decreases to around 40\%. The purity (defined as the number of events
reconstructed and generated in a bin after all selection cuts divided by the
total number of events reconstructed in that bin) ranged from 50\% to 80\%. The
efficiency and purity in double-differential bins are shown in
Fig.~\ref{fig-KinePlan}. 

The value of the cross section in a particular bin, for example for $d^2
\sigma/dx dQ^2$, was determined according to:
\begin{equation}
  \frac{d^2 \sigma}{dx dQ^2} = \frac{N_{\rm data}-N_{\rm bg}}{N_{\rm MC}} \cdot \frac{d^2
  \sigma^{\rm SM}_{\rm Born}}{dx dQ^2} \nonumber \ ,
  \label{eq-xsect}
\end{equation}
where $N_{\rm data}$ is the number of data events in the bin, $N_{\rm bg}$ is the number
of background events estimated from the photoproduction MC and $N_{\rm MC}$ is the
number of signal MC events normalised to the luminosity of the data. 
The SM prediction, $d^2 \sigma^{\rm SM}_{\rm Born}/dx dQ^2$, 
was evaluated according to \eq{Born} using CTEQ5D
PDFs~\cite{epj:c12:375} and using the PDG values~\cite{epj:c15:1} for the fine-structure 
constant, the mass of the $Z$ boson and the weak mixing angle. This procedure 
implicitly takes the acceptance, bin-centring and
radiative corrections from the MC simulation. A similar procedure was used for
$d \sigma / dx$, $d \sigma / dy$ and $d \sigma / dQ^2$. In this way, the cross
sections $d \sigma / dx$ and $d \sigma / dQ^2$ were extrapolated to the full $y$
range.

The statistical uncertainties on the cross sections were calculated from the
numbers of events observed in the bins, taking into account the statistical
uncertainty from the MC simulation (signal and background). Poisson
statistics were used for all bins.

\subsection{Systematic uncertainties}
\label{sec-SysErr}

Systematic uncertainties associated with the MC simulation were estimated by
re-calculating the cross section after modifying the simulation to account for
known uncertainties. Cut values were varied where this method
was not applicable. The positive and negative deviations from the nominal
cross-section values were added in quadrature separately to obtain the total
positive and negative systematic uncertainty.  The uncertainty on the luminosity
of the combined 1999/2000 $e^+p$ sample is 2.5\% and was not included in the
total systematic uncertainty. The other uncertainties are discussed in detail
below.

\subsubsection{Uncorrelated systematic uncertainties}

The following systematic uncertainties are either small or exhibit no bin-to-bin
correlations:
\begin{itemize}
\item positron energy resolution in the MC simulation: the effect on the cross 
  sections of changing the CAL energy resolution for the scattered positron in
  the MC by $\pm 1\%$ was negligible over almost the full kinematic range. The
  effect increased to $\sim \pm 1\%$ only for $d\sigma/dy$ bins at high $y$ 
  and for double-differential bins at high $Q^2$;
\item positron angle: differences between data and MC simulation in
  the positron scattering angle due to uncertainties in the simulation of the
  CTD were at most $1 \mrad$. Typically, the deviations were within $\pm
  1\%$; the effect increased to as much as $\pm 2\%$ only in a few high-$\qq$
  double-differential bins, but was small compared to the statistical
  uncertainty;
\item hadronic angle: the uncertainty associated with the
  reconstruction of $\gh$\ was investigated by varying the calorimeter energy
  scale for the hadronic final state separately for R/B/FCAL as described in
  \sect{HadEne} and by varying $\gamma_{\rm max}$ in a range for which the
  reconstructed value of $\gamma_h$ remained close to optimal.  This resulted in
  an estimated systematic uncertainty in the single-differential cross sections
  of less than $\pm 1\%$ in most bins, increasing to $\sim \pm5\%$ in a few
  high-$\qq$ bins. For $d^2\sigma/dxdQ^2$, the effect is generally
  below $\pm 2\%$ at low and medium $Q^2$, but is relevant at low $\qq$ and low $y$ due
  to the small statistical uncertainty in this region;
\item FCAL circle cut:
  the FCAL circle cut at $20 \cm$ was varied by $\pm 3 \cm$.  The resulting
  changes in the cross sections were typically below $\pm 1\%$. Only for the
  highest $x$ bins of the double-differential cross section did the effect
  increase to $\pm 4\%$;
\item background estimation: 
  \begin{itemize}

  \item systematic uncertainties arising from the normalisation of the
  photoproduction background were estimated by changing the background
  normalisation by a factor of $\pm 40\%$, resulting in negligible changes in
  the single-differential cross sections over the full kinematic range and
  variations of less than $\pm 1\%$ in the double-differential bins;

  \item the cut on the distance of closest approach between the extrapolated
  positron track and the calorimeter cluster associated with the positron was
  changed to $8 \cm$ to estimate the background contamination from wrongly
  identified positrons. The uncertainties in the cross sections associated with
  this variation were below $\pm 1\%$ over the full kinematic range and small
  compared to the statistical uncertainty;
    
  \item the uncertainty due to ``overlay'' events, in which a normal DIS event
  coincided with additional energy deposits in the RCAL from some other interaction,
  was estimated by narrowing or widening the $38\gev < \delta < 65\gev$ interval 
  symmetrically by $\pm 4\gev$. The effect on the cross sections was typically below 1\%; in a
  few high-$\qq$ double-differential bins, the uncertainty was as large as 6\%
  but nevertheless small compared to the statistical uncertainty;

  \item the systematic uncertainty associated with the cosmic-ray rejection was
  evaluated by varying the $P_T/\sqrt{E_T}$ cut by $\pm 1 \sqrt{\Gev}$. The
  cross-section uncertainties were below $\pm 1\%$ over the full kinematic
  range;
  \end{itemize}

\item diffractive contribution:
  the fraction of diffractive events was varied within the uncertainty
  determined from the fit described in \sect{MC}. The resulting uncertainties were
  typically below $\pm 1\%$.
\end{itemize}

The positron identification efficiency
was checked with a data sample of NC DIS events selected using independent
requirements~\cite{thesis:moritz:2001}.  The efficiency curves from data and MC
simulation agreed to better than $0.5\%$. An alternative positron-finding
algorithm \cite{epj:c21:443} was also used: differences in the measured cross
sections were less than 0.5\%. Systematic uncertainties from both of these effects
were neglected.

\subsubsection{Correlated systematic uncertainties}

The following systematic uncertainties were correlated bin-to-bin:
\begin{itemize}
\item $\{\delta_1\}$ positron energy scale: the uncertainty in the
  positron energy scale (as described in \sect{RecoEFinder}) resulted in
  systematic variations in the $d\sigma/dy$ cross section that were comparable to
  the statistical uncertainty at high $y$ and small elsewhere;

\item $\{\delta_2\}$ background estimation: systematic
  uncertainties arising from the estimation of the photoproduction background
  were also estimated by reducing the cut on $y_e$ to $y_e < 0.9$. The resulting
  changes in the cross sections were typically below $\pm 2\%$. In the
  highest-$Q^2$ region at low $x$, where the statistics were low, an average
  uncertainty of $-0.3\%$ was estimated;

\item $\{\delta_3\}$ variation of selection cuts (I): varying the
  positron isolation requirement by $\pm2\gev$ caused a small systematic
  uncertainty in the cross sections at the lower end of the $\qq$ range and up
  to $\pm 4\%$ in the highest-$\qq$ region, where it was still small compared to
  the statistical uncertainty;

\item $\{\delta_4\}$ variation of selection cuts (II): the MC description of the
  positron momentum as measured from the positron track, $p_\trk$, was not perfect. Varying
  the $p_\trk$ requirement by $\pm5 \gev$ resulted in a variation of the
  cross section of the order $\pm 2\%$ over most of the kinematic
  range. The effect was comparable to the statistical uncertainty in a few
  double-differential bins at low $x$ and low $\qq$ and in low-$x$ or high-$y$
  bins in the single-differential cross sections;

\item $\{\delta_5\}$ vertex distribution:
  the uncertainty in the cross sections arising from the measurement of the shape of the
  distribution of the $Z$ coordinate of the event vertex was obtained by varying
  the contribution of events from the satellite bunches, visible as small peaks
  at $|Z|>50 \cm$ in \fig{CtrlPlts}d, within their uncertainties in the MC
  simulation. The effect on the cross sections was within $\sim \pm 0.5\%$ and
  approximately constant over the full kinematic range;

\item $\{\delta_6\}$ uncertainty in the parton-shower scheme:   
  the systematic uncertainty arising from the choice of parton-shower scheme was
  estimated by using the MEPS model of {\sc lepto} to calculate the
  acceptance instead of {\sc ariadne}. The upper and lower limits of the
  systematic uncertainty were determined by studies of
  the hadronic energy flow comparing both MC models with
  data~\cite{thesis:goncalo:2003}. The uncertainty was comparable to the
  statistical uncertainty in a few single-differential bins at high $x$ or low
  $y$. It was also significant in a few low-$\qq$ double-differential bins at
  high $x$, where the statistical uncertainty is small;

  \item $\{\delta_7\}$ formation of hadronic-energy clusters in the
  neighbourhood of the FCAL beam-hole: particles created between the current jet
  and the proton remnant can leave large energy deposits in the forward
  calorimeter.  Uncertainties in the simulation of the energy flow lead to
  differences between the reconstructed $\gamma_h$ in the data and in the
  simulation, especially at low $y$.  To estimate the systematic uncertainty
  associated with this effect, the algorithm employed in the measurement of
  $\gamma_h$ was modified.  In the modified algorithm, energy clusters
  reconstructed in the forward calorimeter within 30~cm of the beam line were
  split into their constituent cells and the hadronic quantities recalculated.
  The effect of the modified algorithm was to give higher values of $\gamma_h$
  in the affected region. The uncertainty obtained was generally small but
  became comparable to the statistical uncertainty at high $x$ or low $y$ in
  the single-differential cross sections and in high-$x$ double-differential bins
  for $\qq$ smaller than around $650 \gev^2$;

\item $\{\delta_8\}$ choice of parton distribution functions:
  the NC MC events were generated with CTEQ5D PDFs. A set of parton density
  functions obtained from a ZEUS NLO QCD fit~\cite{pr:d67:012007}, denoted by
  ZEUS-S, was used to examine the influence of variations in the PDFs on the
  cross-section measurement. The uncertainties associated with the fit were used to obtain
  sets of PDFs that correspond to upper and lower uncertainties on the ZEUS-S
  PDF set. Monte Carlo events were re-weighted to the nominal ZEUS-S parton
  densities and also to the PDFs corresponding to the fit uncertainties, and the
  cross-section extraction was repeated. The differences between the cross
  sections obtained using the upper- and lower-uncertainty PDFs and the nominal
  ZEUS-S parton density function set was taken as the systematic uncertainty
  arising from the choice of PDFs.  The resulting uncertainty was smaller than 
  $1\%$ over the full kinematic range.

\end{itemize}

\subsection{Single-differential cross sections}
\label{sec-Single}

\begin{figure}
  \begin{center}
    \includegraphics[width=.9\textwidth]{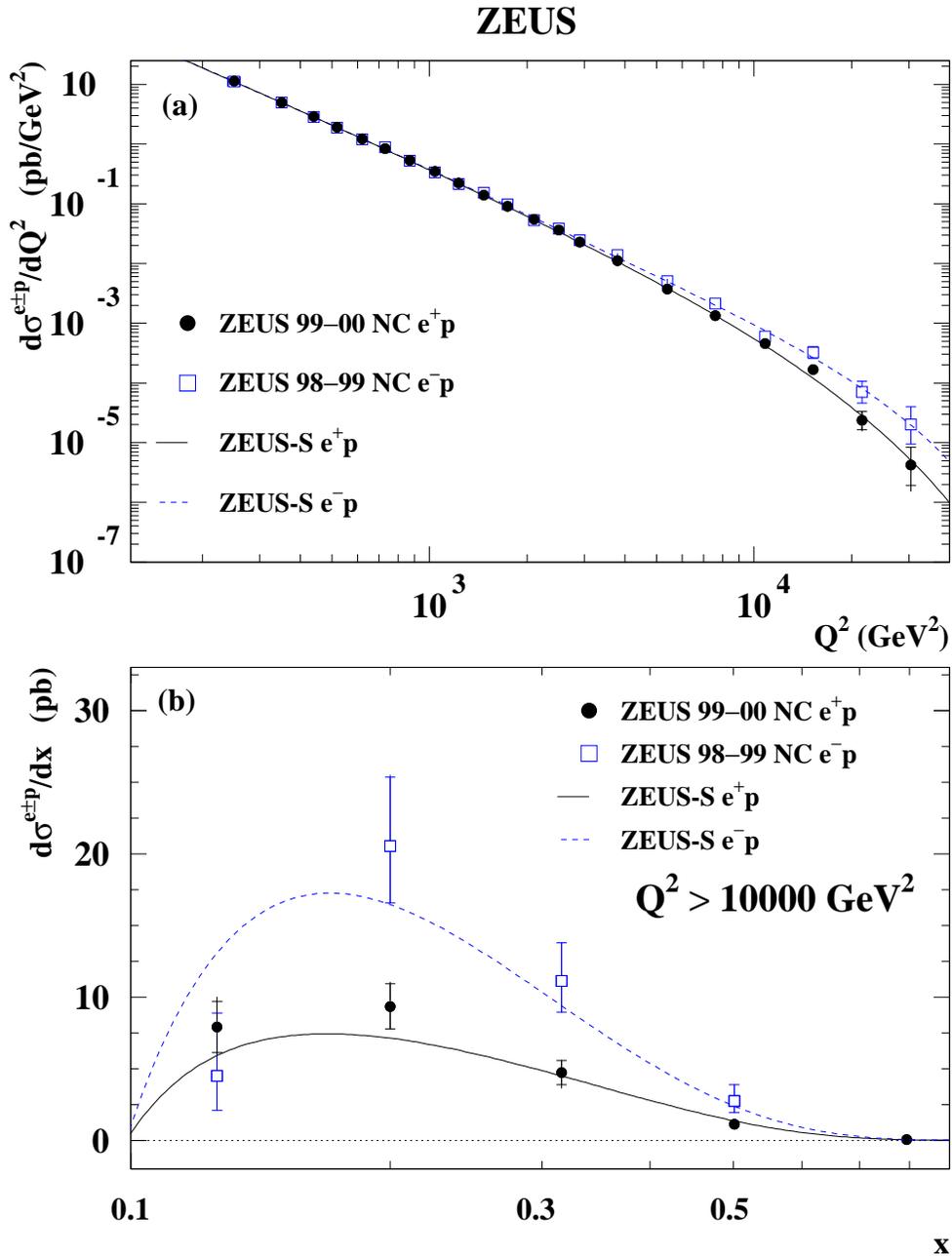}
  \end{center}
  \caption{
    (a) The differential $e^\pm p$ cross-section $d\sigma/dQ^2$
    compared to the Standard Model expectation evaluated using the 
    ZEUS-S PDFs.
    (b) The differential $e^+p$ cross-section $d\sigma/dx$ for
    $Q^2>10\,000\gev^2$ as a function of $x$. 
    The inner bars show the statistical uncertainty, while the
    outer ones show the statistical and systematic uncertainties added
    in quadrature.
  }
  \label{fig-dsdQ2}
\end{figure}

The single-differential cross section $d\sigma/ d Q^2$ is shown in
\fig{dsdQ2}a and tabulated in \tab{dsdQ2}. 
The systematic uncertainties are collected in \tab{dsdQ2_c}.  The SM cross
section, evaluated using the ZEUS-S PDFs, gives a good description of the data.
The figure also shows the recent ZEUS measurement of $d\sigma / d Q^2$ in $e^-p$ NC
DIS, which was also obtained at a centre-of-mass energy of 318~GeV~\cite{epj:c28:175}.
For $Q^2 \gtrsim 3\,000$~GeV$^2$, the $e^-p$ cross section is
larger than that for $e^+p$.  The relative enhancement of the $e^-p$ cross
section over that for $e^+p$ is also clearly demonstrated in \fig{dsdQ2}b, where
$d\sigma/dx$ is plotted for $\qq > 10\,000 \gev^2$ for both $e^+p$ and $e^-p$ NC
DIS.  This effect is due to the parity-violating part of the $Z$-exchange
contribution which enhances (suppresses) the $e^-p$ ($e^+p$) cross section over
that expected under the assumption of single-photon exchange.
 
\begin{figure}
  \begin{center}
\includegraphics[width=.8\textwidth]{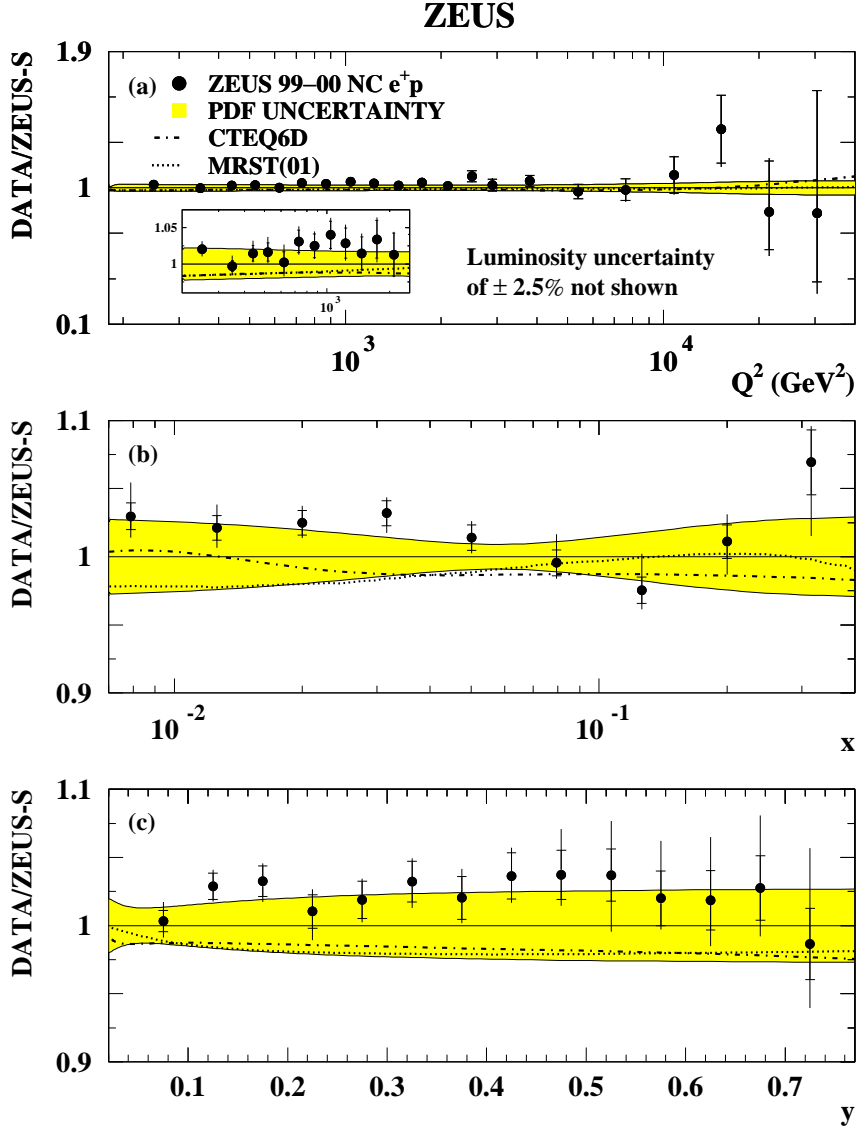}
  \end{center}
  \caption{ 
    Ratios of the single-differential $e^+p$ cross sections to the 
    Standard Model expectation evaluated using the ZEUS-S PDFs:  
    (a) $d\sigma/dQ^2$ (the inset shows the low $Q^2$ region); 
    (b) $d\sigma/dx$ for $Q^2 > 200 \gev^2$,
    and (c) $d\sigma/dy$ for $Q^2 > 200 \gev^2$.
    The shaded band indicates the uncertainty on the calculated cross
    sections due to the uncertainty in the ZEUS-S PDFs.  
    The inner bars show the statistical uncertainty, while the
    outer ones show the statistical and systematic uncertainties added
    in quadrature. 
    The results obtained using the CTEQ6D and the MRST(01) PDFs are
    shown as the dash-dotted and dotted lines respectively.
  }
  \label{fig-ratios}
\end{figure}

The ratio of $d\sigma/dQ^2$ to the cross section obtained
using the ZEUS-S PDFs, as well as the ratios for $d\sigma/dx$ and $d\sigma/dy$
(both for $Q^2 > 200 \gev^2$) are shown in \fig{ratios}. The plots also contain 
the SM predictions using the CTEQ6D~\cite{jhep:07:012} And MRST(01)~\cite{epj:c23:73} 
PDF sets. The data are well described by the SM using the 
ZEUS-S PDFs but systematically higher than the predictions of CTEQ6D and 
MRST(01), although consistent given the luminosity uncertainty of $\pm 2.5\%$.  
The cross-sections $d\sigma/dx$ and $d\sigma/dy$ are tabulated in
\taband{dsdx}{dsdy} (with systematic uncertainties listed in
\taband{dsdx_c}{dsdy_c}). 

\subsection{Reduced cross section and the structure function \boldmath{$F_2$}}
\label{sec-Double}

\begin{figure}
  \begin{center}
    \includegraphics[width=.9\textwidth]{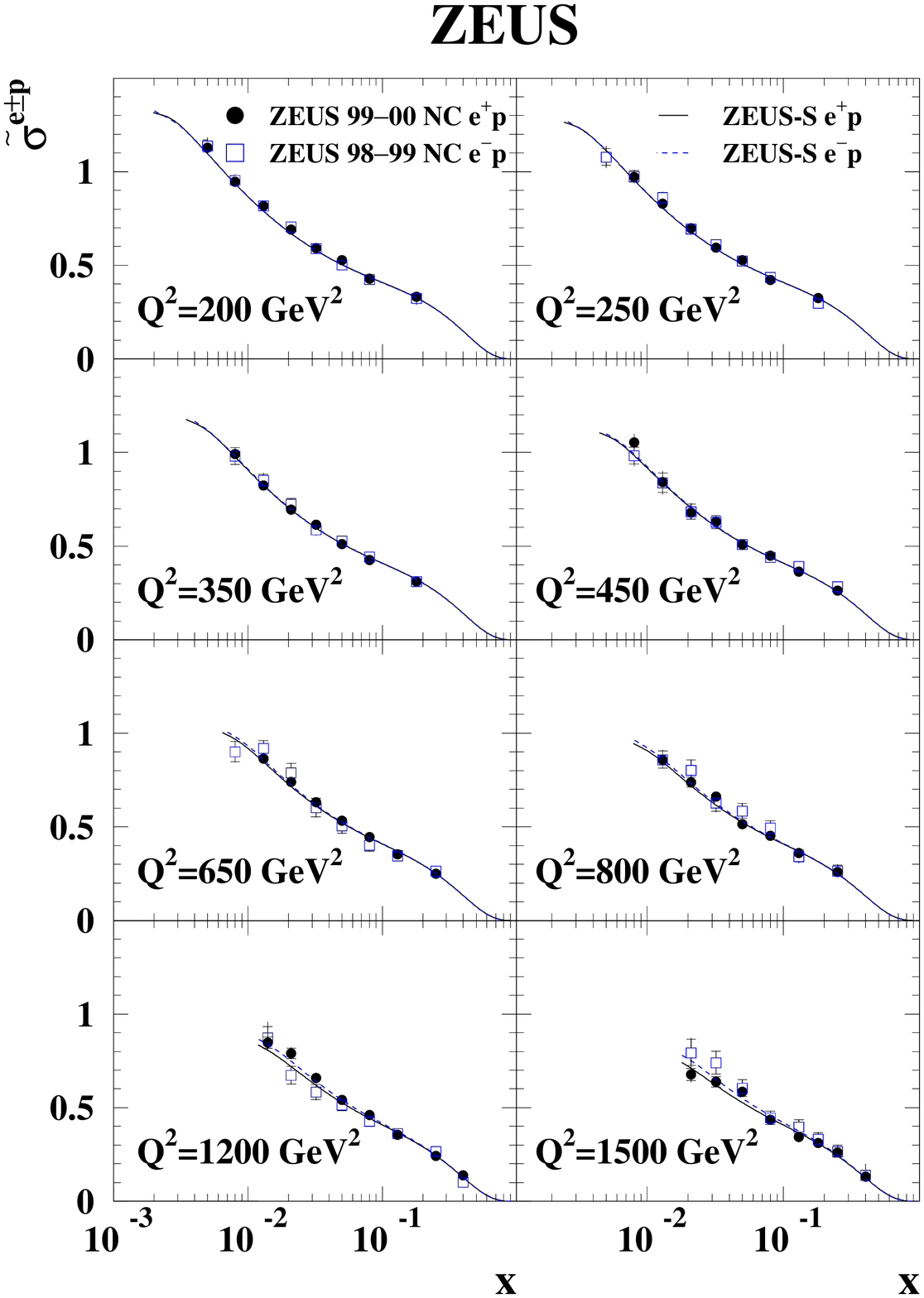}
  \end{center}
  \caption{
    The $e^+ p$ reduced cross section, $\tilde{\sigma}^{e^+ p}$, (solid
    points) plotted as a function of $x$ at fixed $Q^2$ between 
    $200 \gev^2$ and $1\,500 \gev^2$ compared  to $\tilde{\sigma}^{e^-
    p}$ (open squares).  
    The inner bars show the statistical uncertainty, while the
    outer ones show the statistical and systematic uncertainties added
    in quadrature.
    The Standard Model expectations, evaluated using the ZEUS-S PDFs,
    are shown as the solid ($e^+p$) and dashed ($e^-p$) lines.
  }
  \label{fig-RedCross1}
\end{figure}

The reduced cross section, $\tilde{\sigma}^{e^+ p}$, tabulated in
\taband{dsdqdx_1}{dsdqdx_2} (with systematic uncertainties listed in
\taband{dsdxq_c1}{dsdxq_c2}), is shown in
\figand{RedCross1}{RedCross2} as a function of $x$ for various values
of $Q^2$. The rise of $\tilde{\sigma}^{e^+p}$ at fixed $Q^2$ as $x$ decreases
reflects the strong rise of $F_2$\cite{epj:c21:443}.  The SM gives a good
description of the data.  Also shown are the ZEUS measurements of
$\tilde{\sigma}^{e^-p}$.  For $Q^2 \lesssim 3000 \gev^2$, the reduced
cross sections $\tilde{\sigma}^{e^-p}$ and $\tilde{\sigma}^{e^+p}$ are
approximately equal.  For $Q^2 \gtrsim 3000 \gev^2$, the $Z$-boson-exchange
contribution causes $\tilde{\sigma}^{e^+p}$ to be smaller than
$\tilde{\sigma}^{e^-p}$.

To compare the present data to measurements
from other experiments, the structure function $F_2^{\rm em}$ was extracted from the
present data. This was combined with a previous ZEUS measurement of $F_2^{em}$
obtained from data collected at $\sqrt{s}=300 \gev$~\cite{epj:c21:443} 
in 1996 and 1997.

The reduced cross section includes transverse- and longitudinal-photon as well
as $Z$-boson contributions, which can be expressed as relative corrections in
the following way:
\begin{equation}
  \tilde{\sigma}^{e^+p} = F_2^{em}
  \left( 1 + \Delta_{F_2} + \Delta_{xF_3} + \Delta_{F_L} \right) 
                        = F_2^{em} ( 1 + \Delta_{\rm all} ), \nonumber
\end{equation}
where $\Delta_{F_2}$, $\Delta_{xF_3}$, and
$\Delta_{F_L}$ correspond to corrections necessary to account for the weak
contribution to $F_2$ and the contributions of the $xF_3$ and $F_L$ structure
functions to the cross section, respectively. The structure function $F_2^{\rm em}$
was obtained by correcting $\tilde{\sigma}^{e^+p}$ for the relative
contributions, $\Delta_{\rm all}$, using the CTEQ5D PDFs. The size of the 
corrections $\Delta_{\rm all}$ was typically less than 1\% but became as large 
as 50\% at the highest $Q^{2}$. The values of $F_2^{\rm em}$
obtained at the two different centre-of-mass energies were combined using:
\begin{equation}
  F_2^{\rm em} = \frac{ {\cal{L}}_{96/97} F^{\rm em}_{2,96/97} + {\cal{L}}_{99/00}
   F^{\rm em}_{2,99/00} } {{\cal{L}}_{96/97}+{\cal{L}}_{99/00}}, \nonumber
\end{equation}
where the subscripts on the luminosities ($\cal{L}$) and the measured values of
$F_2^{\rm em}$ indicate the data-taking periods to which the values correspond.  The
uncertainties are dominated by the statistical uncertainty. Therefore,
correlations between systematic uncertainties were not taken into account when
evaluating the uncertainty on the combined $F_2^{\rm em}$. The separate ZEUS
measurements of $F_2^{\rm em}$ from data collected at
$\sqrt{s} = 300 \gev$ and $\sqrt{s} = 318 \gev$ were found to be consistent 
within their uncertainties.

\begin{figure}
  \begin{center}
    \includegraphics[width=.85\textwidth]{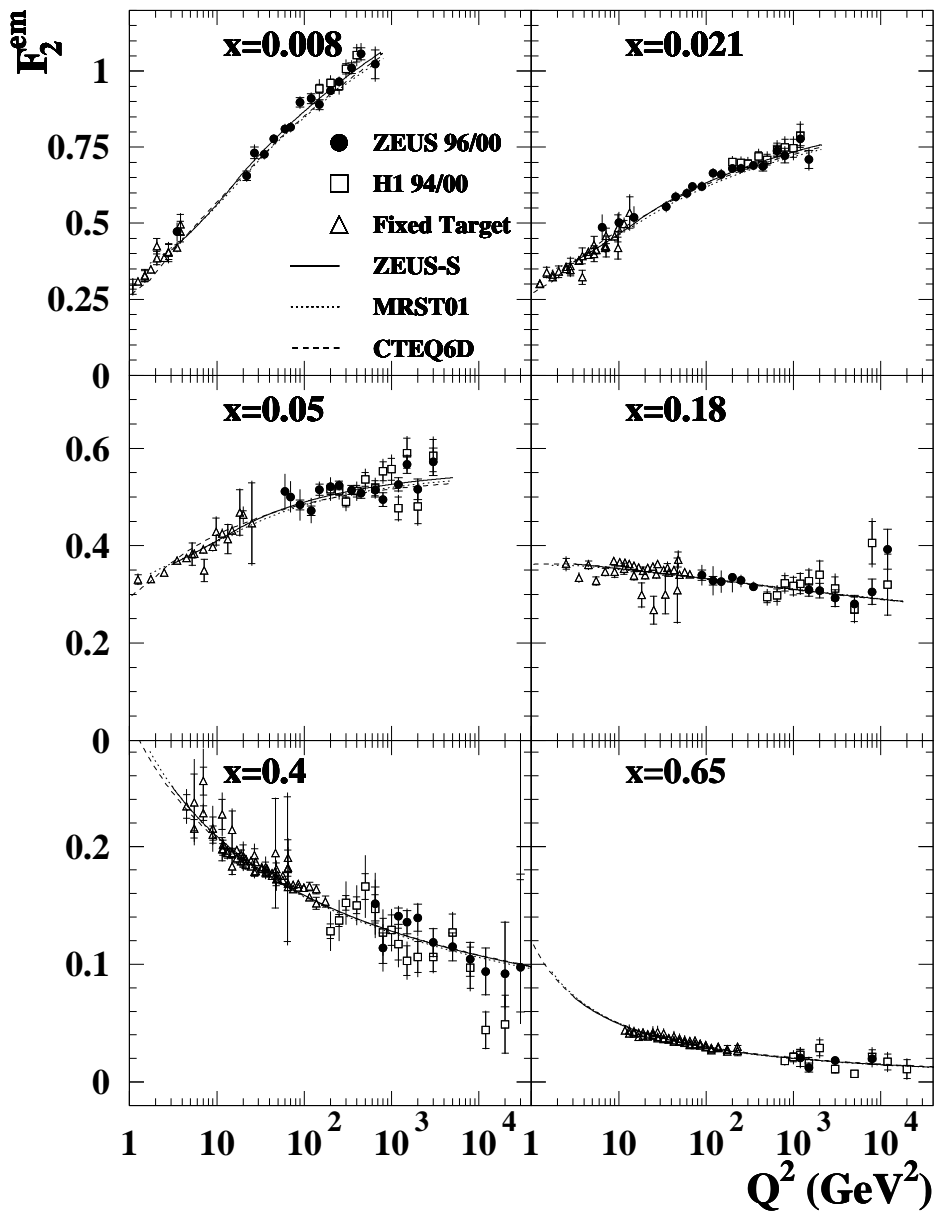}
  \end{center}
  \caption{
    The structure function $F_2^{\rm em}$ obtained by combining the data
    presented here with the previous ZEUS measurements as described
    in the text.
    The inner bars show the statistical uncertainty, while the 
    outer ones show the statistical and systematic uncertainties added
    in quadrature.    
    The results of the fixed-target experiments NMC, BCDMS and E665
    are plotted as the open triangles while those of the H1 experiment
    are shown as the open squares.
   }
  \label{fig-F2}
\end{figure}

\Fig{F2} shows the combined $F_2^{\rm em}$ plotted as a function of
$Q^2$ for several values of $x$.   The results agree
well with those obtained by the H1 collaboration~\cite{epj:c13:609,epj:c21:33}
and with the predictions obtained using the ZEUS-S, CTEQ6D and MRST01 PDFs.  The
results are also in good agreement with the results obtained at lower $Q^2$ in
fixed-target experiments
\cite{np:b483:3,pl:b223:485,pr:d54:3006}. The combined 1996 to 2000 data set
corresponds to a luminosity of 93.2~pb$^{-1}$ which is a factor of $\sim 3$
larger than the luminosity of the previously published data set.

%
%
\section{Summary}
\label{Sect:Summary}

The cross sections for neutral current deep inelastic scattering, $e^+p
\rightarrow e^+X$, have been measured using $63.2\pb^{-1}$ of data collected
with the ZEUS detector during 1999 and 2000.  The
single-differential cross-sections $d\sigma / dQ^2$, $d\sigma / dx$ and $d\sigma
/ dy$ have been measured for $Q^2 > 200 \gev^2$. The effect
of $Z$-boson exchange can be clearly seen in $d\sigma / dx$ measured for $Q^2 >
10\,000 \gev^2$.  The reduced cross section has been measured in the kinematic
range $200 \, \gevv< Q^2 < 30\,000 \,\gevv$ and $0.005 < x < 0.65$.  The Standard
Model predictions including both $\gamma$ and $Z$ exchange and using the 
parton density functions CTEQ6D, ZEUS-S and MRST(01), are in good
agreement with the data. The proton structure function $F_2^{\rm em}$ was extracted using
the combined $e^+ p$ data sample of $93.2 \pb ^{-1}$ taken between 1996 and 2000.

%
\section{Acknowledgements}

We would like to thank the DESY Directorate for their strong support and
encouragement. The remarkable achievements of the HERA machine group were vital
for the successful completion of this work and are greatly appreciated. We are
grateful for the support of the DESY computing and network services. We are
indebted to R. Roberts and M. Seymour for valuable help in the verification of
the Standard Model calculations used in the extraction of the cross sections.

\vfill\eject
{
\def\bibname{\Large\bf References}
\def\refname{\Large\bf References}
\pagestyle{plain}
\ifzeusbst
  \bibliographystyle{./BiBTeX/bst/l4z_default}
\fi
\ifzdrftbst
  \bibliographystyle{./BiBTeX/bst/l4z_draft}
\fi
\ifzbstepj
  \bibliographystyle{./BiBTeX/bst/l4z_epj}
\fi
\ifzbstnp
  \bibliographystyle{./BiBTeX/bst/l4z_np}
\fi
\ifzbstpl
  \bibliographystyle{./BiBTeX/bst/l4z_pl}
\fi
{\raggedright
\bibliography{./BiBTeX/user/syn,%
              ./BiBTeX/bib/l4z_articles,%
              ./BiBTeX/bib/l4z_books,%
              ./BiBTeX/bib/l4z_conferences,%
              ./BiBTeX/bib/l4z_h1,%
              ./BiBTeX/bib/l4z_misc,%
              ./BiBTeX/bib/l4z_old,%
              ./BiBTeX/bib/l4z_preprints,%
              ./BiBTeX/bib/l4z_replaced,%
              ./BiBTeX/bib/l4z_temporary,%
              ./BiBTeX/bib/l4z_zeus}}
}
\vfill\eject

%
%
\clearpage
\begin{table} [!ht] 
  \begin{center}
    {\footnotesize
\renewcommand{\arraystretch}{1.2}
\begin{tabular}{|r@{ -- }r|r|l@{}l@{$\,$}l@{$\,$}l@{$\,$}l|}
\hline
\multicolumn{2}{|c|}{ $Q^2$ range ($\Gev^2$)}&
\multicolumn{1}{c|}{$Q^2_c$ ($\Gev^2$)} & 
\multicolumn{5}{c|} {$d\sigma / dQ^2\ (\! \pb \, / \gev^2$)} \\
\hline
\hline
$  200.0$ & $  300.0$ & $  250$ && $11.310$ & $\pm 0.055$ & $^{+0.107}_{-0.097}$ & $
$ \\
$  300.0$ & $  400.0$ & $  350$ && $ 4.932$ & $\pm 0.037$ & $^{+0.063}_{-0.045}$ & $
$ \\
$  400.0$ & $  475.7$ & $  440$ && $ 2.880$ & $\pm 0.031$ & $^{+0.038}_{-0.019}$ & $
$ \\
$  475.7$ & $  565.7$ & $  520$ && $ 1.917$ & $\pm 0.024$ & $^{+0.031}_{-0.018}$ & $
$ \\
$  565.7$ & $  672.7$ & $  620$ && $ 1.225$ & $\pm 0.018$ & $^{+0.024}_{-0.015}$ & $
$ \\
$  672.7$ & $  800.0$ & $  730$ && $\left( \right.\!  8.39$ & $\pm  0.13$ & $^{+ 0.12}_{- 0.08}$ & $ \left. \!\!\!\! \right) \cdot 10^{-1}
$ \\
$  800.0$ & $  951.4$ & $  870$ && $\left( \right.\!  5.38$ & $\pm  0.09$ & $^{+ 0.06}_{- 0.04}$ & $ \left. \!\!\!\! \right) \cdot 10^{-1}
$ \\
$  951.4$ & $ 1131.0$ & $ 1040$ && $\left( \right.\!  3.47$ & $\pm  0.06$ & $^{+ 0.05}_{- 0.03}$ & $ \left. \!\!\!\! \right) \cdot 10^{-1}
$ \\
$ 1131.0$ & $ 1345.0$ & $ 1230$ && $\left( \right.\!  2.24$ & $\pm  0.05$ & $^{+ 0.03}_{- 0.03}$ & $ \left. \!\!\!\! \right) \cdot 10^{-1}
$ \\
$ 1345.0$ & $ 1600.0$ & $ 1470$ && $\left( \right.\!  1.39$ & $\pm  0.03$ & $^{+ 0.02}_{- 0.01}$ & $ \left. \!\!\!\! \right) \cdot 10^{-1}
$ \\
$ 1600.0$ & $ 1903.0$ & $ 1740$ && $\left( \right.\!  9.15$ & $\pm  0.23$ & $^{+ 0.13}_{- 0.16}$ & $ \left. \!\!\!\! \right) \cdot 10^{-2}
$ \\
$ 1903.0$ & $ 2263.0$ & $ 2100$ && $\left( \right.\!  5.46$ & $\pm  0.16$ & $^{+ 0.06}_{- 0.06}$ & $ \left. \!\!\!\! \right) \cdot 10^{-2}
$ \\
$ 2263.0$ & $ 2691.0$ & $ 2500$ && $\left( \right.\!  3.64$ & $\pm  0.12$ & $^{+ 0.04}_{- 0.06}$ & $ \left. \!\!\!\! \right) \cdot 10^{-2}
$ \\
$ 2691.0$ & $ 3200.0$ & $ 2900$ && $\left( \right.\!  2.30$ & $\pm  0.09$ & $^{+ 0.06}_{- 0.03}$ & $ \left. \!\!\!\! \right) \cdot 10^{-2}
$ \\
$ 3200.0$ & $ 4525.0$ & $ 3800$ && $\left( \right.\!  1.11$ & $\pm  0.04$ & $^{+ 0.01}_{- 0.02}$ & $ \left. \!\!\!\! \right) \cdot 10^{-2}
$ \\
$ 4525.0$ & $ 6400.0$ & $ 5400$ && $\left( \right.\!  3.76$ & $\pm  0.18$ & $^{+ 0.07}_{- 0.05}$ & $ \left. \!\!\!\! \right) \cdot 10^{-3}
$ \\
$ 6400.0$ & $ 9051.0$ & $ 7600$ && $\left( \right.\!  1.33$ & $\pm  0.10$ & $^{+ 0.02}_{- 0.05}$ & $ \left. \!\!\!\! \right) \cdot 10^{-3}
$ \\
$ 9051.0$ & $12800.0$ & $10800$ && $\left( \right.\!  4.55$ & $\pm  0.51$ & $^{+ 0.16}_{- 0.18}$ & $ \left. \!\!\!\! \right) \cdot 10^{-4}
$ \\
$12800.0$ & $18100.0$ & $15200$ && $\left( \right.\!  1.65$ & $\pm  0.27$ & $^{+ 0.04}_{- 0.13}$ & $ \left. \!\!\!\! \right) \cdot 10^{-4}
$ \\
$18100.0$ & $25600.0$ & $21500$ && $\left( \right.\!  2.35$ & $^{+  0.94}_{- 0.70}$ & $^{+ 0.31}_{- 0.41}$ & $ \left. \!\!\!\! \right) \cdot 10^{-5}
$ \\
$25600.0$ & $36200.0$ & $30400$ && $\left( \right.\!   4.2$ & $^{+   4.1}_{-  2.3}$ & $^{+  0.3}_{-  1.4}$ & $ \left. \!\!\!\! \right) \cdot 10^{-6}
$ \\
\hline
\end{tabular}
}

  \end{center}    
  \caption[this space for rent]{
    The single-differential cross-section \sigqq\ for the reaction 
    $e^{+} p \rightarrow e^{+} X$.  
    The following quantities are given for each bin: the \qq\ range,
    the value at which the cross section is quoted, \qqc, and the measured
    cross-section \sigqq\ corrected to the electroweak Born level.
    The first uncertainty on the measured cross section is the
    statistical uncertainty and the second is the systematic
    uncertainty. The uncertainty on the measured luminosity of $2.5\%$ 
    is not included in the total systematic uncertainty. 
  } 
  \label{tab-dsdQ2}
\end{table}
\clearpage
\begin{sidewaystable} [!ht]
  \begin{center}
    {\footnotesize
\renewcommand{\arraystretch}{1.2}
\begin{tabular}{|r|l|c|c||c|c|c|c|c|c|c|c|c|}
\hline
\multicolumn{1}{|c|}{$Q^2_c$} & 
\multicolumn{1}{c|}{$d\sigma / dQ^2$} &
stat.  & 
total sys.  & 
uncor. sys. &
$\delta_1$ &
$\delta_2$ &
$\delta_3$ &
$\delta_4$ &
$\delta_5$ &
$\delta_6$ &
$\delta_7$ &
$\delta_8$\\
\multicolumn{1}{|c|}{$(\Gev^2)$} &
\multicolumn{1}{c|}{$(\pb \, / \gev^2)$} &
(\%) & 
(\%) &
(\%) &
(\%) &
(\%) &
(\%) &
(\%) &
(\%) &
(\%) &
(\%) &
(\%) \\
\hline \hline
$  250$ & $
11.310$ &
$^{+ 0.5  } _{-0.5  }  $ & $^{+ 0.9  } _{-0.9  }  $ & $^{+ 0.1  } _{-0.3  }  $ & $^{-0.4  }  _{+ 0.4  } $ & $^{}_{+ 0.0  } $ & $^{- 0.0  } _{- 0.0  } $ & $^{-0.6  }  _{+ 0.7  } $ & $^{-0.3  }  _{+ 0.3  } $ & $^{+ 0.2  } _{-0.5  }  
$                                 & $^{}_{+ 0.1  } $& $^{+ 0.0  } _{- 0.0  } $ \\
$  350$ & $
 4.932$ &
$^{+ 0.7  } _{-0.7  }  $ & $^{+ 1.3  } _{-0.9  }  $ & $^{+ 0.2  } _{-0.2  }  $ & $^{-0.4  }  _{+ 0.4  } $ & $^{}_{+ 0.0  } $ & $^{+ 0.0  } _{+ 0.0  } $ & $^{-0.7  }  _{+ 1.1  } $ & $^{-0.3  }  _{+ 0.3  } $ & $^{+ 0.1  } _{-0.2  }  
$                                 & $^{}_{+ 0.2  } $& $^{+ 0.1  } _{-0.1  }  $ \\
$  440$ & $
 2.880$ &
$^{+ 1.1  } _{-1.1  }  $ & $^{+ 1.3  } _{-0.6  }  $ & $^{+ 0.5  } _{-0.3  }  $ & $^{-0.4  }  _{+ 0.4  } $ & $^{}_{+ 0.0  } $ & $^{- 0.0  } _{+ 0.0  } $ & $^{-0.2  }  _{+ 1.1  } $ & $^{-0.3  }  _{+ 0.3  } $ & $^{+ 0.1  } _{-0.2  }  
$                                 & $^{}_{- 0.0  } $& $^{+ 0.1  } _{-0.1  }  $ \\
$  520$ & $
 1.917$ &
$^{+ 1.2  } _{-1.2  }  $ & $^{+ 1.6  } _{-0.9  }  $ & $^{+ 0.4  } _{-0.6  }  $ & $^{-0.4  }  _{+ 0.4  } $ & $^{}_{- 0.0  } $ & $^{+ 0.0  } _{- 0.0  } $ & $^{-0.5  }  _{+ 1.4  } $ & $^{-0.3  }  _{+ 0.3  } $ & $^{+ 0.2  } _{-0.4  }  
$                                 & $^{}_{+ 0.5  } $& $^{+ 0.0  } _{- 0.0  } $ \\
$  620$ & $
 1.225$ &
$^{+ 1.5  } _{-1.5  }  $ & $^{+ 1.9  } _{-1.2  }  $ & $^{+ 0.3  } _{-0.5  }  $ & $^{-0.4  }  _{+ 0.4  } $ & $^{}_{+ 0.1  } $ & $^{- 0.0  } _{+ 0.0  } $ & $^{-0.5  }  _{+ 1.9  } $ & $^{-0.2  }  _{+ 0.2  } $ & $^{+ 0.1  } _{-0.2  }  
$                                 & $^{}_{-0.9  }  $& $^{- 0.0  } _{+ 0.0  } $ \\
$  730$ & $
 8.393\cdot 10^{-1}$ &
$^{+ 1.5  } _{-1.5  }  $ & $^{+ 1.4  } _{-0.9  }  $ & $^{+ 0.4  } _{-0.4  }  $ & $^{-0.3  }  _{+ 0.3  } $ & $^{}_{+ 0.0  } $ & $^{+ 0.1  } _{+ 0.1  } $ & $^{-0.7  }  _{+ 1.1  } $ & $^{-0.3  }  _{+ 0.3  } $ & $^{+ 0.2  } _{-0.5  }  
$                                 & $^{}_{+ 0.5  } $& $^{+ 0.0  } _{-0.1  }  $ \\
$  870$ & $
 5.377\cdot 10^{-1}$ &
$^{+ 1.6  } _{-1.6  }  $ & $^{+ 1.1  } _{-0.8  }  $ & $^{+ 0.6  } _{-0.5  }  $ & $^{-0.2  }  _{+ 0.3  } $ & $^{}_{-0.2  }  $ & $^{- 0.0  } _{+ 0.1  } $ & $^{- 0.0  } _{+ 0.7  } $ & $^{-0.4  }  _{+ 0.4  } $ & $^{+ 0.1  } _{-0.2  }  
$                                 & $^{}_{-0.3  }  $& $^{+ 0.1  } _{-0.1  }  $ \\
$ 1040$ & $
 3.474\cdot 10^{-1}$ &
$^{+ 1.8  } _{-1.8  }  $ & $^{+ 1.4  } _{-0.9  }  $ & $^{+ 0.2  } _{-0.5  }  $ & $^{-0.2  }  _{+ 0.2  } $ & $^{}_{- 0.0  } $ & $^{+ 0.0  } _{-0.2  }  $ & $^{-0.6  }  _{+ 1.3  } $ & $^{-0.3  }  _{+ 0.3  } $ & $^{+ 0.1  } _{-0.3  }  
$                                 & $^{}_{+ 0.0  } $& $^{+ 0.1  } _{-0.1  }  $ \\
$ 1230$ & $
 2.238\cdot 10^{-1}$ &
$^{+ 2.0  } _{-2.0  }  $ & $^{+ 1.5  } _{-1.2  }  $ & $^{+ 0.4  } _{-0.5  }  $ & $^{-0.1  }  _{+ 0.2  } $ & $^{}_{-0.5  }  $ & $^{+ 0.1  } _{+ 0.1  } $ & $^{-0.8  }  _{+ 1.4  } $ & $^{-0.3  }  _{+ 0.3  } $ & $^{+ 0.0  } _{-0.1  }  
$                                 & $^{}_{+ 0.2  } $& $^{+ 0.1  } _{-0.1  }  $ \\
$ 1470$ & $
 1.394\cdot 10^{-1}$ &
$^{+ 2.2  } _{-2.2  }  $ & $^{+ 1.5  } _{-0.9  }  $ & $^{+ 0.6  } _{-0.5  }  $ & $^{-0.1  }  _{+ 0.2  } $ & $^{}_{+ 0.1  } $ & $^{-0.1  }  _{-0.1  }  $ & $^{-0.6  }  _{+ 1.1  } $ & $^{-0.4  }  _{+ 0.3  } $ & $^{+ 0.0  } _{-0.1  }  
$                                 & $^{}_{+ 0.6  } $& $^{+ 0.1  } _{-0.1  }  $ \\
$ 1740$ & $
 9.154\cdot 10^{-2}$ &
$^{+ 2.5  } _{-2.5  }  $ & $^{+ 1.4  } _{-1.8  }  $ & $^{+ 0.6  } _{-0.7  }  $ & $^{-0.1  }  _{+ 0.1  } $ & $^{}_{+ 0.2  } $ & $^{- 0.0  } _{+ 0.1  } $ & $^{-0.1  }  _{+ 1.2  } $ & $^{-0.4  }  _{+ 0.4  } $ & $^{+ 0.2  } _{-0.5  }  
$                                 & $^{}_{-1.5  }  $& $^{+ 0.1  } _{-0.1  }  $ \\
$ 2100$ & $
 5.458\cdot 10^{-2}$ &
$^{+ 2.9  } _{-2.9  }  $ & $^{+ 1.1  } _{-1.1  }  $ & $^{+ 0.5  } _{-0.8  }  $ & $^{-0.1  }  _{+ 0.2  } $ & $^{}_{+ 0.5  } $ & $^{+ 0.1  } _{+ 0.0  } $ & $^{+ 0.2  } _{+ 0.5  } $ & $^{-0.3  }  _{+ 0.3  } $ & $^{+ 0.2  } _{-0.4  }  
$                                 & $^{}_{-0.5  }  $& $^{+ 0.1  } _{-0.1  }  $ \\
$ 2500$ & $
 3.635\cdot 10^{-2}$ &
$^{+ 3.3  } _{-3.3  }  $ & $^{+ 1.1  } _{-1.6  }  $ & $^{+ 0.5  } _{-0.7  }  $ & $^{-0.1  }  _{+ 0.2  } $ & $^{}_{+ 0.5  } $ & $^{-0.1  }  _{+ 0.1  } $ & $^{+ 0.3  } _{+ 0.6  } $ & $^{-0.3  }  _{+ 0.3  } $ & $^{+ 0.4  } _{-1.0  }  
$                                 & $^{}_{-1.0  }  $& $^{+ 0.1  } _{-0.1  }  $ \\
$ 2900$ & $
 2.298\cdot 10^{-2}$ &
$^{+ 3.8  } _{-3.8  }  $ & $^{+ 2.5  } _{-1.3  }  $ & $^{+ 1.2  } _{-1.0  }  $ & $^{-0.1  }  _{+ 0.2  } $ & $^{}_{-0.3  }  $ & $^{+ 0.1  } _{+ 0.2  } $ & $^{-0.6  }  _{+ 0.8  } $ & $^{-0.3  }  _{+ 0.3  } $ & $^{+ 0.3  } _{-0.8  }  
$                                 & $^{}_{+ 1.7  } $& $^{+ 0.1  } _{-0.1  }  $ \\
$ 3800$ & $
 1.113\cdot 10^{-2}$ &
$^{+ 3.4  } _{-3.4  }  $ & $^{+ 0.5  } _{-1.8  }  $ & $^{+ 0.4  } _{-0.9  }  $ & $^{-0.1  }  _{+ 0.2  } $ & $^{}_{-1.0  }  $ & $^{- 0.0  } _{-0.1  }  $ & $^{+ 0.0  } _{-0.3  }  $ & $^{-0.3  }  _{+ 0.3  } $ & $^{+ 0.0  } _{- 0.0  } 
$                                 & $^{}_{-1.0  }  $& $^{+ 0.0  } _{- 0.0  } $ \\
$ 5400$ & $
 3.756\cdot 10^{-3}$ &
$^{+ 4.8  } _{-4.8  }  $ & $^{+ 2.0  } _{-1.3  }  $ & $^{+ 1.3  } _{-0.5  }  $ & $^{-0.1  }  _{+ 0.1  } $ & $^{}_{-1.0  }  $ & $^{-0.1  }  _{-0.5  }  $ & $^{+ 0.2  } _{+ 0.3  } $ & $^{-0.4  }  _{+ 0.4  } $ & $^{+ 0.3  } _{-0.6  }  
$                                 & $^{}_{+ 1.2  } $& $^{- 0.0  } _{+ 0.0  } $ \\
$ 7600$ & $
 1.331\cdot 10^{-3}$ &
$^{+ 7.2  } _{-7.2  }  $ & $^{+ 1.3  } _{-3.4  }  $ & $^{+ 0.6  } _{-2.2  }  $ & $^{-0.3  }  _{+ 0.2  } $ & $^{}_{-1.0  }  $ & $^{-0.3  }  _{-0.1  }  $ & $^{+ 0.9  } _{+ 0.4  } $ & $^{-0.4  }  _{+ 0.4  } $ & $^{+ 0.2  } _{-0.5  }  
$                                 & $^{}_{-2.3  }  $& $^{- 0.0  } _{+ 0.0  } $ \\
$10800$ & $
 4.551\cdot 10^{-4}$ &
$^{+  11. } _{ -11. }  $ & $^{+ 3.4  } _{-3.9  }  $ & $^{+ 2.8  } _{-2.6  }  $ & $^{-0.4  }  _{+ 0.5  } $ & $^{}_{-1.0  }  $ & $^{-0.9  }  _{-2.3  }  $ & $^{-0.9  }  _{+ 0.8  } $ & $^{-0.4  }  _{+ 0.4  } $ & $^{+ 0.1  } _{-0.3  }  
$                                 & $^{}_{+ 1.8  } $& $^{+ 0.1  } _{-0.1  }  $ \\
$15200$ & $
 1.655\cdot 10^{-4}$ &
$^{+  16. } _{ -16. }  $ & $^{+ 2.1  } _{-8.1  }  $ & $^{+ 0.4  } _{-7.9  }  $ & $^{-0.3  }  _{+ 0.6  } $ & $^{}_{-1.0  }  $ & $^{-1.0  }  _{+ 0.5  } $ & $^{+ 1.9  } _{-0.5  }  $ & $^{-0.3  }  _{+ 0.3  } $ & $^{+ 0.2  } _{-0.4  }  
$                                 & $^{}_{- 0.0  } $& $^{-0.1  }  _{+ 0.0  } $ \\
$21500$ & $
 2.353\cdot 10^{-5}$ &
$^{+  40. } _{ -30. }  $ & $^{+  13. } _{ -18. }  $ & $^{+ 9.6  } _{-9.0  }  $ & $^{-0.5  }  _{+ 0.4  } $ & $^{}_{-1.0  }  $ & $^{-1.1  }  _{ -15. }  $ & $^{+ 3.0  } _{+ 8.2  } $ & $^{-0.4  }  _{+ 0.4  } $ & $^{+ 0.5  } _{-1.1  }  
$                                 & $^{}_{+ 0.9  } $& $^{+ 0.2  } _{-0.2  }  $ \\
$30400$ & $
 4.217\cdot 10^{-6}$ &
$^{+  97. } _{ -54. }  $ & $^{+ 6.5  } _{ -33. }  $ & $^{+ 4.2  } _{ -33. }  $ & $^{-0.7  }  _{+ 0.3  } $ & $^{}_{-1.0  }  $ & $^{-1.1  }  _{+ 3.7  } $ & $^{+ 2.4  } _{-0.5  }  $ & $^{-0.4  }  _{+ 0.4  } $ & $^{+ 0.9  } _{-2.0  }  
$                                 & $^{}_{+ 0.5  } $& $^{+ 0.4  } _{-0.5  }  $ \\
\hline
\end{tabular}
}

  \end{center}    
  \normalsize
  \caption[this space for rent]{
    Systematic uncertainties with bin-to-bin correlations for the
    single-differential cross-section \sigqq. 
    The left part of the table contains the value at which the cross
    section is quoted, \qqc, the measured cross-section \sigqq\
    corrected to the electroweak Born level, the statistical uncertainty and the total
    systematic uncertainty. The uncertainty on the measured luminosity of $2.5\%$ is 
    not included in the total systematic uncertainty.
    The right part of the table lists the total uncorrelated
    systematic uncertainty followed by the bin-to-bin correlated
    systematic uncertainties $\delta_1$--\,$\delta_8$ defined in
    the text. 
    For the latter, the upper (lower) numbers refer to 
    positive (negative) variation of e.g. the cut value, whereas the
    signs of the numbers reflect the direction of change in the cross
    sections. 
    }
  \label{tab-dsdQ2_c}
\end{sidewaystable}
%
%
\clearpage
\begin{table} [!ht] 
\begin{center}
  {\footnotesize
\renewcommand{\arraystretch}{1.2}
\begin{tabular}{|c|r@{ -- }l@{$\,$}r|l@{$\,$}r|l@{}l@{$\,$}l@{$\,$}l@{$\,$}l|}
\hline
{$Q^2$ cut ($\Gev^2$)} & 
\multicolumn{3}{c|}{ $x$ range} & 
\multicolumn{2}{c|}{$x_c$} & 
\multicolumn{5}{c|} {$d\sigma / dx \ $(pb)} \\ 
\hline
\hline
$200$ & $\left( \right.\!0.63$ & $1.00$ & $ \left. \! \right) \cdot 10^{-2}$ & $ 0.790$ & $\,\cdot\,10^{-2}$ &$ \left( \right.$ \hspace{-1mm} & $  8.13$ & $\pm  0.08$ & $^{+ 0.18}_{- 0.10}$ & $ \left. \! \right) \cdot 10^{ 4}
$ \\
      & $\left( \right.\!0.10$ & $0.16$ & $ \left. \! \right) \cdot 10^{-1}$ & $ 0.126$ & $\,\cdot\,10^{-1}$ &$ \left( \right.$ \hspace{-1mm} & $  5.42$ & $\pm  0.05$ & $^{+ 0.08}_{- 0.06}$ & $ \left. \! \right) \cdot 10^{ 4}
$ \\
      & $\left( \right.\!0.16$ & $0.25$ & $ \left. \! \right) \cdot 10^{-1}$ & $ 0.200$ & $\,\cdot\,10^{-1}$ &$ \left( \right.$ \hspace{-1mm} & $  3.38$ & $\pm  0.03$ & $^{+ 0.03}_{- 0.02}$ & $ \left. \! \right) \cdot 10^{ 4}
$ \\
      & $\left( \right.\!0.25$ & $0.40$ & $ \left. \! \right) \cdot 10^{-1}$ & $ 0.316$ & $\,\cdot\,10^{-1}$ &$ \left( \right.$ \hspace{-1mm} & $  2.03$ & $\pm  0.02$ & $^{+ 0.01}_{- 0.02}$ & $ \left. \! \right) \cdot 10^{ 4}
$ \\
      & $\left( \right.\!0.40$ & $0.63$ & $ \left. \! \right) \cdot 10^{-1}$ & $ 0.501$ & $\,\cdot\,10^{-1}$ &$ \left( \right.$ \hspace{-1mm} & $  1.15$ & $\pm  0.01$ & $^{+ 0.01}_{- 0.01}$ & $ \left. \! \right) \cdot 10^{ 4}
$ \\
      & $\left( \right.\!0.63$ & $1.00$ & $ \left. \! \right) \cdot 10^{-1}$ & $ 0.794$ & $\,\cdot\,10^{-1}$ &$ \left( \right.$ \hspace{-1mm} & $  6.44$ & $\pm  0.06$ & $^{+ 0.12}_{- 0.05}$ & $ \left. \! \right) \cdot 10^{ 3}
$ \\
      & $0.10$ & $0.16$ &  & $ 0.126$ &  &$ \left( \right.$ \hspace{-1mm} & $  3.49$ & $\pm  0.03$ & $^{+ 0.09}_{- 0.04}$ & $ \left. \! \right) \cdot 10^{ 3}
$ \\
      & $0.16$ & $0.25$ &  & $ 0.200$ &  &$ \left( \right.$ \hspace{-1mm} & $  1.87$ & $\pm  0.02$ & $^{+ 0.03}_{- 0.04}$ & $ \left. \! \right) \cdot 10^{ 3}
$ \\
      & $0.25$ & $0.40$ &  & $ 0.316$ &  &$ \left( \right.$ \hspace{-1mm} & $  8.47$ & $\pm  0.19$ & $^{+ 0.09}_{- 0.38}$ & $ \left. \! \right) \cdot 10^{ 2}
$ \\
\hline
$10\,000$ & $0.10$ & $0.16$ &  & $ 0.126$ &  && $  7.91$ & $\pm  1.78$ & $^{+ 1.15}_{- 0.95}$ & $
$ \\
          & $0.16$ & $0.25$ &  & $ 0.200$ &  && $  9.35$ & $\pm  1.57$ & $^{+ 0.66}_{- 0.34}$ & $
$ \\
          & $0.25$ & $0.40$ &  & $ 0.316$ &  && $  4.72$ & $\pm  0.84$ & $^{+ 0.25}_{- 0.74}$ & $
$ \\
          & $0.40$ & $0.63$ &  & $ 0.501$ &  && $  1.13$ & $\pm  0.31$ & $^{+ 0.05}_{- 0.14}$ & $
$ \\
          & $0.63$ & $1.00$ &  & $ 0.794$ &  && $  0.05$ & $^{+  0.06}_{- 0.03}$ & $^{+ 0.00}_{- 0.03}$ & $
$ \\
\hline
\end{tabular}
}

\end{center}    
  \caption[this space for rent]{
    The single-differential cross-section \sigx\ for the reaction 
    $e^{+} p \rightarrow e^{+} X$. 
    The following quantities are given for each bin: the lower $Q^2$
    cut, the $x$ range, the value at which the cross section is
    quoted, $x_c$, and the measured cross-section \sigx\ corrected to the
    electroweak Born level.
    The first uncertainty on the measured cross section is the
    statistical uncertainty and the second is the systematic uncertainty. 
    The uncertainty on the measured luminosity of $2.5\%$ is not 
    included in the total systematic uncertainty.
  }
  \label{tab-dsdx}
\end{table}
\clearpage
\begin{sidewaystable} [!ht] 
  \begin{center}
    \hspace*{-5.5mm}{\footnotesize
\renewcommand{\arraystretch}{1.2}
\begin{tabular}{|c|l|l|c|c||c|c|c|c|c|c|c|c|c|}
\hline
{$Q^2$ cut} & 
\multicolumn{1}{c|}{$x_c$} & 
\multicolumn{1}{c|}{$d\sigma / dx$} &
stat.  & 
total sys.  & 
uncor. sys. &
$\delta_1$ &
$\delta_2$ &
$\delta_3$ &
$\delta_4$ &
$\delta_5$ &
$\delta_6$ &
$\delta_7$ &
$\delta_8$ \\
{($\Gev^2$)} & 
$$ &
\multicolumn{1}{c|}{(pb)} &
(\%) & 
(\%) &
(\%) &
(\%) &
(\%) &
(\%) &
(\%) &
(\%) &
(\%) &
(\%) &
(\%) \\
\hline \hline
$200$ & $0.790\,\cdot\,10^{-2}$ & $
  8.13\cdot 10^{ 4}$ &
$^{+ 0.9  } _{-0.9  }  $ & $^{+ 2.2  } _{-1.2  }  $ & $^{+ 0.7  } _{-0.2  }  $ & $^{-0.4  }  _{+ 0.4  } $ & $^{}_{+ 0.0  } $ & $^{+ 0.0  } _{+ 0.1  } $ & $^{-0.9  }  _{+ 2.1  } $ & $^{-0.3  }  _{+ 0.3  } $ & $^{+ 0.1  } _{-0.2  }  
$                                 & $^{}_{-0.2  }  $& $^{+ 0.0  } _{- 0.0  } $ \\
      & $0.126\,\cdot\,10^{-1}$ & $
  5.42\cdot 10^{ 4}$ &
$^{+ 0.9  } _{-0.9  }  $ & $^{+ 1.5  } _{-1.1  }  $ & $^{+ 0.5  } _{-0.2  }  $ & $^{-0.1  }  _{+ 0.2  } $ & $^{}_{-0.1  }  $ & $^{- 0.0  } _{- 0.0  } $ & $^{-0.8  }  _{+ 1.3  } $ & $^{-0.3  }  _{+ 0.3  } $ & $^{+ 0.3  } _{-0.6  }  
$                                 & $^{}_{+ 0.3  } $& $^{- 0.0  } _{+ 0.0  } $ \\
      & $0.200\,\cdot\,10^{-1}$ & $
  3.38\cdot 10^{ 4}$ &
$^{+ 0.9  } _{-0.9  }  $ & $^{+ 0.8  } _{-0.7  }  $ & $^{+ 0.4  } _{-0.2  }  $ & $^{-0.2  }  _{+ 0.1  } $ & $^{}_{+ 0.1  } $ & $^{+ 0.0  } _{+ 0.0  } $ & $^{-0.2  }  _{+ 0.6  } $ & $^{-0.3  }  _{+ 0.3  } $ & $^{+ 0.1  } _{-0.3  }  
$                                 & $^{}_{-0.1  }  $& $^{-0.1  }  _{+ 0.1  } $ \\
      & $0.316\,\cdot\,10^{-1}$ & $
  2.03\cdot 10^{ 4}$ &
$^{+ 0.9  } _{-0.9  }  $ & $^{+ 0.7  } _{-1.1  }  $ & $^{+ 0.4  } _{-0.3  }  $ & $^{-0.1  }  _{+ 0.2  } $ & $^{}_{- 0.0  } $ & $^{- 0.0  } _{+ 0.0  } $ & $^{-0.6  }  _{+ 0.2  } $ & $^{-0.3  }  _{+ 0.3  } $ & $^{+ 0.2  } _{-0.4  }  
$                                 & $^{}_{-0.8  }  $& $^{-0.2  }  _{+ 0.2  } $ \\
      & $0.501\,\cdot\,10^{-1}$ & $
  1.15\cdot 10^{ 4}$ &
$^{+ 0.9  } _{-0.9  }  $ & $^{+ 0.7  } _{-0.7  }  $ & $^{+ 0.6  } _{-0.1  }  $ & $^{-0.1  }  _{+ 0.2  } $ & $^{}_{- 0.0  } $ & $^{- 0.0  } _{+ 0.0  } $ & $^{-0.3  }  _{+ 0.3  } $ & $^{-0.3  }  _{+ 0.3  } $ & $^{+ 0.0  } _{-0.1  }  
$                                 & $^{}_{+ 0.2  } $& $^{-0.3  }  _{+ 0.3  } $ \\
      & $0.794\,\cdot\,10^{-1}$ & $
  6.44\cdot 10^{ 3}$ &
$^{+ 0.9  } _{-0.9  }  $ & $^{+ 1.9  } _{-0.7  }  $ & $^{+ 0.3  } _{-0.2  }  $ & $^{-0.2  }  _{+ 0.2  } $ & $^{}_{- 0.0  } $ & $^{+ 0.0  } _{- 0.0  } $ & $^{-0.1  }  _{+ 0.4  } $ & $^{-0.3  }  _{+ 0.3  } $ & $^{+ 0.5  } _{-1.1  }  
$                                 & $^{}_{+ 1.3  } $& $^{-0.2  }  _{+ 0.2  } $ \\
      & $0.126$ & $
  3.49\cdot 10^{ 3}$ &
$^{+ 1.0  } _{-1.0  }  $ & $^{+ 2.5  } _{-1.0  }  $ & $^{+ 0.3  } _{-0.4  }  $ & $^{-0.2  }  _{+ 0.2  } $ & $^{}_{+ 0.0  } $ & $^{- 0.0  } _{- 0.0  } $ & $^{-0.3  }  _{+ 0.7  } $ & $^{-0.3  }  _{+ 0.3  } $ & $^{+ 0.8  } _{-1.9  }  
$                                 & $^{}_{+ 1.5  } $& $^{+ 0.0  } _{+ 0.0  } $ \\
      & $0.200$ & $
  1.87\cdot 10^{ 3}$ &
$^{+ 1.2  } _{-1.2  }  $ & $^{+ 1.6  } _{-2.0  }  $ & $^{+ 0.6  } _{-1.5  }  $ & $^{-0.2  }  _{+ 0.2  } $ & $^{}_{+ 0.0  } $ & $^{- 0.0  } _{- 0.0  } $ & $^{-0.4  }  _{+ 0.0  } $ & $^{-0.4  }  _{+ 0.4  } $ & $^{+ 0.6  } _{-1.4  }  
$                                 & $^{}_{-1.0  }  $& $^{+ 0.3  } _{-0.3  }  $ \\
      & $0.316$ & $
  8.47\cdot 10^{ 2}$ &
$^{+ 2.2  } _{-2.2  }  $ & $^{+ 1.1  } _{-4.5  }  $ & $^{+ 0.5  } _{-2.3  }  $ & $^{-0.2  }  _{+ 0.2  } $ & $^{}_{-0.2  }  $ & $^{- 0.0  } _{-0.1  }  $ & $^{+ 0.1  } _{+ 0.0  } $ & $^{-0.4  }  _{+ 0.4  } $ & $^{+ 0.3  } _{-0.8  }  
$                                 & $^{}_{-3.8  }  $& $^{+ 0.3  } _{-0.4  }  $ \\
\hline
$10\,000$ & $0.126$ & $
  7.91$ &
$^{+  23. } _{ -23. }  $ & $^{+  14. } _{ -12. }  $ & $^{+ 4.5  } _{-9.4  }  $ & $^{-0.9  }  _{+ 1.7  } $ & $^{}_{+  13. } $ & $^{-3.1  }  _{-5.9  }  $ & $^{+ 3.0  } _{-1.0  }  $ & $^{-0.4  }  _{+ 0.4  } $ & $^{+ 1.3  } _{-3.1  }  
$                                 & $^{}_{-0.6  }  $& $^{-0.2  }  _{+ 0.2  } $ \\
          & $0.200$ & $
  9.35$ &
$^{+  17. } _{ -17. }  $ & $^{+ 7.1  } _{-3.6  }  $ & $^{+ 0.8  } _{-3.4  }  $ & $^{-0.2  }  _{+ 0.2  } $ & $^{}_{+ 3.1  } $ & $^{-1.1  }  _{+ 0.3  } $ & $^{+ 2.1  } _{-0.5  }  $ & $^{-0.3  }  _{+ 0.3  } $ & $^{+ 0.2  } _{-0.5  }  
$                                 & $^{}_{+ 6.0  } $& $^{-0.1  }  _{+ 0.1  } $ \\
          & $0.316$ & $
  4.72$ &
$^{+  18. } _{ -18. }  $ & $^{+ 5.3  } _{ -16. }  $ & $^{+ 4.3  } _{-3.7  }  $ & $^{-0.3  }  _{+ 0.3  } $ & $^{}_{ -14. }  $ & $^{-0.6  }  _{-4.7  }  $ & $^{+ 1.6  } _{+ 2.6  } $ & $^{-0.4  }  _{+ 0.4  } $ & $^{+ 0.1  } _{-0.3  }  
$                                 & $^{}_{-3.3  }  $& $^{- 0.0  } _{+ 0.0  } $ \\
          & $0.501$ & $
  1.13$ &
$^{+  28. } _{ -28. }  $ & $^{+ 4.6  } _{ -12. }  $ & $^{+ 3.5  } _{ -10. }  $ & $^{-0.2  }  _{+ 0.3  } $ & $^{}_{+ 0.6  } $ & $^{-0.4  }  _{+ 1.3  } $ & $^{-6.4  }  _{-0.3  }  $ & $^{-0.4  }  _{+ 0.4  } $ & $^{+ 0.2  } _{-0.4  }  
$                                 & $^{}_{+ 2.6  } $& $^{-0.1  }  _{+ 0.1  } $ \\
          & $0.794$ & $
  0.05$ &
$^{+ 132. } _{ -65. }  $ & $^{+ 7.3  } _{ -67. }  $ & $^{+ 4.9  } _{ -47. }  $ & $^{-0.3  }  _{+ 0.6  } $ & $^{}_{+ 0.2  } $ & $^{-0.1  }  _{+ 1.1  } $ & $^{+ 0.7  } _{-0.6  }  $ & $^{-0.3  }  _{+ 0.3  } $ & $^{+ 1.7  } _{-4.1  }  
$                                 & $^{}_{ -47. }  $& $^{+ 3.3  } _{-3.6  }  $ \\
\hline
\end{tabular}
}

  \end{center}    
  \normalsize
  \caption[this space for rent]{
    Systematic uncertainties with bin-to-bin correlations
    for the single-differential cross-section \sigx. 
    The left part of the table contains the lower $Q^2$ cut, the
    value at which the cross section is quoted, $x_c$, the measured
    cross-section \sigx\ corrected to the electroweak Born level, the
    statistical uncertainty and the total systematic uncertainty. 
    The uncertainty on the measured luminosity of $2.5\%$ is 
    not included in the total systematic uncertainty.
    The right part of the table lists the total uncorrelated
    systematic uncertainty followed by the bin-to-bin correlated
    systematic uncertainties $\delta_1$--\,$\delta_8$ defined in
    the text. 
    For the latter, the upper (lower) numbers refer to positive
    (negative) variation of e.g. the cut value, whereas the signs of
    the numbers reflect the direction of change in the cross sections. 
    }
  \label{tab-dsdx_c}
\end{sidewaystable}
%
%
\clearpage
\begin{table} [!ht] 
\begin{center}
  {\footnotesize
\renewcommand{\arraystretch}{1.2}
\begin{tabular}{|c|r@{ -- }r|r|l@{}l@{$\,$}l@{$\,$}l@{$\,$}l|}
\hline
{$Q^2$ cut ($\Gev^2$)} & 
\multicolumn{2}{c|}{ $y$ range} & 
\multicolumn{1}{c|}{$y_c$} & 
\multicolumn{5}{c|} {$d\sigma / dy \ $ (pb)} \\
\hline
\hline
$200$ & $0.05$ & $0.10$ & $0.075$ &$ \left( \right.$ \hspace{-1mm} & $  7.39$ & $\pm  0.06$ & $^{+ 0.08}_{- 0.07}$ & $ \left. \! \right) \cdot 10^{ 3}
$ \\
      & $0.10$ & $0.15$ & $0.125$ &$ \left( \right.$ \hspace{-1mm} & $  5.25$ & $\pm  0.05$ & $^{+ 0.04}_{- 0.03}$ & $ \left. \! \right) \cdot 10^{ 3}
$ \\
      & $0.15$ & $0.20$ & $0.175$ &$ \left( \right.$ \hspace{-1mm} & $  4.06$ & $\pm  0.04$ & $^{+ 0.03}_{- 0.04}$ & $ \left. \! \right) \cdot 10^{ 3}
$ \\
      & $0.20$ & $0.25$ & $0.225$ &$ \left( \right.$ \hspace{-1mm} & $  3.23$ & $\pm  0.04$ & $^{+ 0.03}_{- 0.05}$ & $ \left. \! \right) \cdot 10^{ 3}
$ \\
      & $0.25$ & $0.30$ & $0.275$ &$ \left( \right.$ \hspace{-1mm} & $  2.74$ & $\pm  0.04$ & $^{+ 0.02}_{- 0.03}$ & $ \left. \! \right) \cdot 10^{ 3}
$ \\
      & $0.30$ & $0.35$ & $0.325$ &$ \left( \right.$ \hspace{-1mm} & $  2.38$ & $\pm  0.03$ & $^{+ 0.02}_{- 0.03}$ & $ \left. \! \right) \cdot 10^{ 3}
$ \\
      & $0.35$ & $0.40$ & $0.375$ &$ \left( \right.$ \hspace{-1mm} & $  2.05$ & $\pm  0.03$ & $^{+ 0.03}_{- 0.02}$ & $ \left. \! \right) \cdot 10^{ 3}
$ \\
      & $0.40$ & $0.45$ & $0.425$ &$ \left( \right.$ \hspace{-1mm} & $  1.83$ & $\pm  0.03$ & $^{+ 0.02}_{- 0.02}$ & $ \left. \! \right) \cdot 10^{ 3}
$ \\
      & $0.45$ & $0.50$ & $0.475$ &$ \left( \right.$ \hspace{-1mm} & $  1.63$ & $\pm  0.03$ & $^{+ 0.04}_{- 0.02}$ & $ \left. \! \right) \cdot 10^{ 3}
$ \\
      & $0.50$ & $0.55$ & $0.525$ &$ \left( \right.$ \hspace{-1mm} & $  1.46$ & $\pm  0.03$ & $^{+ 0.05}_{- 0.05}$ & $ \left. \! \right) \cdot 10^{ 3}
$ \\
      & $0.55$ & $0.60$ & $0.575$ &$ \left( \right.$ \hspace{-1mm} & $  1.30$ & $\pm  0.03$ & $^{+ 0.05}_{- 0.01}$ & $ \left. \! \right) \cdot 10^{ 3}
$ \\
      & $0.60$ & $0.65$ & $0.625$ &$ \left( \right.$ \hspace{-1mm} & $  1.19$ & $\pm  0.03$ & $^{+ 0.05}_{- 0.03}$ & $ \left. \! \right) \cdot 10^{ 3}
$ \\
      & $0.65$ & $0.70$ & $0.675$ &$ \left( \right.$ \hspace{-1mm} & $  1.10$ & $\pm  0.03$ & $^{+ 0.05}_{- 0.03}$ & $ \left. \! \right) \cdot 10^{ 3}
$ \\
      & $0.70$ & $0.75$ & $0.725$ &$ \left( \right.$ \hspace{-1mm} & $  9.72$ & $\pm  0.26$ & $^{+ 0.64}_{- 0.38}$ & $ \left. \! \right) \cdot 10^{ 2}
$ \\
\hline
\end{tabular}
}

\end{center}    
  \caption[this space for rent]{
    The single-differential cross-section \sigy\ for the reaction 
    $e^{+} p \rightarrow e^{+} X$. 
    The following quantities are given for each bin: the lower $Q^2$
    cut, the $y$ range, the value at which the cross section is
    quoted, $y_c$, and the measured cross section \sigy\ corrected to the
    electroweak Born level.
    The first uncertainty on the measured cross section is the
    statistical uncertainty and the second is the systematic
    uncertainty. The uncertainty on the measured luminosity of $2.5\%$ 
    is not included in the total systematic uncertainty.
  }
\label{tab-dsdy}
\end{table}
\clearpage
\begin{sidewaystable} [!ht] 
  \begin{center}
    {\footnotesize
\renewcommand{\arraystretch}{1.2}
\begin{tabular}{|c|l|c|c|c||c|c|c|c|c|c|c|c|c|}
\hline
{$Q^2$ cut} & 
\multicolumn{1}{c|}{$y_c$} & 
$d\sigma / dy$ &
stat.  & 
total sys.  & 
uncor. sys. &
$\delta_1$ &
$\delta_2$ &
$\delta_3$ &
$\delta_4$ &
$\delta_5$ &
$\delta_6$ &
$\delta_7$ &
$\delta_8$ \\
{($\Gev^2$)} & 
$$ &
(pb) &
(\%) & 
(\%) &
(\%) &
(\%) &
(\%) &
(\%) &
(\%) &
(\%) &
(\%) &
(\%) &
(\%) \\
\hline \hline
$200$ & $0.075$ & $
  7.39\cdot 10^{ 3}$ &
$^{+ 0.8  } _{-0.8  }  $ & $^{+ 1.1  } _{-0.9  }  $ & $^{+ 0.7  } _{-0.2  }  $ & $^{-0.2  }  _{+ 0.2  } $ & $^{}_{+ 0.0  } $ & $^{- 0.0  } _{- 0.0  } $ & $^{-0.6  }  _{- 0.0  } $ & $^{-0.3  }  _{+ 0.3  } $ & $^{+ 0.4  } _{-0.9  }  
$                                 & $^{}_{+ 0.0  } $& $^{-0.2  }  _{+ 0.2  } $ \\
      & $0.125$ & $
  5.25\cdot 10^{ 3}$ &
$^{+ 0.9  } _{-0.9  }  $ & $^{+ 0.7  } _{-0.5  }  $ & $^{+ 0.3  } _{-0.2  }  $ & $^{-0.2  }  _{+ 0.1  } $ & $^{}_{+ 0.0  } $ & $^{+ 0.0  } _{+ 0.0  } $ & $^{-0.2  }  _{+ 0.5  } $ & $^{-0.3  }  _{+ 0.3  } $ & $^{+ 0.0  } _{- 0.0  } 
$                                 & $^{}_{-0.3  }  $& $^{-0.1  }  _{+ 0.1  } $ \\
      & $0.175$ & $
  4.06\cdot 10^{ 3}$ &
$^{+ 1.1  } _{-1.1  }  $ & $^{+ 0.7  } _{-1.0  }  $ & $^{+ 0.3  } _{-0.3  }  $ & $^{-0.2  }  _{+ 0.2  } $ & $^{}_{+ 0.0  } $ & $^{+ 0.0  } _{+ 0.0  } $ & $^{-0.3  }  _{+ 0.4  } $ & $^{-0.3  }  _{+ 0.3  } $ & $^{+ 0.3  } _{-0.8  }  
$                                 & $^{}_{-0.1  }  $& $^{-0.1  }  _{+ 0.1  } $ \\
      & $0.225$ & $
  3.23\cdot 10^{ 3}$ &
$^{+ 1.2  } _{-1.2  }  $ & $^{+ 1.0  } _{-1.7  }  $ & $^{+ 0.7  } _{-0.3  }  $ & $^{-0.1  }  _{+ 0.1  } $ & $^{}_{+ 0.0  } $ & $^{+ 0.0  } _{+ 0.0  } $ & $^{-0.4  }  _{+ 0.4  } $ & $^{-0.3  }  _{+ 0.3  } $ & $^{+ 0.6  } _{-1.5  }  
$                                 & $^{}_{-0.1  }  $& $^{- 0.0  } _{+ 0.1  } $ \\
      & $0.275$ & $
  2.74\cdot 10^{ 3}$ &
$^{+ 1.3  } _{-1.3  }  $ & $^{+ 0.7  } _{-0.9  }  $ & $^{+ 0.7  } _{-0.3  }  $ & $^{-0.1  }  _{+ 0.1  } $ & $^{}_{+ 0.0  } $ & $^{+ 0.0  } _{+ 0.0  } $ & $^{-0.5  }  _{+ 0.4  } $ & $^{-0.3  }  _{+ 0.3  } $ & $^{+ 0.0  } _{- 0.0  } 
$                                 & $^{}_{-0.3  }  $& $^{- 0.0  } _{+ 0.0  } $ \\
      & $0.325$ & $
  2.38\cdot 10^{ 3}$ &
$^{+ 1.4  } _{-1.4  }  $ & $^{+ 0.8  } _{-1.2  }  $ & $^{+ 0.5  } _{-0.4  }  $ & $^{-0.1  }  _{+ 0.1  } $ & $^{}_{+ 0.0  } $ & $^{- 0.0  } _{+ 0.1  } $ & $^{-0.7  }  _{+ 0.7  } $ & $^{-0.3  }  _{+ 0.3  } $ & $^{+ 0.2  } _{-0.5  }  
$                                 & $^{}_{-0.1  }  $& $^{- 0.0  } _{+ 0.0  } $ \\
      & $0.375$ & $
  2.05\cdot 10^{ 3}$ &
$^{+ 1.5  } _{-1.5  }  $ & $^{+ 1.4  } _{-1.0  }  $ & $^{+ 0.8  } _{-0.2  }  $ & $^{-0.1  }  _{+ 0.2  } $ & $^{}_{+ 0.0  } $ & $^{+ 0.0  } _{+ 0.0  } $ & $^{-0.8  }  _{+ 0.9  } $ & $^{-0.3  }  _{+ 0.3  } $ & $^{+ 0.1  } _{-0.1  }  
$                                 & $^{}_{+ 0.7  } $& $^{- 0.0  } _{- 0.0  } $ \\
      & $0.425$ & $
  1.83\cdot 10^{ 3}$ &
$^{+ 1.6  } _{-1.6  }  $ & $^{+ 1.2  } _{-0.9  }  $ & $^{+ 0.9  } _{-0.2  }  $ & $^{-0.1  }  _{+ 0.2  } $ & $^{}_{+ 0.0  } $ & $^{- 0.0  } _{+ 0.1  } $ & $^{-0.7  }  _{+ 0.8  } $ & $^{-0.3  }  _{+ 0.3  } $ & $^{+ 0.0  } _{-0.1  }  
$                                 & $^{}_{-0.4  }  $& $^{+ 0.0  } _{- 0.0  } $ \\
      & $0.475$ & $
  1.63\cdot 10^{ 3}$ &
$^{+ 1.7  } _{-1.7  }  $ & $^{+ 2.7  } _{-1.3  }  $ & $^{+ 0.5  } _{-0.6  }  $ & $^{-0.1  }  _{+ 0.2  } $ & $^{}_{+ 0.0  } $ & $^{-0.1  }  _{+ 0.1  } $ & $^{-0.6  }  _{+ 2.3  } $ & $^{-0.3  }  _{+ 0.3  } $ & $^{+ 0.4  } _{-0.9  }  
$                                 & $^{}_{+ 1.3  } $& $^{+ 0.0  } _{- 0.0  } $ \\
      & $0.525$ & $
  1.46\cdot 10^{ 3}$ &
$^{+ 1.8  } _{-1.8  }  $ & $^{+ 3.3  } _{-3.5  }  $ & $^{+ 0.8  } _{-0.2  }  $ & $^{-0.3  }  _{+ 0.4  } $ & $^{}_{+ 0.0  } $ & $^{-0.1  }  _{-0.2  }  $ & $^{-1.4  }  _{+ 3.3  } $ & $^{-0.3  }  _{+ 0.3  } $ & $^{+ 0.2  } _{-0.5  }  
$                                 & $^{}_{-3.1  }  $& $^{+ 0.0  } _{- 0.0  } $ \\
      & $0.575$ & $
  1.30\cdot 10^{ 3}$ &
$^{+ 2.0  } _{-2.0  }  $ & $^{+ 3.6  } _{-1.1  }  $ & $^{+ 0.5  } _{-0.7  }  $ & $^{-0.5  }  _{+ 0.5  } $ & $^{}_{+ 0.0  } $ & $^{+ 0.0  } _{- 0.0  } $ & $^{-0.5  }  _{+ 3.3  } $ & $^{-0.3  }  _{+ 0.3  } $ & $^{+ 0.0  } _{-0.1  }  
$                                 & $^{}_{+ 1.2  } $& $^{+ 0.0  } _{- 0.0  } $ \\
      & $0.625$ & $
  1.19\cdot 10^{ 3}$ &
$^{+ 2.1  } _{-2.1  }  $ & $^{+ 4.0  } _{-2.5  }  $ & $^{+ 1.0  } _{-0.4  }  $ & $^{-1.1  }  _{+ 1.2  } $ & $^{}_{+ 0.0  } $ & $^{-0.1  }  _{-0.3  }  $ & $^{-1.9  }  _{+ 3.7  } $ & $^{-0.3  }  _{+ 0.3  } $ & $^{+ 0.3  } _{-0.7  }  
$                                 & $^{}_{-0.5  }  $& $^{+ 0.1  } _{-0.1  }  $ \\
      & $0.675$ & $
  1.10\cdot 10^{ 3}$ &
$^{+ 2.3  } _{-2.3  }  $ & $^{+ 4.6  } _{-2.6  }  $ & $^{+ 0.9  } _{-0.5  }  $ & $^{-2.2  }  _{+ 2.0  } $ & $^{}_{+ 0.0  } $ & $^{+ 0.1  } _{+ 0.2  } $ & $^{-1.1  }  _{+ 3.9  } $ & $^{-0.3  }  _{+ 0.3  } $ & $^{+ 0.2  } _{-0.4  }  
$                                 & $^{}_{+ 1.3  } $& $^{+ 0.1  } _{-0.1  }  $ \\
      & $0.725$ & $
  9.72\cdot 10^{ 2}$ &
$^{+ 2.7  } _{-2.7  }  $ & $^{+ 6.6  } _{-3.9  }  $ & $^{+ 1.2  } _{-1.0  }  $ & $^{-3.1  }  _{+ 4.4  } $ & $^{}_{+ 0.0  } $ & $^{-0.1  }  _{+ 0.3  } $ & $^{-1.3  }  _{+ 4.2  } $ & $^{-0.3  }  _{+ 0.3  } $ & $^{+ 0.6  } _{-1.4  }  
$                                 & $^{}_{+ 2.4  } $& $^{+ 0.1  } _{-0.1  }  $ \\
\hline
\end{tabular}
}

  \end{center} 
  \normalsize
  \caption[this space for rent]{
    Systematic uncertainties with bin-to-bin correlations for the
    single-differential cross-section \sigy. 
    The left part of the table contains the lower $Q^2$ cut, the
    value at which the cross section is quoted, $y_c$, the measured
    cross section \sigy\ corrected to the electroweak Born level, the 
    statistical uncertainty and the total systematic uncertainty. 
    The uncertainty on the measured luminosity of $2.5\%$ is not 
    included in the total systematic uncertainty.
    The right part of the table lists the total uncorrelated
    systematic uncertainty followed by the bin-to-bin correlated
    systematic uncertainties $\delta_1$--\,$\delta_8$ defined in
    the text.
    For the latter, the upper (lower) numbers refer to positive
    (negative) variation of e.g. the cut value, whereas the signs of
    the numbers reflect the direction of change in the cross
    sections.
    }
  \label{tab-dsdy_c}
\end{sidewaystable}
%
%
\clearpage
\begin{table} [!ht] 
\begin{center}
  {\tiny
\renewcommand{\arraystretch}{1.2}
\begin{tabular}{|r@{ -- }l|r@{ -- }l@{$\,$}r|r|l|l@{}l@{$\,$}l@{$\,$}l@{$\,$}l|}
\hline
\multicolumn{2}{|c|}{$Q^2$ range ($\Gev^2$)} &
\multicolumn{3}{c|}{ $x$ range} & 
\multicolumn{1}{c|}{$Q^2_c$ ($\Gev^2$)} & 
\multicolumn{1}{c|}{$x_c$} &
\multicolumn{5}{c|}{$\tilde{\sigma}(e^+p) $} \\
\hline
\hline $  185.$ & $  240.$ & $\left( \right.\!0.37$ & $0.60$ & $ \left. \! \right) \cdot 10^{-2}$ & $  200$ & $0.50$$\,\cdot\,10^{-2}$ & $ $ & $ 1.127$ & $\pm 0.017$ & $^{+0.020}_{-0.030}$ & $$ \\
\multicolumn{2}{|c|}{} & $\left( \right.\!0.60$ & $1.00$ & $ \left. \! \right) \cdot 10^{-2}$ &  & $0.80$$\,\cdot\,10^{-2}$ & $\left( \right.\! $ \hspace{-0.5mm} & $  9.45$ & $\pm  0.14$ & $^{+ 0.15}_{- 0.11}$ & $ \left. \!\!\!\! \right) \cdot 10^{-1}$ \\
\multicolumn{2}{|c|}{} & $\left( \right.\!0.10$ & $0.17$ & $ \left. \! \right) \cdot 10^{-1}$ &  & $0.13$$\,\cdot\,10^{-1}$ & $\left( \right.\! $ \hspace{-0.5mm} & $  8.16$ & $\pm  0.12$ & $^{+ 0.06}_{- 0.09}$ & $ \left. \!\!\!\! \right) \cdot 10^{-1}$ \\
\multicolumn{2}{|c|}{} & $\left( \right.\!0.17$ & $0.25$ & $ \left. \! \right) \cdot 10^{-1}$ &  & $0.21$$\,\cdot\,10^{-1}$ & $\left( \right.\! $ \hspace{-0.5mm} & $  6.90$ & $\pm  0.12$ & $^{+ 0.07}_{- 0.10}$ & $ \left. \!\!\!\! \right) \cdot 10^{-1}$ \\
\multicolumn{2}{|c|}{} & $\left( \right.\!0.25$ & $0.37$ & $ \left. \! \right) \cdot 10^{-1}$ &  & $0.32$$\,\cdot\,10^{-1}$ & $\left( \right.\! $ \hspace{-0.5mm} & $  5.93$ & $\pm  0.11$ & $^{+ 0.11}_{- 0.09}$ & $ \left. \!\!\!\! \right) \cdot 10^{-1}$ \\
\multicolumn{2}{|c|}{} & $\left( \right.\!0.37$ & $0.60$ & $ \left. \! \right) \cdot 10^{-1}$ &  & $0.50$$\,\cdot\,10^{-1}$ & $\left( \right.\! $ \hspace{-0.5mm} & $  5.28$ & $\pm  0.09$ & $^{+ 0.04}_{- 0.11}$ & $ \left. \!\!\!\! \right) \cdot 10^{-1}$ \\
\multicolumn{2}{|c|}{} & $\left( \right.\!0.60$ & $1.20$ & $ \left. \! \right) \cdot 10^{-1}$ &  & $0.80$$\,\cdot\,10^{-1}$ & $\left( \right.\! $ \hspace{-0.5mm} & $  4.30$ & $\pm  0.07$ & $^{+ 0.23}_{- 0.05}$ & $ \left. \!\!\!\! \right) \cdot 10^{-1}$ \\
\multicolumn{2}{|c|}{} & $0.12$ & $0.25$ & $$ &  & $0.18$$$ & $\left( \right.\! $ \hspace{-0.5mm} & $  3.33$ & $\pm  0.06$ & $^{+ 0.07}_{- 0.44}$ & $ \left. \!\!\!\! \right) \cdot 10^{-1}$ \\
\hline $  240.$ & $  310.$ & $\left( \right.\!0.60$ & $1.00$ & $ \left. \! \right) \cdot 10^{-2}$ & $  250$ & $0.80$$\,\cdot\,10^{-2}$ & $\left( \right.\! $ \hspace{-0.5mm} & $  9.72$ & $\pm  0.16$ & $^{+ 0.10}_{- 0.14}$ & $ \left. \!\!\!\! \right) \cdot 10^{-1}$ \\
\multicolumn{2}{|c|}{} & $\left( \right.\!0.10$ & $0.17$ & $ \left. \! \right) \cdot 10^{-1}$ &  & $0.13$$\,\cdot\,10^{-1}$ & $\left( \right.\! $ \hspace{-0.5mm} & $  8.30$ & $\pm  0.14$ & $^{+ 0.10}_{- 0.07}$ & $ \left. \!\!\!\! \right) \cdot 10^{-1}$ \\
\multicolumn{2}{|c|}{} & $\left( \right.\!0.17$ & $0.25$ & $ \left. \! \right) \cdot 10^{-1}$ &  & $0.21$$\,\cdot\,10^{-1}$ & $\left( \right.\! $ \hspace{-0.5mm} & $  6.97$ & $\pm  0.14$ & $^{+ 0.12}_{- 0.05}$ & $ \left. \!\!\!\! \right) \cdot 10^{-1}$ \\
\multicolumn{2}{|c|}{} & $\left( \right.\!0.25$ & $0.37$ & $ \left. \! \right) \cdot 10^{-1}$ &  & $0.32$$\,\cdot\,10^{-1}$ & $\left( \right.\! $ \hspace{-0.5mm} & $  5.95$ & $\pm  0.12$ & $^{+ 0.09}_{- 0.09}$ & $ \left. \!\!\!\! \right) \cdot 10^{-1}$ \\
\multicolumn{2}{|c|}{} & $\left( \right.\!0.37$ & $0.60$ & $ \left. \! \right) \cdot 10^{-1}$ &  & $0.50$$\,\cdot\,10^{-1}$ & $\left( \right.\! $ \hspace{-0.5mm} & $  5.28$ & $\pm  0.10$ & $^{+ 0.06}_{- 0.09}$ & $ \left. \!\!\!\! \right) \cdot 10^{-1}$ \\
\multicolumn{2}{|c|}{} & $\left( \right.\!0.60$ & $1.20$ & $ \left. \! \right) \cdot 10^{-1}$ &  & $0.80$$\,\cdot\,10^{-1}$ & $\left( \right.\! $ \hspace{-0.5mm} & $  4.22$ & $\pm  0.07$ & $^{+ 0.22}_{- 0.07}$ & $ \left. \!\!\!\! \right) \cdot 10^{-1}$ \\
\multicolumn{2}{|c|}{} & $0.12$ & $0.25$ & $$ &  & $0.18$$$ & $\left( \right.\! $ \hspace{-0.5mm} & $  3.24$ & $\pm  0.07$ & $^{+ 0.11}_{- 0.17}$ & $ \left. \!\!\!\! \right) \cdot 10^{-1}$ \\
\hline $  310.$ & $  410.$ & $\left( \right.\!0.60$ & $1.00$ & $ \left. \! \right) \cdot 10^{-2}$ & $  350$ & $0.80$$\,\cdot\,10^{-2}$ & $\left( \right.\! $ \hspace{-0.5mm} & $  9.92$ & $\pm  0.21$ & $^{+ 0.27}_{- 0.12}$ & $ \left. \!\!\!\! \right) \cdot 10^{-1}$ \\
\multicolumn{2}{|c|}{} & $\left( \right.\!0.10$ & $0.17$ & $ \left. \! \right) \cdot 10^{-1}$ &  & $0.13$$\,\cdot\,10^{-1}$ & $\left( \right.\! $ \hspace{-0.5mm} & $  8.23$ & $\pm  0.16$ & $^{+ 0.10}_{- 0.14}$ & $ \left. \!\!\!\! \right) \cdot 10^{-1}$ \\
\multicolumn{2}{|c|}{} & $\left( \right.\!0.17$ & $0.25$ & $ \left. \! \right) \cdot 10^{-1}$ &  & $0.21$$\,\cdot\,10^{-1}$ & $\left( \right.\! $ \hspace{-0.5mm} & $  6.94$ & $\pm  0.16$ & $^{+ 0.06}_{- 0.11}$ & $ \left. \!\!\!\! \right) \cdot 10^{-1}$ \\
\multicolumn{2}{|c|}{} & $\left( \right.\!0.25$ & $0.37$ & $ \left. \! \right) \cdot 10^{-1}$ &  & $0.32$$\,\cdot\,10^{-1}$ & $\left( \right.\! $ \hspace{-0.5mm} & $  6.14$ & $\pm  0.14$ & $^{+ 0.07}_{- 0.06}$ & $ \left. \!\!\!\! \right) \cdot 10^{-1}$ \\
\multicolumn{2}{|c|}{} & $\left( \right.\!0.37$ & $0.60$ & $ \left. \! \right) \cdot 10^{-1}$ &  & $0.50$$\,\cdot\,10^{-1}$ & $\left( \right.\! $ \hspace{-0.5mm} & $  5.10$ & $\pm  0.11$ & $^{+ 0.05}_{- 0.08}$ & $ \left. \!\!\!\! \right) \cdot 10^{-1}$ \\
\multicolumn{2}{|c|}{} & $\left( \right.\!0.60$ & $1.20$ & $ \left. \! \right) \cdot 10^{-1}$ &  & $0.80$$\,\cdot\,10^{-1}$ & $\left( \right.\! $ \hspace{-0.5mm} & $  4.26$ & $\pm  0.08$ & $^{+ 0.07}_{- 0.03}$ & $ \left. \!\!\!\! \right) \cdot 10^{-1}$ \\
\multicolumn{2}{|c|}{} & $0.12$ & $0.25$ & $$ &  & $0.18$$$ & $\left( \right.\! $ \hspace{-0.5mm} & $  3.10$ & $\pm  0.07$ & $^{+ 0.13}_{- 0.04}$ & $ \left. \!\!\!\! \right) \cdot 10^{-1}$ \\
\hline $  410.$ & $  530.$ & $\left( \right.\!0.60$ & $1.00$ & $ \left. \! \right) \cdot 10^{-2}$ & $  450$ & $0.80$$\,\cdot\,10^{-2}$ & $ $ & $  1.05$ & $\pm  0.02$ & $^{+ 0.04}_{- 0.02}$ & $$ \\
\multicolumn{2}{|c|}{} & $\left( \right.\!0.10$ & $0.17$ & $ \left. \! \right) \cdot 10^{-1}$ &  & $0.13$$\,\cdot\,10^{-1}$ & $\left( \right.\! $ \hspace{-0.5mm} & $  8.42$ & $\pm  0.22$ & $^{+ 0.11}_{- 0.11}$ & $ \left. \!\!\!\! \right) \cdot 10^{-1}$ \\
\multicolumn{2}{|c|}{} & $\left( \right.\!0.17$ & $0.25$ & $ \left. \! \right) \cdot 10^{-1}$ &  & $0.21$$\,\cdot\,10^{-1}$ & $\left( \right.\! $ \hspace{-0.5mm} & $  6.79$ & $\pm  0.20$ & $^{+ 0.17}_{- 0.09}$ & $ \left. \!\!\!\! \right) \cdot 10^{-1}$ \\
\multicolumn{2}{|c|}{} & $\left( \right.\!0.25$ & $0.37$ & $ \left. \! \right) \cdot 10^{-1}$ &  & $0.32$$\,\cdot\,10^{-1}$ & $\left( \right.\! $ \hspace{-0.5mm} & $  6.30$ & $\pm  0.18$ & $^{+ 0.06}_{- 0.16}$ & $ \left. \!\!\!\! \right) \cdot 10^{-1}$ \\
\multicolumn{2}{|c|}{} & $\left( \right.\!0.37$ & $0.60$ & $ \left. \! \right) \cdot 10^{-1}$ &  & $0.50$$\,\cdot\,10^{-1}$ & $\left( \right.\! $ \hspace{-0.5mm} & $  5.07$ & $\pm  0.13$ & $^{+ 0.09}_{- 0.05}$ & $ \left. \!\!\!\! \right) \cdot 10^{-1}$ \\
\multicolumn{2}{|c|}{} & $\left( \right.\!0.60$ & $1.00$ & $ \left. \! \right) \cdot 10^{-1}$ &  & $0.80$$\,\cdot\,10^{-1}$ & $\left( \right.\! $ \hspace{-0.5mm} & $  4.49$ & $\pm  0.11$ & $^{+ 0.06}_{- 0.03}$ & $ \left. \!\!\!\! \right) \cdot 10^{-1}$ \\
\multicolumn{2}{|c|}{} & $0.10$ & $0.17$ & $$ &  & $0.13$$$ & $\left( \right.\! $ \hspace{-0.5mm} & $  3.64$ & $\pm  0.10$ & $^{+ 0.10}_{- 0.04}$ & $ \left. \!\!\!\! \right) \cdot 10^{-1}$ \\
\multicolumn{2}{|c|}{} & $0.17$ & $0.30$ & $$ &  & $0.25$$$ & $\left( \right.\! $ \hspace{-0.5mm} & $  2.62$ & $\pm  0.08$ & $^{+ 0.04}_{- 0.12}$ & $ \left. \!\!\!\! \right) \cdot 10^{-1}$ \\
\hline $  530.$ & $  710.$ & $\left( \right.\!0.10$ & $0.17$ & $ \left. \! \right) \cdot 10^{-1}$ & $  650$ & $0.13$$\,\cdot\,10^{-1}$ & $\left( \right.\! $ \hspace{-0.5mm} & $  8.64$ & $\pm  0.20$ & $^{+ 0.24}_{- 0.10}$ & $ \left. \!\!\!\! \right) \cdot 10^{-1}$ \\
\multicolumn{2}{|c|}{} & $\left( \right.\!0.17$ & $0.25$ & $ \left. \! \right) \cdot 10^{-1}$ &  & $0.21$$\,\cdot\,10^{-1}$ & $\left( \right.\! $ \hspace{-0.5mm} & $  7.39$ & $\pm  0.22$ & $^{+ 0.05}_{- 0.09}$ & $ \left. \!\!\!\! \right) \cdot 10^{-1}$ \\
\multicolumn{2}{|c|}{} & $\left( \right.\!0.25$ & $0.37$ & $ \left. \! \right) \cdot 10^{-1}$ &  & $0.32$$\,\cdot\,10^{-1}$ & $\left( \right.\! $ \hspace{-0.5mm} & $  6.32$ & $\pm  0.21$ & $^{+ 0.05}_{- 0.19}$ & $ \left. \!\!\!\! \right) \cdot 10^{-1}$ \\
\multicolumn{2}{|c|}{} & $\left( \right.\!0.37$ & $0.60$ & $ \left. \! \right) \cdot 10^{-1}$ &  & $0.50$$\,\cdot\,10^{-1}$ & $\left( \right.\! $ \hspace{-0.5mm} & $  5.33$ & $\pm  0.18$ & $^{+ 0.05}_{- 0.03}$ & $ \left. \!\!\!\! \right) \cdot 10^{-1}$ \\
\multicolumn{2}{|c|}{} & $\left( \right.\!0.60$ & $1.00$ & $ \left. \! \right) \cdot 10^{-1}$ &  & $0.80$$\,\cdot\,10^{-1}$ & $\left( \right.\! $ \hspace{-0.5mm} & $  4.46$ & $\pm  0.15$ & $^{+ 0.07}_{- 0.08}$ & $ \left. \!\!\!\! \right) \cdot 10^{-1}$ \\
\multicolumn{2}{|c|}{} & $0.10$ & $0.17$ & $$ &  & $0.13$$$ & $\left( \right.\! $ \hspace{-0.5mm} & $  3.53$ & $\pm  0.13$ & $^{+ 0.08}_{- 0.04}$ & $ \left. \!\!\!\! \right) \cdot 10^{-1}$ \\
\multicolumn{2}{|c|}{} & $0.17$ & $0.30$ & $$ &  & $0.25$$$ & $\left( \right.\! $ \hspace{-0.5mm} & $  2.50$ & $\pm  0.09$ & $^{+ 0.08}_{- 0.05}$ & $ \left. \!\!\!\! \right) \cdot 10^{-1}$ \\
\hline $  710.$ & $  900.$ & $\left( \right.\!0.90$ & $1.70$ & $ \left. \! \right) \cdot 10^{-2}$ & $  800$ & $1.30$$\,\cdot\,10^{-2}$ & $\left( \right.\! $ \hspace{-0.5mm} & $  8.58$ & $\pm  0.24$ & $^{+ 0.25}_{- 0.14}$ & $ \left. \!\!\!\! \right) \cdot 10^{-1}$ \\
\multicolumn{2}{|c|}{} & $\left( \right.\!0.17$ & $0.25$ & $ \left. \! \right) \cdot 10^{-1}$ &  & $0.21$$\,\cdot\,10^{-1}$ & $\left( \right.\! $ \hspace{-0.5mm} & $  7.39$ & $\pm  0.26$ & $^{+ 0.28}_{- 0.07}$ & $ \left. \!\!\!\! \right) \cdot 10^{-1}$ \\
\multicolumn{2}{|c|}{} & $\left( \right.\!0.25$ & $0.37$ & $ \left. \! \right) \cdot 10^{-1}$ &  & $0.32$$\,\cdot\,10^{-1}$ & $\left( \right.\! $ \hspace{-0.5mm} & $  6.61$ & $\pm  0.23$ & $^{+ 0.06}_{- 0.15}$ & $ \left. \!\!\!\! \right) \cdot 10^{-1}$ \\
\multicolumn{2}{|c|}{} & $\left( \right.\!0.37$ & $0.60$ & $ \left. \! \right) \cdot 10^{-1}$ &  & $0.50$$\,\cdot\,10^{-1}$ & $\left( \right.\! $ \hspace{-0.5mm} & $  5.15$ & $\pm  0.18$ & $^{+ 0.11}_{- 0.04}$ & $ \left. \!\!\!\! \right) \cdot 10^{-1}$ \\
\multicolumn{2}{|c|}{} & $\left( \right.\!0.60$ & $1.00$ & $ \left. \! \right) \cdot 10^{-1}$ &  & $0.80$$\,\cdot\,10^{-1}$ & $\left( \right.\! $ \hspace{-0.5mm} & $  4.52$ & $\pm  0.16$ & $^{+ 0.07}_{- 0.05}$ & $ \left. \!\!\!\! \right) \cdot 10^{-1}$ \\
\multicolumn{2}{|c|}{} & $0.10$ & $0.17$ & $$ &  & $0.13$$$ & $\left( \right.\! $ \hspace{-0.5mm} & $  3.60$ & $\pm  0.14$ & $^{+ 0.06}_{- 0.05}$ & $ \left. \!\!\!\! \right) \cdot 10^{-1}$ \\
\multicolumn{2}{|c|}{} & $0.17$ & $0.30$ & $$ &  & $0.25$$$ & $\left( \right.\! $ \hspace{-0.5mm} & $  2.59$ & $\pm  0.12$ & $^{+ 0.14}_{- 0.05}$ & $ \left. \!\!\!\! \right) \cdot 10^{-1}$ \\
\hline
\end{tabular}
}

\end{center}    
  \normalsize
  \caption[this space for rent]{
    The reduced cross-section $\tilde{\sigma}^{e^+ p}$ for the
    reaction  $e^{+} p \rightarrow e^{+} X$.  
    The following quantities are given for each bin: the $Q^2$ and $x$
    ranges, the values at which the cross section is quoted, \qqc\ and
    $x_c$, and the measured reduced cross section, $\tilde{\sigma}^{e^+ p}$, 
    corrected to the electroweak Born level. 
    The first uncertainty on the measured cross section is the
    statistical uncertainty and the second is the systematic uncertainty. 
    The uncertainty on the measured luminosity of $2.5\%$ is not included 
    in the total systematic uncertainty.
  }
  \label{tab-dsdqdx_1}
\end{table}
\clearpage
\begin{table} [ht]
  \strut \vspace{-2cm}
  \begin{center}
    {\tiny
\renewcommand{\arraystretch}{1.2}
\begin{tabular}{|r@{ -- }l|r@{ -- }l@{$\,$}r|r|l|l@{}l@{$\,$}l@{$\,$}l@{$\,$}l|}
\hline
\multicolumn{2}{|c|}{$Q^2$ range ($\Gev^2$)} &
\multicolumn{3}{c|}{ $x$ range} & 
\multicolumn{1}{c|}{$Q^2_c$ ($\Gev^2$)} & 
\multicolumn{1}{c|}{$x_c$} &
\multicolumn{5}{c|}{$\tilde{\sigma}(e^+p) $} \\
\hline
\hline $  900.$ & $ 1300.$ & $\left( \right.\!0.10$ & $0.17$ & $ \left. \! \right) \cdot 10^{-1}$ & $ 1200$ & $0.14$$\,\cdot\,10^{-1}$ & $\left( \right.\! $ \hspace{-0.5mm} & $  8.47$ & $\pm  0.29$ & $^{+ 0.42}_{- 0.26}$ & $ \left. \! \right) \cdot 10^{-1}$ \\
\multicolumn{2}{|c|}{} & $\left( \right.\!0.17$ & $0.25$ & $ \left. \! \right) \cdot 10^{-1}$ &  & $0.21$$\,\cdot\,10^{-1}$ & $\left( \right.\! $ \hspace{-0.5mm} & $  7.90$ & $\pm  0.27$ & $^{+ 0.20}_{- 0.14}$ & $ \left. \! \right) \cdot 10^{-1}$ \\
\multicolumn{2}{|c|}{} & $\left( \right.\!0.25$ & $0.37$ & $ \left. \! \right) \cdot 10^{-1}$ &  & $0.32$$\,\cdot\,10^{-1}$ & $\left( \right.\! $ \hspace{-0.5mm} & $  6.59$ & $\pm  0.23$ & $^{+ 0.08}_{- 0.20}$ & $ \left. \! \right) \cdot 10^{-1}$ \\
\multicolumn{2}{|c|}{} & $\left( \right.\!0.37$ & $0.60$ & $ \left. \! \right) \cdot 10^{-1}$ &  & $0.50$$\,\cdot\,10^{-1}$ & $\left( \right.\! $ \hspace{-0.5mm} & $  5.41$ & $\pm  0.17$ & $^{+ 0.05}_{- 0.11}$ & $ \left. \! \right) \cdot 10^{-1}$ \\
\multicolumn{2}{|c|}{} & $\left( \right.\!0.60$ & $1.00$ & $ \left. \! \right) \cdot 10^{-1}$ &  & $0.80$$\,\cdot\,10^{-1}$ & $\left( \right.\! $ \hspace{-0.5mm} & $  4.60$ & $\pm  0.15$ & $^{+ 0.05}_{- 0.04}$ & $ \left. \! \right) \cdot 10^{-1}$ \\
\multicolumn{2}{|c|}{} & $0.10$ & $0.17$ & $$ &  & $0.13$$$ & $\left( \right.\! $ \hspace{-0.5mm} & $  3.56$ & $\pm  0.12$ & $^{+ 0.11}_{- 0.03}$ & $ \left. \! \right) \cdot 10^{-1}$ \\
\multicolumn{2}{|c|}{} & $0.17$ & $0.30$ & $$ &  & $0.25$$$ & $\left( \right.\! $ \hspace{-0.5mm} & $  2.42$ & $\pm  0.09$ & $^{+ 0.05}_{- 0.04}$ & $ \left. \! \right) \cdot 10^{-1}$ \\
\multicolumn{2}{|c|}{} & $0.30$ & $0.53$ & $$ &  & $0.40$$$ & $\left( \right.\! $ \hspace{-0.5mm} & $  1.38$ & $\pm  0.09$ & $^{+ 0.02}_{- 0.15}$ & $ \left. \! \right) \cdot 10^{-1}$ \\
\hline $ 1300.$ & $ 1800.$ & $\left( \right.\!0.17$ & $0.25$ & $ \left. \! \right) \cdot 10^{-1}$ & $ 1500$ & $0.21$$\,\cdot\,10^{-1}$ & $\left( \right.\! $ \hspace{-0.5mm} & $  6.77$ & $\pm  0.32$ & $^{+ 0.31}_{- 0.26}$ & $ \left. \! \right) \cdot 10^{-1}$ \\
\multicolumn{2}{|c|}{} & $\left( \right.\!0.25$ & $0.37$ & $ \left. \! \right) \cdot 10^{-1}$ &  & $0.32$$\,\cdot\,10^{-1}$ & $\left( \right.\! $ \hspace{-0.5mm} & $  6.38$ & $\pm  0.28$ & $^{+ 0.13}_{- 0.06}$ & $ \left. \! \right) \cdot 10^{-1}$ \\
\multicolumn{2}{|c|}{} & $\left( \right.\!0.37$ & $0.60$ & $ \left. \! \right) \cdot 10^{-1}$ &  & $0.50$$\,\cdot\,10^{-1}$ & $\left( \right.\! $ \hspace{-0.5mm} & $  5.85$ & $\pm  0.23$ & $^{+ 0.13}_{- 0.08}$ & $ \left. \! \right) \cdot 10^{-1}$ \\
\multicolumn{2}{|c|}{} & $\left( \right.\!0.60$ & $1.00$ & $ \left. \! \right) \cdot 10^{-1}$ &  & $0.80$$\,\cdot\,10^{-1}$ & $\left( \right.\! $ \hspace{-0.5mm} & $  4.36$ & $\pm  0.18$ & $^{+ 0.04}_{- 0.12}$ & $ \left. \! \right) \cdot 10^{-1}$ \\
\multicolumn{2}{|c|}{} & $0.10$ & $0.15$ & $$ &  & $0.13$$$ & $\left( \right.\! $ \hspace{-0.5mm} & $  3.44$ & $\pm  0.17$ & $^{+ 0.11}_{- 0.03}$ & $ \left. \! \right) \cdot 10^{-1}$ \\
\multicolumn{2}{|c|}{} & $0.15$ & $0.23$ & $$ &  & $0.18$$$ & $\left( \right.\! $ \hspace{-0.5mm} & $  3.10$ & $\pm  0.16$ & $^{+ 0.03}_{- 0.09}$ & $ \left. \! \right) \cdot 10^{-1}$ \\
\multicolumn{2}{|c|}{} & $0.23$ & $0.35$ & $$ &  & $0.25$$$ & $\left( \right.\! $ \hspace{-0.5mm} & $  2.61$ & $\pm  0.15$ & $^{+ 0.05}_{- 0.08}$ & $ \left. \! \right) \cdot 10^{-1}$ \\
\multicolumn{2}{|c|}{} & $0.35$ & $0.53$ & $$ &  & $0.40$$$ & $\left( \right.\! $ \hspace{-0.5mm} & $  1.32$ & $\pm  0.12$ & $^{+ 0.03}_{- 0.08}$ & $ \left. \! \right) \cdot 10^{-1}$ \\
\hline $ 1800.$ & $ 2500.$ & $\left( \right.\!0.23$ & $0.37$ & $ \left. \! \right) \cdot 10^{-1}$ & $ 2000$ & $0.32$$\,\cdot\,10^{-1}$ & $\left( \right.\! $ \hspace{-0.5mm} & $  6.24$ & $\pm  0.33$ & $^{+ 0.23}_{- 0.08}$ & $ \left. \! \right) \cdot 10^{-1}$ \\
\multicolumn{2}{|c|}{} & $\left( \right.\!0.37$ & $0.60$ & $ \left. \! \right) \cdot 10^{-1}$ &  & $0.50$$\,\cdot\,10^{-1}$ & $\left( \right.\! $ \hspace{-0.5mm} & $  5.22$ & $\pm  0.26$ & $^{+ 0.06}_{- 0.08}$ & $ \left. \! \right) \cdot 10^{-1}$ \\
\multicolumn{2}{|c|}{} & $\left( \right.\!0.60$ & $1.00$ & $ \left. \! \right) \cdot 10^{-1}$ &  & $0.80$$\,\cdot\,10^{-1}$ & $\left( \right.\! $ \hspace{-0.5mm} & $  4.44$ & $\pm  0.21$ & $^{+ 0.09}_{- 0.03}$ & $ \left. \! \right) \cdot 10^{-1}$ \\
\multicolumn{2}{|c|}{} & $0.10$ & $0.15$ & $$ &  & $0.13$$$ & $\left( \right.\! $ \hspace{-0.5mm} & $  4.00$ & $\pm  0.22$ & $^{+ 0.07}_{- 0.04}$ & $ \left. \! \right) \cdot 10^{-1}$ \\
\multicolumn{2}{|c|}{} & $0.15$ & $0.23$ & $$ &  & $0.18$$$ & $\left( \right.\! $ \hspace{-0.5mm} & $  3.22$ & $\pm  0.19$ & $^{+ 0.03}_{- 0.05}$ & $ \left. \! \right) \cdot 10^{-1}$ \\
\multicolumn{2}{|c|}{} & $0.23$ & $0.35$ & $$ &  & $0.25$$$ & $\left( \right.\! $ \hspace{-0.5mm} & $  2.45$ & $\pm  0.17$ & $^{+ 0.04}_{- 0.05}$ & $ \left. \! \right) \cdot 10^{-1}$ \\
\multicolumn{2}{|c|}{} & $0.35$ & $0.53$ & $$ &  & $0.40$$$ & $\left( \right.\! $ \hspace{-0.5mm} & $  1.27$ & $\pm  0.13$ & $^{+ 0.02}_{- 0.14}$ & $ \left. \! \right) \cdot 10^{-1}$ \\
\hline $ 2500.$ & $ 3500.$ & $\left( \right.\!0.37$ & $0.60$ & $ \left. \! \right) \cdot 10^{-1}$ & $ 3000$ & $0.50$$\,\cdot\,10^{-1}$ & $\left( \right.\! $ \hspace{-0.5mm} & $  5.43$ & $\pm  0.33$ & $^{+ 0.09}_{- 0.09}$ & $ \left. \! \right) \cdot 10^{-1}$ \\
\multicolumn{2}{|c|}{} & $\left( \right.\!0.60$ & $1.00$ & $ \left. \! \right) \cdot 10^{-1}$ &  & $0.80$$\,\cdot\,10^{-1}$ & $\left( \right.\! $ \hspace{-0.5mm} & $  4.12$ & $\pm  0.25$ & $^{+ 0.06}_{- 0.08}$ & $ \left. \! \right) \cdot 10^{-1}$ \\
\multicolumn{2}{|c|}{} & $0.10$ & $0.15$ & $$ &  & $0.13$$$ & $\left( \right.\! $ \hspace{-0.5mm} & $  3.50$ & $\pm  0.24$ & $^{+ 0.15}_{- 0.04}$ & $ \left. \! \right) \cdot 10^{-1}$ \\
\multicolumn{2}{|c|}{} & $0.15$ & $0.23$ & $$ &  & $0.18$$$ & $\left( \right.\! $ \hspace{-0.5mm} & $  3.17$ & $\pm  0.22$ & $^{+ 0.05}_{- 0.11}$ & $ \left. \! \right) \cdot 10^{-1}$ \\
\multicolumn{2}{|c|}{} & $0.23$ & $0.35$ & $$ &  & $0.25$$$ & $\left( \right.\! $ \hspace{-0.5mm} & $  2.29$ & $\pm  0.19$ & $^{+ 0.02}_{- 0.11}$ & $ \left. \! \right) \cdot 10^{-1}$ \\
\multicolumn{2}{|c|}{} & $0.35$ & $0.53$ & $$ &  & $0.40$$$ & $\left( \right.\! $ \hspace{-0.5mm} & $  1.35$ & $\pm  0.15$ & $^{+ 0.06}_{- 0.05}$ & $ \left. \! \right) \cdot 10^{-1}$ \\
\multicolumn{2}{|c|}{} & $0.53$ & $1.00$ & $$ &  & $0.65$$$ & $\left( \right.\! $ \hspace{-0.5mm} & $  2.02$ & $\pm  0.35$ & $^{+ 0.23}_{- 0.13}$ & $ \left. \! \right) \cdot 10^{-2}$ \\
\hline $ 3500.$ & $ 5600.$ & $\left( \right.\!0.40$ & $1.00$ & $ \left. \! \right) \cdot 10^{-1}$ & $ 5000$ & $0.80$$\,\cdot\,10^{-1}$ & $\left( \right.\! $ \hspace{-0.5mm} & $  4.17$ & $\pm  0.24$ & $^{+ 0.09}_{- 0.07}$ & $ \left. \! \right) \cdot 10^{-1}$ \\
\multicolumn{2}{|c|}{} & $0.10$ & $0.15$ & $$ &  & $0.13$$$ & $\left( \right.\! $ \hspace{-0.5mm} & $  3.61$ & $\pm  0.26$ & $^{+ 0.04}_{- 0.04}$ & $ \left. \! \right) \cdot 10^{-1}$ \\
\multicolumn{2}{|c|}{} & $0.15$ & $0.23$ & $$ &  & $0.18$$$ & $\left( \right.\! $ \hspace{-0.5mm} & $  2.70$ & $\pm  0.21$ & $^{+ 0.04}_{- 0.08}$ & $ \left. \! \right) \cdot 10^{-1}$ \\
\multicolumn{2}{|c|}{} & $0.23$ & $0.35$ & $$ &  & $0.25$$$ & $\left( \right.\! $ \hspace{-0.5mm} & $  2.17$ & $\pm  0.19$ & $^{+ 0.06}_{- 0.06}$ & $ \left. \! \right) \cdot 10^{-1}$ \\
\multicolumn{2}{|c|}{} & $0.35$ & $0.53$ & $$ &  & $0.40$$$ & $\left( \right.\! $ \hspace{-0.5mm} & $  1.12$ & $\pm  0.14$ & $^{+ 0.04}_{- 0.01}$ & $ \left. \! \right) \cdot 10^{-1}$ \\
\hline $ 5600.$ & $ 9000.$ & $\left( \right.\!0.70$ & $1.50$ & $ \left. \! \right) \cdot 10^{-1}$ & $ 8000$ & $1.30$$\,\cdot\,10^{-1}$ & $\left( \right.\! $ \hspace{-0.5mm} & $  3.06$ & $\pm  0.29$ & $^{+ 0.08}_{- 0.07}$ & $ \left. \! \right) \cdot 10^{-1}$ \\
\multicolumn{2}{|c|}{} & $0.15$ & $0.23$ & $$ &  & $0.18$$$ & $\left( \right.\! $ \hspace{-0.5mm} & $  2.73$ & $\pm  0.28$ & $^{+ 0.03}_{- 0.07}$ & $ \left. \! \right) \cdot 10^{-1}$ \\
\multicolumn{2}{|c|}{} & $0.23$ & $0.35$ & $$ &  & $0.25$$$ & $\left( \right.\! $ \hspace{-0.5mm} & $  2.01$ & $\pm  0.24$ & $^{+ 0.08}_{- 0.05}$ & $ \left. \! \right) \cdot 10^{-1}$ \\
\multicolumn{2}{|c|}{} & $0.35$ & $0.53$ & $$ &  & $0.40$$$ & $\left( \right.\! $ \hspace{-0.5mm} & $   9.7$ & $\pm   1.7$ & $^{+  0.5}_{-  0.3}$ & $ \left. \! \right) \cdot 10^{-2}$ \\
\multicolumn{2}{|c|}{} & $0.53$ & $1.00$ & $$ &  & $0.65$$$ & $\left( \right.\! $ \hspace{-0.5mm} & $   1.3$ & $^{+   0.5}_{-  0.4}$ & $^{+  0.1}_{-  0.1}$ & $ \left. \! \right) \cdot 10^{-2}$ \\
\hline $ 9000.$ & $15000.$ & $\left( \right.\!0.90$ & $2.30$ & $ \left. \! \right) \cdot 10^{-1}$ & $12000$ & $1.80$$\,\cdot\,10^{-1}$ & $\left( \right.\! $ \hspace{-0.5mm} & $   3.2$ & $\pm   0.4$ & $^{+  0.1}_{-  0.2}$ & $ \left. \! \right) \cdot 10^{-1}$ \\
\multicolumn{2}{|c|}{} & $0.23$ & $0.35$ & $$ &  & $0.25$$$ & $\left( \right.\! $ \hspace{-0.5mm} & $   2.0$ & $\pm   0.4$ & $^{+  0.0}_{-  0.1}$ & $ \left. \! \right) \cdot 10^{-1}$ \\
\multicolumn{2}{|c|}{} & $0.35$ & $0.53$ & $$ &  & $0.40$$$ & $\left( \right.\! $ \hspace{-0.5mm} & $   9.0$ & $\pm   2.3$ & $^{+  0.6}_{-  0.8}$ & $ \left. \! \right) \cdot 10^{-2}$ \\
\hline $15000.$ & $25000.$ & $0.15$ & $0.35$ & $$ & $20000$ & $0.25$$$ & $\left( \right.\! $ \hspace{-0.5mm} & $   8.8$ & $\pm   2.5$ & $^{+  0.8}_{-  0.2}$ & $ \left. \! \right) \cdot 10^{-2}$ \\
\multicolumn{2}{|c|}{} & $0.35$ & $1.00$ & $$ &  & $0.40$$$ & $\left( \right.\! $ \hspace{-0.5mm} & $   6.3$ & $^{+   3.8}_{-  2.5}$ & $^{+  1.0}_{-  1.1}$ & $ \left. \! \right) \cdot 10^{-2}$ \\
\hline $25000.$ & $50000.$ & $0.25$ & $1.00$ & $$ & $30000$ & $0.40$$$ & $\left( \right.\! $ \hspace{-0.5mm} & $   5.4$ & $^{+   3.6}_{-  2.3}$ & $^{+  0.3}_{-  2.3}$ & $ \left. \! \right) \cdot 10^{-2}$ \\
\hline
\end{tabular}
}

  \end{center}    
  \normalsize
  \caption[this space for rent]{
    The reduced cross-section $\tilde{\sigma}^{e^+ p}$ for the
    reaction $e^{+} p \rightarrow e^{+} X$.  
    The following quantities are given for each bin: the $Q^2$ and $x$
    ranges, the values at which the cross section is quoted, \qqc\ and
    $x_c$, and the measured reduced cross-section, $\tilde{\sigma}^{e^+ p}$,
    corrected to electroweak Born level. 
    The first uncertainty on the measured cross section is the
    statistical uncertainty and the second is the systematic
    uncertainty. The uncertainty on the measured luminosity of $2.5\%$ 
    is not included in the total systematic uncertainty.  
  }
  \label{tab-dsdqdx_2}
\end{table}
\clearpage
\begin{table} [!ht] 
  \begin{center}
    {\tiny
\renewcommand{\arraystretch}{1.2}
\begin{tabular}{|c|l|c|c|c||c|c|c|c|c|c|c|c|c|}
\hline
{$Q^2_c$} & 
\multicolumn{1}{c|}{$x_c$} & 
 $\tilde{\sigma}(e^+p) $&
stat.  & 
total sys.  & 
uncor. sys. &
$\delta_1$ &
$\delta_2$ &
$\delta_3$ &
$\delta_4$ &
$\delta_5$ &
$\delta_6$ &
$\delta_7$ &
$\delta_8$ \\
{($\gev^2$)} & 
$$ &
 &
(\%) & 
(\%) &
(\%) &
(\%) &
(\%) &
(\%) &
(\%) &
(\%) &
(\%) &
(\%) &
(\%) \\
\hline
\hline
  200      & $0.50\,\cdot\,10^{-2}$ & $
  1.13$ &
$^{+ 1.5  } _{-1.5  }  $ & $^{+ 1.7  } _{-2.7  }  $ & $^{+ 0.5  } _{-0.4  }  $ & $^{-0.3  }  _{+ 0.4  } $ & $^{}_{+ 0.0  } $ & $^{-0.1  }  _{-0.1  }  $ & $^{-1.0  }  _{+ 1.6  } $ & $^{-0.3  }  _{+ 0.3  } $ & $^{+ 0.2  } _{-0.4  }  
$                                 & $^{}_{-2.3  }  $& $^{+ 0.0  } _{- 0.0  } $ \\
           & $0.80\,\cdot\,10^{-2}$ & $
  0.95$ &
$^{+ 1.4  } _{-1.4  }  $ & $^{+ 1.5  } _{-1.2  }  $ & $^{+ 0.8  } _{-0.4  }  $ & $^{-0.1  }  _{+ 0.1  } $ & $^{}_{+ 0.0  } $ & $^{- 0.0  } _{+ 0.1  } $ & $^{-0.8  }  _{+ 1.0  } $ & $^{-0.3  }  _{+ 0.3  } $ & $^{+ 0.2  } _{-0.6  }  
$                                 & $^{}_{+ 0.8  } $& $^{- 0.0  } _{+ 0.0  } $ \\
           & $0.13\,\cdot\,10^{-1}$ & $
  0.82$ &
$^{+ 1.4  } _{-1.4  }  $ & $^{+ 0.7  } _{-1.1  }  $ & $^{+ 0.3  } _{-0.3  }  $ & $^{-0.1  }  _{+ 0.1  } $ & $^{}_{+ 0.0  } $ & $^{+ 0.0  } _{+ 0.0  } $ & $^{-0.6  }  _{+ 0.5  } $ & $^{-0.3  }  _{+ 0.3  } $ & $^{+ 0.3  } _{-0.7  }  
$                                 & $^{}_{-0.3  }  $& $^{- 0.0  } _{+ 0.1  } $ \\
           & $0.21\,\cdot\,10^{-1}$ & $
  0.69$ &
$^{+ 1.8  } _{-1.8  }  $ & $^{+ 1.0  } _{-1.5  }  $ & $^{+ 0.4  } _{-0.6  }  $ & $^{-0.2  }  _{+ 0.1  } $ & $^{}_{+ 0.0  } $ & $^{+ 0.0  } _{+ 0.0  } $ & $^{-0.3  }  _{+ 0.1  } $ & $^{-0.3  }  _{+ 0.3  } $ & $^{+ 0.4  } _{-0.8  }  
$                                 & $^{}_{-1.2  }  $& $^{-0.2  }  _{+ 0.2  } $ \\
           & $0.32\,\cdot\,10^{-1}$ & $
  0.59$ &
$^{+ 1.8  } _{-1.8  }  $ & $^{+ 1.9  } _{-1.5  }  $ & $^{+ 1.2  } _{-1.0  }  $ & $^{-0.2  }  _{+ 0.1  } $ & $^{}_{+ 0.0  } $ & $^{+ 0.0  } _{+ 0.0  } $ & $^{-0.2  }  _{-0.2  }  $ & $^{-0.3  }  _{+ 0.3  } $ & $^{+ 0.6  } _{-1.5  }  
$                                 & $^{}_{+ 0.5  } $& $^{-0.3  }  _{+ 0.3  } $ \\
           & $0.50\,\cdot\,10^{-1}$ & $
  0.53$ &
$^{+ 1.7  } _{-1.7  }  $ & $^{+ 0.8  } _{-2.0  }  $ & $^{+ 0.5  } _{-1.3  }  $ & $^{-0.1  }  _{+ 0.2  } $ & $^{}_{+ 0.0  } $ & $^{+ 0.0  } _{+ 0.0  } $ & $^{-0.4  }  _{- 0.0  } $ & $^{-0.3  }  _{+ 0.3  } $ & $^{+ 0.2  } _{-0.5  }  
$                                 & $^{}_{-1.4  }  $& $^{-0.4  }  _{+ 0.4  } $ \\
           & $0.80\,\cdot\,10^{-1}$ & $
  0.43$ &
$^{+ 1.5  } _{-1.5  }  $ & $^{+ 5.3  } _{-1.1  }  $ & $^{+ 0.5  } _{-0.4  }  $ & $^{-0.1  }  _{+ 0.1  } $ & $^{}_{+ 0.0  } $ & $^{+ 0.0  } _{+ 0.0  } $ & $^{+ 0.2  } _{+ 0.1  } $ & $^{-0.3  }  _{+ 0.3  } $ & $^{+ 0.8  } _{-1.9  }  
$                                 & $^{}_{+ 4.8  } $& $^{-0.4  }  _{+ 0.4  } $ \\
           & $0.18$ & $
  0.33$ &
$^{+ 1.9  } _{-1.9  }  $ & $^{+ 2.0  } _{ -13. }  $ & $^{+ 1.3  } _{-3.4  }  $ & $^{-0.1  }  _{+ 0.2  } $ & $^{}_{+ 0.0  } $ & $^{+ 0.0  } _{+ 0.0  } $ & $^{+ 0.1  } _{+ 0.5  } $ & $^{-0.4  }  _{+ 0.4  } $ & $^{+ 0.6  } _{-1.4  }  
$                                 & $^{}_{ -13. }  $& $^{+ 0.3  } _{-0.3  }  $ \\
\hline
  250      & $0.80\,\cdot\,10^{-2}$ & $
  0.97$ &
$^{+ 1.7  } _{-1.7  }  $ & $^{+ 1.0  } _{-1.4  }  $ & $^{+ 0.8  } _{-0.7  }  $ & $^{-0.2  }  _{+ 0.1  } $ & $^{}_{+ 0.0  } $ & $^{+ 0.0  } _{+ 0.0  } $ & $^{-0.9  }  _{+ 0.7  } $ & $^{-0.3  }  _{+ 0.3  } $ & $^{+ 0.1  } _{-0.2  }  
$                                 & $^{}_{-0.7  }  $& $^{- 0.0  } _{+ 0.0  } $ \\
           & $0.13\,\cdot\,10^{-1}$ & $
  0.83$ &
$^{+ 1.7  } _{-1.7  }  $ & $^{+ 1.2  } _{-0.9  }  $ & $^{+ 0.6  } _{-0.2  }  $ & $^{-0.1  }  _{+ 0.1  } $ & $^{}_{+ 0.0  } $ & $^{+ 0.0  } _{- 0.0  } $ & $^{-0.5  }  _{+ 0.5  } $ & $^{-0.3  }  _{+ 0.3  } $ & $^{+ 0.1  } _{-0.3  }  
$                                 & $^{}_{+ 0.9  } $& $^{- 0.0  } _{+ 0.0  } $ \\
           & $0.21\,\cdot\,10^{-1}$ & $
  0.70$ &
$^{+ 2.0  } _{-2.0  }  $ & $^{+ 1.8  } _{-0.7  }  $ & $^{+ 0.9  } _{-0.3  }  $ & $^{-0.2  }  _{+ 0.2  } $ & $^{}_{+ 0.0  } $ & $^{+ 0.0  } _{+ 0.0  } $ & $^{+ 0.3  } _{+ 0.2  } $ & $^{-0.3  }  _{+ 0.3  } $ & $^{+ 0.2  } _{-0.5  }  
$                                 & $^{}_{+ 1.4  } $& $^{-0.1  }  _{+ 0.1  } $ \\
           & $0.32\,\cdot\,10^{-1}$ & $
  0.59$ &
$^{+ 2.1  } _{-2.1  }  $ & $^{+ 1.6  } _{-1.5  }  $ & $^{+ 0.8  } _{-0.5  }  $ & $^{-0.2  }  _{+ 0.2  } $ & $^{}_{+ 0.0  } $ & $^{+ 0.0  } _{+ 0.0  } $ & $^{-1.1  }  _{- 0.0  } $ & $^{-0.3  }  _{+ 0.3  } $ & $^{+ 0.5  } _{-1.3  }  
$                                 & $^{}_{+ 0.0  } $& $^{-0.3  }  _{+ 0.3  } $ \\
           & $0.50\,\cdot\,10^{-1}$ & $
  0.53$ &
$^{+ 2.0  } _{-2.0  }  $ & $^{+ 1.2  } _{-1.6  }  $ & $^{+ 1.0  } _{-0.4  }  $ & $^{-0.1  }  _{+ 0.1  } $ & $^{}_{+ 0.0  } $ & $^{+ 0.0  } _{+ 0.0  } $ & $^{-1.0  }  _{+ 0.3  } $ & $^{-0.3  }  _{+ 0.3  } $ & $^{+ 0.3  } _{-0.7  }  
$                                 & $^{}_{-0.8  }  $& $^{-0.4  }  _{+ 0.4  } $ \\
           & $0.80\,\cdot\,10^{-1}$ & $
  0.42$ &
$^{+ 1.8  } _{-1.8  }  $ & $^{+ 5.3  } _{-1.7  }  $ & $^{+ 0.4  } _{-0.8  }  $ & $^{-0.3  }  _{+ 0.2  } $ & $^{}_{+ 0.0  } $ & $^{+ 0.0  } _{+ 0.0  } $ & $^{- 0.0  } _{+ 0.6  } $ & $^{-0.3  }  _{+ 0.3  } $ & $^{+ 1.4  } _{-3.3  }  
$                                 & $^{}_{+ 4.0  } $& $^{-0.3  }  _{+ 0.4  } $ \\
           & $0.18$ & $
  0.32$ &
$^{+ 2.0  } _{-2.0  }  $ & $^{+ 3.4  } _{-5.3  }  $ & $^{+ 0.4  } _{-1.6  }  $ & $^{-0.2  }  _{+ 0.2  } $ & $^{}_{+ 0.0  } $ & $^{+ 0.0  } _{- 0.0  } $ & $^{-0.6  }  _{- 0.0  } $ & $^{-0.4  }  _{+ 0.4  } $ & $^{+ 1.4  } _{-3.3  }  
$                                 & $^{}_{-4.8  }  $& $^{+ 0.3  } _{-0.3  }  $ \\
\hline
  350      & $0.80\,\cdot\,10^{-2}$ & $
  0.99$ &
$^{+ 2.1  } _{-2.1  }  $ & $^{+ 2.7  } _{-1.2  }  $ & $^{+ 1.0  } _{-0.3  }  $ & $^{-0.2  }  _{+ 0.1  } $ & $^{}_{+ 0.0  } $ & $^{+ 0.0  } _{- 0.0  } $ & $^{-0.7  }  _{+ 2.6  } $ & $^{-0.2  }  _{+ 0.2  } $ & $^{+ 0.1  } _{-0.3  }  
$                                 & $^{}_{-0.5  }  $& $^{+ 0.0  } _{+ 0.0  } $ \\
           & $0.13\,\cdot\,10^{-1}$ & $
  0.82$ &
$^{+ 1.9  } _{-1.9  }  $ & $^{+ 1.3  } _{-1.7  }  $ & $^{+ 1.1  } _{-0.3  }  $ & $^{-0.1  }  _{+ 0.1  } $ & $^{}_{+ 0.0  } $ & $^{- 0.0  } _{+ 0.1  } $ & $^{-0.7  }  _{+ 0.5  } $ & $^{-0.3  }  _{+ 0.3  } $ & $^{+ 0.6  } _{-1.3  }  
$                                 & $^{}_{+ 0.0  } $& $^{- 0.0  } _{+ 0.0  } $ \\
           & $0.21\,\cdot\,10^{-1}$ & $
  0.69$ &
$^{+ 2.3  } _{-2.3  }  $ & $^{+ 0.9  } _{-1.5  }  $ & $^{+ 0.8  } _{-0.9  }  $ & $^{-0.2  }  _{+ 0.2  } $ & $^{}_{+ 0.0  } $ & $^{- 0.0  } _{+ 0.0  } $ & $^{-0.5  }  _{+ 0.4  } $ & $^{-0.4  }  _{+ 0.4  } $ & $^{+ 0.2  } _{-0.4  }  
$                                 & $^{}_{-0.8  }  $& $^{-0.1  }  _{+ 0.1  } $ \\
           & $0.32\,\cdot\,10^{-1}$ & $
  0.61$ &
$^{+ 2.3  } _{-2.3  }  $ & $^{+ 1.2  } _{-0.9  }  $ & $^{+ 0.9  } _{-0.8  }  $ & $^{-0.1  }  _{+ 0.2  } $ & $^{}_{+ 0.0  } $ & $^{+ 0.0  } _{+ 0.0  } $ & $^{+ 0.2  } _{+ 0.6  } $ & $^{-0.3  }  _{+ 0.3  } $ & $^{+ 0.1  } _{-0.2  }  
$                                 & $^{}_{+ 0.3  } $& $^{-0.3  }  _{+ 0.3  } $ \\
           & $0.50\,\cdot\,10^{-1}$ & $
  0.51$ &
$^{+ 2.2  } _{-2.2  }  $ & $^{+ 0.9  } _{-1.5  }  $ & $^{+ 0.9  } _{-0.4  }  $ & $^{-0.2  }  _{+ 0.3  } $ & $^{}_{+ 0.0  } $ & $^{+ 0.0  } _{- 0.0  } $ & $^{-0.5  }  _{+ 0.1  } $ & $^{-0.3  }  _{+ 0.3  } $ & $^{+ 0.2  } _{-0.4  }  
$                                 & $^{}_{-1.1  }  $& $^{-0.4  }  _{+ 0.4  } $ \\
           & $0.80\,\cdot\,10^{-1}$ & $
  0.43$ &
$^{+ 1.9  } _{-1.9  }  $ & $^{+ 1.7  } _{-0.7  }  $ & $^{+ 0.5  } _{-0.3  }  $ & $^{-0.2  }  _{+ 0.1  } $ & $^{}_{+ 0.0  } $ & $^{+ 0.0  } _{+ 0.0  } $ & $^{-0.3  }  _{+ 0.6  } $ & $^{-0.3  }  _{+ 0.3  } $ & $^{+ 0.2  } _{-0.5  }  
$                                 & $^{}_{+ 1.4  } $& $^{-0.3  }  _{+ 0.3  } $ \\
           & $0.18$ & $
  0.31$ &
$^{+ 2.1  } _{-2.1  }  $ & $^{+ 4.1  } _{-1.3  }  $ & $^{+ 1.9  } _{-0.8  }  $ & $^{-0.3  }  _{+ 0.2  } $ & $^{}_{+ 0.0  } $ & $^{+ 0.0  } _{+ 0.0  } $ & $^{-0.4  }  _{+ 0.7  } $ & $^{-0.3  }  _{+ 0.3  } $ & $^{+ 0.8  } _{-1.8  }  
$                                 & $^{}_{+ 2.9  } $& $^{+ 0.3  } _{-0.2  }  $ \\
\hline
  450      & $0.80\,\cdot\,10^{-2}$ & $
  1.05$ &
$^{+ 2.3  } _{-2.3  }  $ & $^{+ 3.7  } _{-1.9  }  $ & $^{+ 0.5  } _{-0.6  }  $ & $^{-0.7  }  _{+ 0.8  } $ & $^{}_{+ 0.0  } $ & $^{- 0.0  } _{+ 0.3  } $ & $^{-0.3  }  _{+ 3.6  } $ & $^{-0.3  }  _{+ 0.3  } $ & $^{+ 0.1  } _{-0.2  }  
$                                 & $^{}_{-1.6  }  $& $^{- 0.0  } _{- 0.0  } $ \\
           & $0.13\,\cdot\,10^{-1}$ & $
  0.84$ &
$^{+ 2.6  } _{-2.6  }  $ & $^{+ 1.2  } _{-1.3  }  $ & $^{+ 1.2  } _{-0.4  }  $ & $^{- 0.0  } _{+ 0.1  } $ & $^{}_{+ 0.0  } $ & $^{+ 0.0  } _{-0.1  }  $ & $^{-0.5  }  _{+ 0.7  } $ & $^{-0.2  }  _{+ 0.2  } $ & $^{+ 0.0  } _{-0.1  }  
$                                 & $^{}_{-0.8  }  $& $^{- 0.0  } _{+ 0.0  } $ \\
           & $0.21\,\cdot\,10^{-1}$ & $
  0.68$ &
$^{+ 3.0  } _{-3.0  }  $ & $^{+ 2.4  } _{-1.4  }  $ & $^{+ 0.7  } _{-0.4  }  $ & $^{-0.2  }  _{+ 0.1  } $ & $^{}_{+ 0.0  } $ & $^{+ 0.1  } _{+ 0.1  } $ & $^{+ 0.4  } _{+ 0.5  } $ & $^{-0.2  }  _{+ 0.2  } $ & $^{+ 0.5  } _{-1.2  }  
$                                 & $^{}_{+ 2.2  } $& $^{-0.1  }  _{+ 0.1  } $ \\
           & $0.32\,\cdot\,10^{-1}$ & $
  0.63$ &
$^{+ 2.8  } _{-2.8  }  $ & $^{+ 0.9  } _{-2.6  }  $ & $^{+ 0.8  } _{-0.8  }  $ & $^{-0.1  }  _{+ 0.2  } $ & $^{}_{+ 0.0  } $ & $^{+ 0.0  } _{+ 0.0  } $ & $^{-1.0  }  _{+ 0.3  } $ & $^{-0.3  }  _{+ 0.3  } $ & $^{+ 0.4  } _{-0.9  }  
$                                 & $^{}_{-1.9  }  $& $^{-0.2  }  _{+ 0.2  } $ \\
           & $0.50\,\cdot\,10^{-1}$ & $
  0.51$ &
$^{+ 2.6  } _{-2.6  }  $ & $^{+ 1.7  } _{-1.0  }  $ & $^{+ 1.0  } _{-0.2  }  $ & $^{-0.2  }  _{+ 0.2  } $ & $^{}_{+ 0.0  } $ & $^{+ 0.0  } _{- 0.0  } $ & $^{-0.7  }  _{+ 0.3  } $ & $^{-0.3  }  _{+ 0.3  } $ & $^{+ 0.1  } _{-0.3  }  
$                                 & $^{}_{+ 1.3  } $& $^{-0.3  }  _{+ 0.3  } $ \\
           & $0.80\,\cdot\,10^{-1}$ & $
  0.45$ &
$^{+ 2.5  } _{-2.5  }  $ & $^{+ 1.3  } _{-0.6  }  $ & $^{+ 0.5  } _{-0.2  }  $ & $^{-0.2  }  _{+ 0.1  } $ & $^{}_{+ 0.0  } $ & $^{+ 0.0  } _{- 0.0  } $ & $^{-0.2  }  _{+ 0.2  } $ & $^{-0.3  }  _{+ 0.3  } $ & $^{+ 0.1  } _{-0.1  }  
$                                 & $^{}_{+ 1.2  } $& $^{-0.2  }  _{+ 0.2  } $ \\
           & $0.13$ & $
  0.36$ &
$^{+ 2.7  } _{-2.7  }  $ & $^{+ 2.7  } _{-1.1  }  $ & $^{+ 0.6  } _{-0.4  }  $ & $^{-0.2  }  _{+ 0.1  } $ & $^{}_{+ 0.0  } $ & $^{+ 0.0  } _{-0.1  }  $ & $^{+ 0.0  } _{+ 0.1  } $ & $^{-0.4  }  _{+ 0.4  } $ & $^{+ 0.9  } _{-2.1  }  
$                                 & $^{}_{+ 1.6  } $& $^{+ 0.1  } _{-0.1  }  $ \\
           & $0.25$ & $
  0.26$ &
$^{+ 3.0  } _{-3.0  }  $ & $^{+ 1.6  } _{-4.5  }  $ & $^{+ 0.3  } _{-3.2  }  $ & $^{-0.3  }  _{+ 0.1  } $ & $^{}_{+ 0.0  } $ & $^{+ 0.0  } _{+ 0.0  } $ & $^{+ 0.1  } _{+ 0.2  } $ & $^{-0.4  }  _{+ 0.4  } $ & $^{+ 0.6  } _{-1.5  }  
$                                 & $^{}_{-3.0  }  $& $^{+ 0.4  } _{-0.5  }  $ \\
\hline
  650      & $0.13\,\cdot\,10^{-1}$ & $
  0.86$ &
$^{+ 2.4  } _{-2.4  }  $ & $^{+ 2.8  } _{-1.1  }  $ & $^{+ 0.7  } _{-0.6  }  $ & $^{-0.1  }  _{+ 0.2  } $ & $^{}_{+ 0.0  } $ & $^{-0.1  }  _{-0.1  }  $ & $^{-0.3  }  _{+ 2.6  } $ & $^{-0.4  }  _{+ 0.4  } $ & $^{+ 0.4  } _{-0.8  }  
$                                 & $^{}_{+ 0.6  } $& $^{- 0.0  } _{+ 0.0  } $ \\
           & $0.21\,\cdot\,10^{-1}$ & $
  0.74$ &
$^{+ 3.0  } _{-3.0  }  $ & $^{+ 0.7  } _{-1.2  }  $ & $^{+ 0.8  } _{-0.6  }  $ & $^{-0.1  }  _{+ 0.1  } $ & $^{}_{+ 0.0  } $ & $^{+ 0.1  } _{- 0.0  } $ & $^{-0.6  }  _{-0.1  }  $ & $^{-0.4  }  _{+ 0.4  } $ & $^{+ 0.1  } _{-0.2  }  
$                                 & $^{}_{+ 0.3  } $& $^{-0.1  }  _{+ 0.1  } $ \\
           & $0.32\,\cdot\,10^{-1}$ & $
  0.63$ &
$^{+ 3.4  } _{-3.4  }  $ & $^{+ 0.7  } _{-3.0  }  $ & $^{+ 0.4  } _{-0.9  }  $ & $^{-0.3  }  _{+ 0.2  } $ & $^{}_{+ 0.0  } $ & $^{- 0.0  } _{+ 0.1  } $ & $^{-0.3  }  _{+ 0.4  } $ & $^{-0.2  }  _{+ 0.2  } $ & $^{+ 0.4  } _{-1.0  }  
$                                 & $^{}_{-2.6  }  $& $^{-0.2  }  _{+ 0.2  } $ \\
           & $0.50\,\cdot\,10^{-1}$ & $
  0.53$ &
$^{+ 3.3  } _{-3.3  }  $ & $^{+ 0.9  } _{-0.6  }  $ & $^{+ 0.6  } _{-0.4  }  $ & $^{-0.2  }  _{+ 0.1  } $ & $^{}_{+ 0.0  } $ & $^{+ 0.0  } _{-0.1  }  $ & $^{+ 0.1  } _{+ 0.2  } $ & $^{-0.2  }  _{+ 0.1  } $ & $^{+ 0.1  } _{-0.2  }  
$                                 & $^{}_{+ 0.5  } $& $^{-0.3  }  _{+ 0.3  } $ \\
           & $0.80\,\cdot\,10^{-1}$ & $
  0.45$ &
$^{+ 3.4  } _{-3.4  }  $ & $^{+ 1.6  } _{-1.9  }  $ & $^{+ 0.8  } _{-0.6  }  $ & $^{-0.2  }  _{+ 0.2  } $ & $^{}_{+ 0.0  } $ & $^{+ 0.0  } _{-0.2  }  $ & $^{-0.4  }  _{-0.4  }  $ & $^{-0.2  }  _{+ 0.2  } $ & $^{+ 0.6  } _{-1.4  }  
$                                 & $^{}_{-1.5  }  $& $^{-0.1  }  _{+ 0.1  } $ \\
           & $0.13$ & $
  0.35$ &
$^{+ 3.6  } _{-3.6  }  $ & $^{+ 2.3  } _{-1.2  }  $ & $^{+ 0.7  } _{-0.8  }  $ & $^{-0.1  }  _{+ 0.2  } $ & $^{}_{+ 0.0  } $ & $^{+ 0.0  } _{+ 0.0  } $ & $^{-0.6  }  _{+ 1.5  } $ & $^{-0.2  }  _{+ 0.2  } $ & $^{+ 0.5  } _{-1.2  }  
$                                 & $^{}_{+ 1.0  } $& $^{+ 0.1  } _{-0.1  }  $ \\
           & $0.25$ & $
  0.25$ &
$^{+ 3.8  } _{-3.8  }  $ & $^{+ 3.0  } _{-2.1  }  $ & $^{+ 2.0  } _{-1.8  }  $ & $^{-0.1  }  _{+ 0.3  } $ & $^{}_{+ 0.0  } $ & $^{+ 0.0  } _{+ 0.0  } $ & $^{-0.5  }  _{+ 0.8  } $ & $^{-0.2  }  _{+ 0.2  } $ & $^{+ 0.8  } _{-1.9  }  
$                                 & $^{}_{+ 0.8  } $& $^{+ 0.4  } _{-0.5  }  $ \\
\hline
  800      & $0.13\,\cdot\,10^{-1}$ & $
  0.86$ &
$^{+ 2.8  } _{-2.8  }  $ & $^{+ 3.0  } _{-1.7  }  $ & $^{+ 1.3  } _{-0.6  }  $ & $^{-0.5  }  _{+ 0.5  } $ & $^{}_{-0.1  }  $ & $^{- 0.0  } _{+ 0.4  } $ & $^{-0.8  }  _{+ 2.5  } $ & $^{-0.3  }  _{+ 0.3  } $ & $^{+ 0.3  } _{-0.8  }  
$                                 & $^{}_{-1.0  }  $& $^{- 0.0  } _{+ 0.0  } $ \\
           & $0.21\,\cdot\,10^{-1}$ & $
  0.74$ &
$^{+ 3.6  } _{-3.6  }  $ & $^{+ 3.8  } _{-1.0  }  $ & $^{+ 1.5  } _{-0.3  }  $ & $^{-0.2  }  _{+ 0.1  } $ & $^{}_{+ 0.0  } $ & $^{- 0.0  } _{+ 0.2  } $ & $^{+ 0.2  } _{+ 1.8  } $ & $^{-0.4  }  _{+ 0.4  } $ & $^{+ 0.3  } _{-0.7  }  
$                                 & $^{}_{+ 2.9  } $& $^{-0.1  }  _{+ 0.1  } $ \\
           & $0.32\,\cdot\,10^{-1}$ & $
  0.66$ &
$^{+ 3.5  } _{-3.5  }  $ & $^{+ 0.9  } _{-2.3  }  $ & $^{+ 0.4  } _{-1.6  }  $ & $^{-0.1  }  _{+ 0.1  } $ & $^{}_{+ 0.0  } $ & $^{-0.1  }  _{+ 0.1  } $ & $^{-0.5  }  _{+ 0.6  } $ & $^{-0.4  }  _{+ 0.4  } $ & $^{+ 0.1  } _{-0.2  }  
$                                 & $^{}_{-1.4  }  $& $^{-0.2  }  _{+ 0.2  } $ \\
           & $0.50\,\cdot\,10^{-1}$ & $
  0.52$ &
$^{+ 3.5  } _{-3.5  }  $ & $^{+ 2.2  } _{-0.7  }  $ & $^{+ 1.4  } _{-0.3  }  $ & $^{-0.2  }  _{+ 0.2  } $ & $^{}_{+ 0.0  } $ & $^{+ 0.0  } _{- 0.0  } $ & $^{+ 0.9  } _{+ 0.1  } $ & $^{-0.4  }  _{+ 0.4  } $ & $^{+ 0.2  } _{-0.5  }  
$                                 & $^{}_{+ 1.2  } $& $^{-0.3  }  _{+ 0.3  } $ \\
           & $0.80\,\cdot\,10^{-1}$ & $
  0.45$ &
$^{+ 3.6  } _{-3.6  }  $ & $^{+ 1.4  } _{-1.2  }  $ & $^{+ 0.8  } _{-1.0  }  $ & $^{-0.1  }  _{+ 0.1  } $ & $^{}_{+ 0.0  } $ & $^{+ 0.0  } _{- 0.0  } $ & $^{+ 0.3  } _{+ 1.1  } $ & $^{-0.4  }  _{+ 0.4  } $ & $^{+ 0.1  } _{-0.3  }  
$                                 & $^{}_{+ 0.1  } $& $^{-0.1  }  _{+ 0.1  } $ \\
           & $0.13$ & $
  0.36$ &
$^{+ 4.0  } _{-4.0  }  $ & $^{+ 1.6  } _{-1.5  }  $ & $^{+ 0.4  } _{-0.7  }  $ & $^{- 0.0  } _{+ 0.2  } $ & $^{}_{+ 0.0  } $ & $^{+ 0.0  } _{+ 0.0  } $ & $^{-1.3  }  _{+ 0.0  } $ & $^{-0.4  }  _{+ 0.4  } $ & $^{+ 0.1  } _{-0.2  }  
$                                 & $^{}_{+ 1.4  } $& $^{+ 0.1  } _{-0.1  }  $ \\
           & $0.25$ & $
  0.26$ &
$^{+ 4.5  } _{-4.5  }  $ & $^{+ 5.5  } _{-2.1  }  $ & $^{+ 0.5  } _{-0.9  }  $ & $^{-0.1  }  _{+ 0.1  } $ & $^{}_{+ 0.0  } $ & $^{+ 0.0  } _{+ 0.0  } $ & $^{+ 0.9  } _{-0.3  }  $ & $^{-0.2  }  _{+ 0.2  } $ & $^{+ 1.8  } _{-4.2  }  
$                                 & $^{}_{+ 3.4  } $& $^{+ 0.4  } _{-0.5  }  $ \\
\hline
\end{tabular}
}

  \end{center}    
  \normalsize
  \caption[this space for rent]{
    Systematic uncertainties with bin-to-bin correlations
    for the reduced cross-section $\tilde{\sigma}^{e^+ p}$. 
    The left part of the table contains the quoted $Q^2$ and $x$
    values, $Q^2_c$ and $x_c$, the measured cross-section 
    $\tilde{\sigma}^{e^+ p}$ corrected to the electroweak Born level,
    the statistical uncertainty and the total systematic uncertainty. 
    The uncertainty on the measured luminosity of $2.5\%$ is not 
    included in the total systematic uncertainty.
    The right part of the table lists the total uncorrelated
    systematic uncertainty followed by the bin-to-bin correlated
    systematic uncertainties $\delta_1$--\,$\delta_8$ defined in
    the text. 
    For the latter, the upper (lower)
    numbers refer to positive (negative) variation of e.g. the cut
    value, whereas the signs of the numbers reflect the direction of
    change in the cross sections. 
    }
  \label{tab-dsdxq_c1}
\end{table}
\clearpage
\begin{table} [!ht] 
  \strut \vspace{-2cm}
  \begin{center}
    {\tiny
\renewcommand{\arraystretch}{1.2}
\begin{tabular}{|c|l|c|c|c||c|c|c|c|c|c|c|c|c|}
\hline
{$Q^2_c$} & 
\multicolumn{1}{c|}{$x_c$} & 
 $\tilde{\sigma}(e^+p) $&
stat.  & 
total sys.  & 
uncor. sys. &
$\delta_1$ &
$\delta_2$ &
$\delta_3$ &
$\delta_4$ &
$\delta_5$ &
$\delta_6$ &
$\delta_7$ &
$\delta_8$ \\
{($\gev^2$)} & 
$$ &
 &
(\%) & 
(\%) &
(\%) &
(\%) &
(\%) &
(\%) &
(\%) &
(\%) &
(\%) &
(\%) &
(\%) \\
\hline
\hline
 1200      & $0.14\,\cdot\,10^{-1}$ & $
  0.85$ &
$^{+ 3.5  } _{-3.5  }  $ & $^{+ 5.0  } _{-3.1  }  $ & $^{+ 0.7  } _{-1.1  }  $ & $^{-0.5  }  _{+ 0.7  } $ & $^{}_{-1.5  }  $ & $^{+ 0.0  } _{-0.3  }  $ & $^{-2.3  }  _{+ 3.9  } $ & $^{-0.3  }  _{+ 0.3  } $ & $^{+ 0.2  } _{-0.4  }  
$                                 & $^{}_{+ 2.8  } $& $^{- 0.0  } _{+ 0.0  } $ \\
           & $0.21\,\cdot\,10^{-1}$ & $
  0.79$ &
$^{+ 3.5  } _{-3.5  }  $ & $^{+ 2.6  } _{-1.8  }  $ & $^{+ 1.2  } _{-0.3  }  $ & $^{-0.1  }  _{+ 0.1  } $ & $^{}_{+ 0.0  } $ & $^{+ 0.1  } _{+ 0.3  } $ & $^{-1.0  }  _{+ 2.2  } $ & $^{-0.3  }  _{+ 0.3  } $ & $^{+ 0.6  } _{-1.4  }  
$                                 & $^{}_{-0.1  }  $& $^{-0.1  }  _{+ 0.1  } $ \\
           & $0.32\,\cdot\,10^{-1}$ & $
  0.66$ &
$^{+ 3.5  } _{-3.5  }  $ & $^{+ 1.3  } _{-3.0  }  $ & $^{+ 0.8  } _{-1.1  }  $ & $^{- 0.0  } _{+ 0.1  } $ & $^{}_{+ 0.0  } $ & $^{-0.1  }  _{-0.2  }  $ & $^{-1.4  }  _{+ 0.6  } $ & $^{-0.3  }  _{+ 0.3  } $ & $^{+ 0.9  } _{-2.1  }  
$                                 & $^{}_{-1.1  }  $& $^{-0.2  }  _{+ 0.2  } $ \\
           & $0.50\,\cdot\,10^{-1}$ & $
  0.54$ &
$^{+ 3.2  } _{-3.2  }  $ & $^{+ 1.0  } _{-2.0  }  $ & $^{+ 0.6  } _{-0.5  }  $ & $^{-0.1  }  _{+ 0.1  } $ & $^{}_{+ 0.0  } $ & $^{+ 0.0  } _{- 0.0  } $ & $^{-0.4  }  _{-0.2  }  $ & $^{-0.3  }  _{+ 0.3  } $ & $^{+ 0.7  } _{-1.6  }  
$                                 & $^{}_{-0.9  }  $& $^{-0.2  }  _{+ 0.2  } $ \\
           & $0.80\,\cdot\,10^{-1}$ & $
  0.46$ &
$^{+ 3.2  } _{-3.2  }  $ & $^{+ 1.1  } _{-0.9  }  $ & $^{+ 0.8  } _{-0.3  }  $ & $^{- 0.0  } _{+ 0.2  } $ & $^{}_{+ 0.0  } $ & $^{+ 0.0  } _{+ 0.0  } $ & $^{-0.4  }  _{+ 0.7  } $ & $^{-0.3  }  _{+ 0.3  } $ & $^{+ 0.1  } _{-0.3  }  
$                                 & $^{}_{-0.6  }  $& $^{- 0.0  } _{+ 0.1  } $ \\
           & $0.13$ & $
  0.36$ &
$^{+ 3.4  } _{-3.4  }  $ & $^{+ 3.2  } _{-1.0  }  $ & $^{+ 0.6  } _{-0.2  }  $ & $^{-0.1  }  _{+ 0.1  } $ & $^{}_{+ 0.0  } $ & $^{+ 0.0  } _{+ 0.0  } $ & $^{+ 0.7  } _{+ 0.1  } $ & $^{-0.3  }  _{+ 0.3  } $ & $^{+ 0.9  } _{-2.0  }  
$                                 & $^{}_{+ 2.3  } $& $^{+ 0.2  } _{-0.2  }  $ \\
           & $0.25$ & $
  0.24$ &
$^{+ 3.9  } _{-3.9  }  $ & $^{+ 2.2  } _{-1.6  }  $ & $^{+ 0.6  } _{-0.8  }  $ & $^{-0.1  }  _{+ 0.2  } $ & $^{}_{+ 0.0  } $ & $^{+ 0.0  } _{+ 0.0  } $ & $^{-0.9  }  _{+ 0.1  } $ & $^{-0.4  }  _{+ 0.4  } $ & $^{+ 0.8  } _{-1.8  }  
$                                 & $^{}_{+ 0.8  } $& $^{+ 0.4  } _{-0.4  }  $ \\
           & $0.40$ & $
  0.14$ &
$^{+ 6.3  } _{-6.3  }  $ & $^{+ 1.5  } _{ -11. }  $ & $^{+ 1.1  } _{-6.3  }  $ & $^{-0.3  }  _{+ 0.2  } $ & $^{}_{+ 0.0  } $ & $^{+ 0.0  } _{+ 0.0  } $ & $^{-1.0  }  _{-0.6  }  $ & $^{-0.5  }  _{+ 0.5  } $ & $^{+ 0.9  } _{-2.1  }  
$                                 & $^{}_{-8.9  }  $& $^{+ 0.4  } _{-0.4  }  $ \\
\hline
 1500      & $0.21\,\cdot\,10^{-1}$ & $
  0.68$ &
$^{+ 4.8  } _{-4.8  }  $ & $^{+ 4.6  } _{-3.8  }  $ & $^{+ 0.9  } _{-0.9  }  $ & $^{-0.1  }  _{+ 0.0  } $ & $^{}_{+ 0.6  } $ & $^{-0.2  }  _{-1.0  }  $ & $^{-1.5  }  _{+ 4.3  } $ & $^{-0.3  }  _{+ 0.3  } $ & $^{+ 0.7  } _{-1.7  }  
$                                 & $^{}_{-3.0  }  $& $^{- 0.0  } _{+ 0.0  } $ \\
           & $0.32\,\cdot\,10^{-1}$ & $
  0.64$ &
$^{+ 4.4  } _{-4.4  }  $ & $^{+ 2.0  } _{-1.0  }  $ & $^{+ 1.1  } _{-0.8  }  $ & $^{- 0.0  } _{+ 0.1  } $ & $^{}_{+ 0.0  } $ & $^{+ 0.2  } _{-0.2  }  $ & $^{-0.1  }  _{+ 1.6  } $ & $^{-0.3  }  _{+ 0.3  } $ & $^{+ 0.1  } _{-0.2  }  
$                                 & $^{}_{-0.1  }  $& $^{-0.1  }  _{+ 0.1  } $ \\
           & $0.50\,\cdot\,10^{-1}$ & $
  0.58$ &
$^{+ 3.9  } _{-3.9  }  $ & $^{+ 2.2  } _{-1.3  }  $ & $^{+ 1.6  } _{-0.2  }  $ & $^{-0.1  }  _{+ 0.1  } $ & $^{}_{+ 0.0  } $ & $^{+ 0.1  } _{+ 0.3  } $ & $^{+ 0.6  } _{+ 0.4  } $ & $^{-0.4  }  _{+ 0.4  } $ & $^{+ 0.5  } _{-1.2  }  
$                                 & $^{}_{+ 1.2  } $& $^{-0.2  }  _{+ 0.2  } $ \\
           & $0.80\,\cdot\,10^{-1}$ & $
  0.44$ &
$^{+ 4.1  } _{-4.1  }  $ & $^{+ 0.8  } _{-2.8  }  $ & $^{+ 0.4  } _{-2.2  }  $ & $^{-0.1  }  _{+ 0.2  } $ & $^{}_{+ 0.0  } $ & $^{- 0.0  } _{- 0.0  } $ & $^{-0.1  }  _{-0.1  }  $ & $^{-0.4  }  _{+ 0.4  } $ & $^{+ 0.3  } _{-0.7  }  
$                                 & $^{}_{-1.6  }  $& $^{- 0.0  } _{+ 0.0  } $ \\
           & $0.13$ & $
  0.34$ &
$^{+ 4.9  } _{-4.9  }  $ & $^{+ 3.2  } _{-0.9  }  $ & $^{+ 1.8  } _{-0.6  }  $ & $^{-0.1  }  _{+ 0.1  } $ & $^{}_{+ 0.0  } $ & $^{+ 0.0  } _{+ 0.0  } $ & $^{-0.5  }  _{+ 1.1  } $ & $^{-0.3  }  _{+ 0.3  } $ & $^{+ 0.0  } _{- 0.0  } 
$                                 & $^{}_{+ 2.3  } $& $^{+ 0.2  } _{-0.2  }  $ \\
           & $0.18$ & $
  0.31$ &
$^{+ 5.0  } _{-5.0  }  $ & $^{+ 1.0  } _{-2.8  }  $ & $^{+ 0.8  } _{-0.5  }  $ & $^{-0.2  }  _{+ 0.2  } $ & $^{}_{+ 0.0  } $ & $^{-0.1  }  _{+ 0.0  } $ & $^{-0.8  }  _{-0.1  }  $ & $^{-0.3  }  _{+ 0.3  } $ & $^{+ 0.2  } _{-0.4  }  
$                                 & $^{}_{-2.6  }  $& $^{+ 0.2  } _{-0.2  }  $ \\
           & $0.25$ & $
  0.26$ &
$^{+ 5.9  } _{-5.9  }  $ & $^{+ 2.0  } _{-3.2  }  $ & $^{+ 1.5  } _{-2.5  }  $ & $^{+ 0.1  } _{+ 0.2  } $ & $^{}_{+ 0.0  } $ & $^{+ 0.0  } _{+ 0.0  } $ & $^{-0.4  }  _{-0.6  }  $ & $^{-0.3  }  _{+ 0.3  } $ & $^{+ 0.6  } _{-1.4  }  
$                                 & $^{}_{-1.5  }  $& $^{+ 0.2  } _{-0.3  }  $ \\
           & $0.40$ & $
  0.13$ &
$^{+ 8.7  } _{-8.7  }  $ & $^{+ 2.5  } _{-6.1  }  $ & $^{+ 2.4  } _{-4.4  }  $ & $^{+ 0.1  } _{+ 0.4  } $ & $^{}_{+ 0.0  } $ & $^{+ 0.0  } _{+ 0.0  } $ & $^{+ 0.1  } _{-0.3  }  $ & $^{-0.4  }  _{+ 0.4  } $ & $^{+ 0.3  } _{-0.6  }  
$                                 & $^{}_{-4.1  }  $& $^{+ 0.3  } _{-0.3  }  $ \\
\hline
 2000      & $0.32\,\cdot\,10^{-1}$ & $
  0.62$ &
$^{+ 5.3  } _{-5.3  }  $ & $^{+ 3.8  } _{-1.2  }  $ & $^{+ 1.0  } _{-1.0  }  $ & $^{+ 0.0  } _{+ 0.1  } $ & $^{}_{+ 1.7  } $ & $^{- 0.0  } _{+ 0.1  } $ & $^{+ 0.1  } _{+ 2.9  } $ & $^{-0.4  }  _{+ 0.4  } $ & $^{+ 0.1  } _{-0.1  }  
$                                 & $^{}_{+ 1.3  } $& $^{-0.1  }  _{+ 0.2  } $ \\
           & $0.50\,\cdot\,10^{-1}$ & $
  0.52$ &
$^{+ 5.0  } _{-5.0  }  $ & $^{+ 1.1  } _{-1.6  }  $ & $^{+ 0.8  } _{-1.4  }  $ & $^{-0.1  }  _{+ 0.2  } $ & $^{}_{+ 0.0  } $ & $^{-0.1  }  _{- 0.0  } $ & $^{+ 0.0  } _{-0.3  }  $ & $^{-0.3  }  _{+ 0.3  } $ & $^{+ 0.2  } _{-0.4  }  
$                                 & $^{}_{+ 0.7  } $& $^{-0.2  }  _{+ 0.2  } $ \\
           & $0.80\,\cdot\,10^{-1}$ & $
  0.44$ &
$^{+ 4.8  } _{-4.8  }  $ & $^{+ 2.0  } _{-0.7  }  $ & $^{+ 0.8  } _{-0.4  }  $ & $^{-0.1  }  _{+ 0.1  } $ & $^{}_{+ 0.0  } $ & $^{- 0.0  } _{- 0.0  } $ & $^{+ 0.7  } _{+ 0.3  } $ & $^{-0.3  }  _{+ 0.3  } $ & $^{+ 0.5  } _{-1.1  }  
$                                 & $^{}_{+ 1.1  } $& $^{- 0.0  } _{+ 0.0  } $ \\
           & $0.13$ & $
  0.40$ &
$^{+ 5.5  } _{-5.5  }  $ & $^{+ 1.8  } _{-1.1  }  $ & $^{+ 1.6  } _{-0.8  }  $ & $^{-0.3  }  _{+ 0.2  } $ & $^{}_{+ 0.0  } $ & $^{- 0.0  } _{+ 0.1  } $ & $^{-0.2  }  _{+ 0.4  } $ & $^{-0.3  }  _{+ 0.3  } $ & $^{+ 0.2  } _{-0.5  }  
$                                 & $^{}_{+ 0.4  } $& $^{+ 0.2  } _{-0.2  }  $ \\
           & $0.18$ & $
  0.32$ &
$^{+ 5.8  } _{-5.8  }  $ & $^{+ 0.9  } _{-1.5  }  $ & $^{+ 0.5  } _{-1.5  }  $ & $^{-0.1  }  _{+ 0.3  } $ & $^{}_{+ 0.0  } $ & $^{+ 0.0  } _{+ 0.0  } $ & $^{+ 0.1  } _{+ 0.5  } $ & $^{-0.4  }  _{+ 0.4  } $ & $^{+ 0.1  } _{-0.3  }  
$                                 & $^{}_{+ 0.3  } $& $^{+ 0.2  } _{-0.2  }  $ \\
           & $0.25$ & $
  0.24$ &
$^{+ 7.1  } _{-7.1  }  $ & $^{+ 1.6  } _{-2.0  }  $ & $^{+ 1.4  } _{-1.4  }  $ & $^{-0.1  }  _{+ 0.2  } $ & $^{}_{+ 0.0  } $ & $^{+ 0.0  } _{+ 0.0  } $ & $^{-1.0  }  _{+ 0.6  } $ & $^{-0.3  }  _{+ 0.3  } $ & $^{+ 0.4  } _{-0.9  }  
$                                 & $^{}_{+ 0.3  } $& $^{+ 0.2  } _{-0.3  }  $ \\
           & $0.40$ & $
  0.13$ &
$^{+  10. } _{ -10. }  $ & $^{+ 1.8  } _{ -11. }  $ & $^{+ 1.8  } _{-5.1  }  $ & $^{-0.4  }  _{-0.2  }  $ & $^{}_{+ 0.0  } $ & $^{+ 0.0  } _{+ 0.0  } $ & $^{-0.7  }  _{+ 0.4  } $ & $^{-0.3  }  _{+ 0.3  } $ & $^{+ 0.1  } _{-0.2  }  
$                                 & $^{}_{-9.3  }  $& $^{+ 0.3  } _{-0.2  }  $ \\
\hline
 3000      & $0.50\,\cdot\,10^{-1}$ & $
  0.54$ &
$^{+ 6.1  } _{-6.1  }  $ & $^{+ 1.6  } _{-1.6  }  $ & $^{+ 1.3  } _{-1.3  }  $ & $^{-0.1  }  _{+ 0.2  } $ & $^{}_{+ 0.1  } $ & $^{+ 0.2  } _{+ 0.2  } $ & $^{+ 0.7  } _{+ 0.7  } $ & $^{-0.3  }  _{+ 0.3  } $ & $^{+ 0.2  } _{-0.5  }  
$                                 & $^{}_{+ 0.3  } $& $^{-0.1  }  _{+ 0.1  } $ \\
           & $0.80\,\cdot\,10^{-1}$ & $
  0.41$ &
$^{+ 6.0  } _{-6.0  }  $ & $^{+ 1.5  } _{-1.9  }  $ & $^{+ 1.3  } _{-0.3  }  $ & $^{-0.1  }  _{+ 0.2  } $ & $^{}_{+ 0.0  } $ & $^{-0.1  }  _{+ 0.3  } $ & $^{-0.4  }  _{-0.6  }  $ & $^{-0.3  }  _{+ 0.3  } $ & $^{+ 0.2  } _{-0.5  }  
$                                 & $^{}_{-1.7  }  $& $^{+ 0.0  } _{- 0.0  } $ \\
           & $0.13$ & $
  0.35$ &
$^{+ 6.8  } _{-6.8  }  $ & $^{+ 4.2  } _{-1.2  }  $ & $^{+ 2.7  } _{-0.9  }  $ & $^{-0.3  }  _{+ 0.1  } $ & $^{}_{+ 0.0  } $ & $^{- 0.0  } _{+ 0.1  } $ & $^{-0.3  }  _{+ 1.5  } $ & $^{-0.3  }  _{+ 0.3  } $ & $^{+ 0.1  } _{-0.3  }  
$                                 & $^{}_{+ 2.9  } $& $^{+ 0.2  } _{-0.2  }  $ \\
           & $0.18$ & $
  0.32$ &
$^{+ 6.8  } _{-6.8  }  $ & $^{+ 1.4  } _{-3.4  }  $ & $^{+ 1.4  } _{-2.9  }  $ & $^{-0.1  }  _{+ 0.2  } $ & $^{}_{+ 0.0  } $ & $^{- 0.0  } _{-0.4  }  $ & $^{- 0.0  } _{+ 0.0  } $ & $^{-0.3  }  _{+ 0.3  } $ & $^{+ 0.3  } _{-0.7  }  
$                                 & $^{}_{-1.6  }  $& $^{+ 0.2  } _{-0.2  }  $ \\
           & $0.25$ & $
  0.23$ &
$^{+ 8.4  } _{-8.4  }  $ & $^{+ 0.7  } _{-4.6  }  $ & $^{+ 0.2  } _{-2.4  }  $ & $^{-0.1  }  _{+ 0.1  } $ & $^{}_{+ 0.0  } $ & $^{+ 0.0  } _{+ 0.1  } $ & $^{-0.3  }  _{+ 0.1  } $ & $^{-0.4  }  _{+ 0.4  } $ & $^{+ 0.2  } _{-0.5  }  
$                                 & $^{}_{-3.9  }  $& $^{+ 0.2  } _{-0.2  }  $ \\
           & $0.40$ & $
  0.14$ &
$^{+  11. } _{ -11. }  $ & $^{+ 4.4  } _{-4.0  }  $ & $^{+ 2.5  } _{-1.1  }  $ & $^{-0.1  }  _{+ 0.0  } $ & $^{}_{+ 0.0  } $ & $^{+ 0.0  } _{+ 0.0  } $ & $^{+ 0.9  } _{+ 0.8  } $ & $^{-0.3  }  _{+ 0.3  } $ & $^{+ 1.4  } _{-3.4  }  
$                                 & $^{}_{-3.6  }  $& $^{+ 0.1  } _{-0.2  }  $ \\
           & $0.65$ & $
  0.02$ &
$^{+  18. } _{ -18. }  $ & $^{+  11. } _{-6.2  }  $ & $^{+ 3.8  } _{-4.9  }  $ & $^{- 0.0  } _{+ 0.2  } $ & $^{}_{+ 0.0  } $ & $^{+ 0.0  } _{+ 0.0  } $ & $^{-1.8  }  _{-1.5  }  $ & $^{-0.4  }  _{+ 0.4  } $ & $^{+ 2.7  } _{-6.3  }  
$                                 & $^{}_{+ 8.6  } $& $^{+ 1.1  } _{-1.2  }  $ \\
\hline
 5000      & $0.80\,\cdot\,10^{-1}$ & $
  0.42$ &
$^{+ 5.7  } _{-5.7  }  $ & $^{+ 2.2  } _{-1.6  }  $ & $^{+ 1.7  } _{-0.8  }  $ & $^{-0.1  }  _{+ 0.2  } $ & $^{}_{-0.3  }  $ & $^{-0.5  }  _{-0.7  }  $ & $^{+ 0.5  } _{+ 0.8  } $ & $^{-0.3  }  _{+ 0.3  } $ & $^{+ 0.4  } _{-1.0  }  
$                                 & $^{}_{+ 0.9  } $& $^{+ 0.1  } _{-0.1  }  $ \\
           & $0.13$ & $
  0.36$ &
$^{+ 7.2  } _{-7.2  }  $ & $^{+ 1.1  } _{-1.2  }  $ & $^{+ 0.5  } _{-0.9  }  $ & $^{-0.1  }  _{+ 0.1  } $ & $^{}_{-0.3  }  $ & $^{+ 0.5  } _{+ 0.2  } $ & $^{-0.1  }  _{-0.5  }  $ & $^{-0.3  }  _{+ 0.3  } $ & $^{+ 0.2  } _{-0.4  }  
$                                 & $^{}_{+ 0.8  } $& $^{+ 0.2  } _{-0.2  }  $ \\
           & $0.18$ & $
  0.27$ &
$^{+ 7.8  } _{-7.8  }  $ & $^{+ 1.5  } _{-3.0  }  $ & $^{+ 0.4  } _{-1.8  }  $ & $^{-0.1  }  _{+ 0.2  } $ & $^{}_{-0.3  }  $ & $^{- 0.0  } _{+ 0.1  } $ & $^{-1.0  }  _{-0.9  }  $ & $^{-0.3  }  _{+ 0.3  } $ & $^{+ 0.6  } _{-1.5  }  
$                                 & $^{}_{-1.7  }  $& $^{+ 0.2  } _{-0.2  }  $ \\
           & $0.25$ & $
  0.22$ &
$^{+ 8.8  } _{-8.8  }  $ & $^{+ 2.8  } _{-2.8  }  $ & $^{+ 0.8  } _{-2.6  }  $ & $^{-0.1  }  _{+ 0.0  } $ & $^{}_{-0.3  }  $ & $^{+ 0.0  } _{+ 0.1  } $ & $^{-0.3  }  _{-0.5  }  $ & $^{-0.3  }  _{+ 0.3  } $ & $^{+ 0.6  } _{-1.4  }  
$                                 & $^{}_{+ 2.3  } $& $^{+ 0.2  } _{-0.1  }  $ \\
           & $0.40$ & $
  0.11$ &
$^{+  12. } _{ -12. }  $ & $^{+ 3.5  } _{-1.1  }  $ & $^{+ 1.4  } _{-0.7  }  $ & $^{+ 0.1  } _{+ 0.4  } $ & $^{}_{-0.3  }  $ & $^{+ 0.0  } _{+ 0.0  } $ & $^{-0.1  }  _{-0.3  }  $ & $^{-0.4  }  _{+ 0.3  } $ & $^{+ 0.3  } _{-0.6  }  
$                                 & $^{}_{+ 3.1  } $& $^{+ 0.1  } _{-0.2  }  $ \\
\hline
 8000      & $0.13$ & $
  0.31$ &
$^{+ 9.4  } _{-9.4  }  $ & $^{+ 2.6  } _{-2.3  }  $ & $^{+ 0.5  } _{-2.1  }  $ & $^{-0.2  }  _{+ 0.2  } $ & $^{}_{-0.3  }  $ & $^{+ 0.2  } _{+ 0.9  } $ & $^{+ 1.7  } _{+ 0.1  } $ & $^{-0.4  }  _{+ 0.4  } $ & $^{+ 0.0  } _{- 0.0  } 
$                                 & $^{}_{+ 1.6  } $& $^{+ 0.5  } _{-0.5  }  $ \\
           & $0.18$ & $
  0.27$ &
$^{+  10. } _{ -10. }  $ & $^{+ 1.1  } _{-2.4  }  $ & $^{+ 0.5  } _{-2.2  }  $ & $^{-0.2  }  _{+ 0.0  } $ & $^{}_{-0.3  }  $ & $^{-0.1  }  _{-0.7  }  $ & $^{-0.2  }  _{+ 0.8  } $ & $^{-0.4  }  _{+ 0.4  } $ & $^{+ 0.2  } _{-0.4  }  
$                                 & $^{}_{-0.4  }  $& $^{+ 0.2  } _{-0.2  }  $ \\
           & $0.25$ & $
  0.20$ &
$^{+  12. } _{ -12. }  $ & $^{+ 3.8  } _{-2.3  }  $ & $^{+ 1.7  } _{-2.1  }  $ & $^{-0.4  }  _{+ 0.3  } $ & $^{}_{-0.3  }  $ & $^{-0.2  }  _{+ 0.2  } $ & $^{+ 0.9  } _{+ 2.4  } $ & $^{-0.3  }  _{+ 0.3  } $ & $^{+ 0.7  } _{-1.7  }  
$                                 & $^{}_{+ 1.2  } $& $^{+ 0.1  } _{-0.2  }  $ \\
           & $0.40$ & $
  0.10$ &
$^{+  17. } _{ -17. }  $ & $^{+ 5.0  } _{-3.5  }  $ & $^{+ 4.8  } _{-3.3  }  $ & $^{- 0.0  } _{+ 0.2  } $ & $^{}_{-0.3  }  $ & $^{- 0.0  } _{+ 0.1  } $ & $^{+ 0.4  } _{-0.7  }  $ & $^{-0.3  }  _{+ 0.3  } $ & $^{+ 0.3  } _{-0.6  }  
$                                 & $^{}_{+ 1.0  } $& $^{+ 0.1  } _{- 0.0  } $ \\
           & $0.65$ & $
  0.01$ &
$^{+  40. } _{ -30. }  $ & $^{+ 5.4  } _{-7.2  }  $ & $^{+ 4.3  } _{-6.8  }  $ & $^{-0.4  }  _{+ 0.6  } $ & $^{}_{-0.3  }  $ & $^{+ 0.0  } _{+ 0.1  } $ & $^{+ 0.3  } _{-0.4  }  $ & $^{-0.3  }  _{+ 0.3  } $ & $^{+ 1.3  } _{-3.0  }  
$                                 & $^{}_{-1.8  }  $& $^{+ 0.8  } _{-0.8  }  $ \\
\hline
12000      & $0.18$ & $
  0.32$ &
$^{+  12. } _{ -12. }  $ & $^{+ 2.7  } _{-6.6  }  $ & $^{+ 2.2  } _{-5.1  }  $ & $^{-0.6  }  _{+ 0.7  } $ & $^{}_{-0.3  }  $ & $^{-1.4  }  _{-3.2  }  $ & $^{+ 0.6  } _{+ 0.7  } $ & $^{-0.3  }  _{+ 0.3  } $ & $^{+ 0.9  } _{-2.0  }  
$                                 & $^{}_{-0.1  }  $& $^{+ 0.3  } _{-0.3  }  $ \\
           & $0.25$ & $
  0.20$ &
$^{+  20. } _{ -20. }  $ & $^{+ 1.7  } _{-6.7  }  $ & $^{+ 1.3  } _{-5.7  }  $ & $^{-0.2  }  _{+ 0.1  } $ & $^{}_{-0.3  }  $ & $^{-0.4  }  _{+ 0.7  } $ & $^{+ 0.8  } _{-0.3  }  $ & $^{-0.4  }  _{+ 0.4  } $ & $^{+ 0.1  } _{-0.3  }  
$                                 & $^{}_{-3.3  }  $& $^{+ 0.3  } _{-0.3  }  $ \\
           & $0.40$ & $
  0.09$ &
$^{+  25. } _{ -25. }  $ & $^{+ 6.7  } _{-8.9  }  $ & $^{+ 6.1  } _{-5.7  }  $ & $^{-0.1  }  _{+ 0.5  } $ & $^{}_{-0.3  }  $ & $^{-0.2  }  _{+ 0.5  } $ & $^{-4.6  }  _{-0.1  }  $ & $^{-0.5  }  _{+ 0.5  } $ & $^{+ 1.1  } _{-2.6  }  
$                                 & $^{}_{-4.9  }  $& $^{- 0.0  } _{+ 0.0  } $ \\
\hline
20000      & $0.25$ & $
  0.09$ &
$^{+  28. } _{ -28. }  $ & $^{+ 9.5  } _{-2.6  }  $ & $^{+ 2.0  } _{-1.9  }  $ & $^{-0.4  }  _{+ 0.5  } $ & $^{}_{-0.3  }  $ & $^{-1.4  }  _{+ 4.3  } $ & $^{+ 2.9  } _{-0.4  }  $ & $^{-0.4  }  _{+ 0.4  } $ & $^{+ 0.2  } _{-0.4  }  
$                                 & $^{}_{+ 7.7  } $& $^{+ 0.7  } _{-0.8  }  $ \\
           & $0.40$ & $
  0.06$ &
$^{+  60. } _{ -40. }  $ & $^{+  16. } _{ -17. }  $ & $^{+ 2.7  } _{ -17. }  $ & $^{-0.6  }  _{+ 0.5  } $ & $^{}_{-0.3  }  $ & $^{-0.4  }  _{+ 1.5  } $ & $^{+ 1.9  } _{+  16. } $ & $^{-0.3  }  _{+ 0.3  } $ & $^{+ 0.8  } _{-1.8  }  
$                                 & $^{}_{+ 0.7  } $& $^{-0.1  }  _{+ 0.2  } $ \\
\hline
30000      & $0.40$ & $
  0.05$ &
$^{+  68. } _{ -43. }  $ & $^{+ 5.3  } _{ -42. }  $ & $^{+ 4.3  } _{-2.8  }  $ & $^{-0.8  }  _{+ 0.3  } $ & $^{}_{-0.3  }  $ & $^{-1.3  }  _{ -38. }  $ & $^{+ 2.5  } _{-0.5  }  $ & $^{-0.4  }  _{+ 0.4  } $ & $^{+ 0.8  } _{-1.8  }  
$                                 & $^{}_{ -19. }  $& $^{+ 0.0  } _{-0.1  }  $ \\
\hline
\end{tabular}
}

  \end{center}    
  \normalsize
  \caption[this space for rent]{
    Systematic uncertainties with bin-to-bin correlations
    for the reduced cross-section $\tilde{\sigma}^{e^+ p}$. 
    The left part of the table contains the quoted $Q^2$ and $x$
    values, $Q^2_c$ and $x_c$, the measured cross-section 
    $\tilde{\sigma}^{e^+ p}$ corrected to the electroweak Born level,
    the statistical uncertainty and the total systematic uncertainty. 
    The uncertainty on the measured luminosity of $2.5\%$ 
    is not included in the total systematic uncertainty.
    The right part of the table lists the total uncorrelated
    systematic uncertainty followed by the bin-to-bin correlated
    systematic uncertainties $\delta_1$--\,$\delta_8$ defined in
    the text.
    For the latter, the upper (lower)
    numbers refer to positive (negative) variation of e.g. the cut
    value, whereas the signs of the numbers reflect the direction of
    change in the cross sections. 
    }
  \label{tab-dsdxq_c2}
\end{table}
\clearpage

%
%
%
%
%
%
\begin{figure}
  \begin{center}
    \includegraphics[width=.85\textwidth]{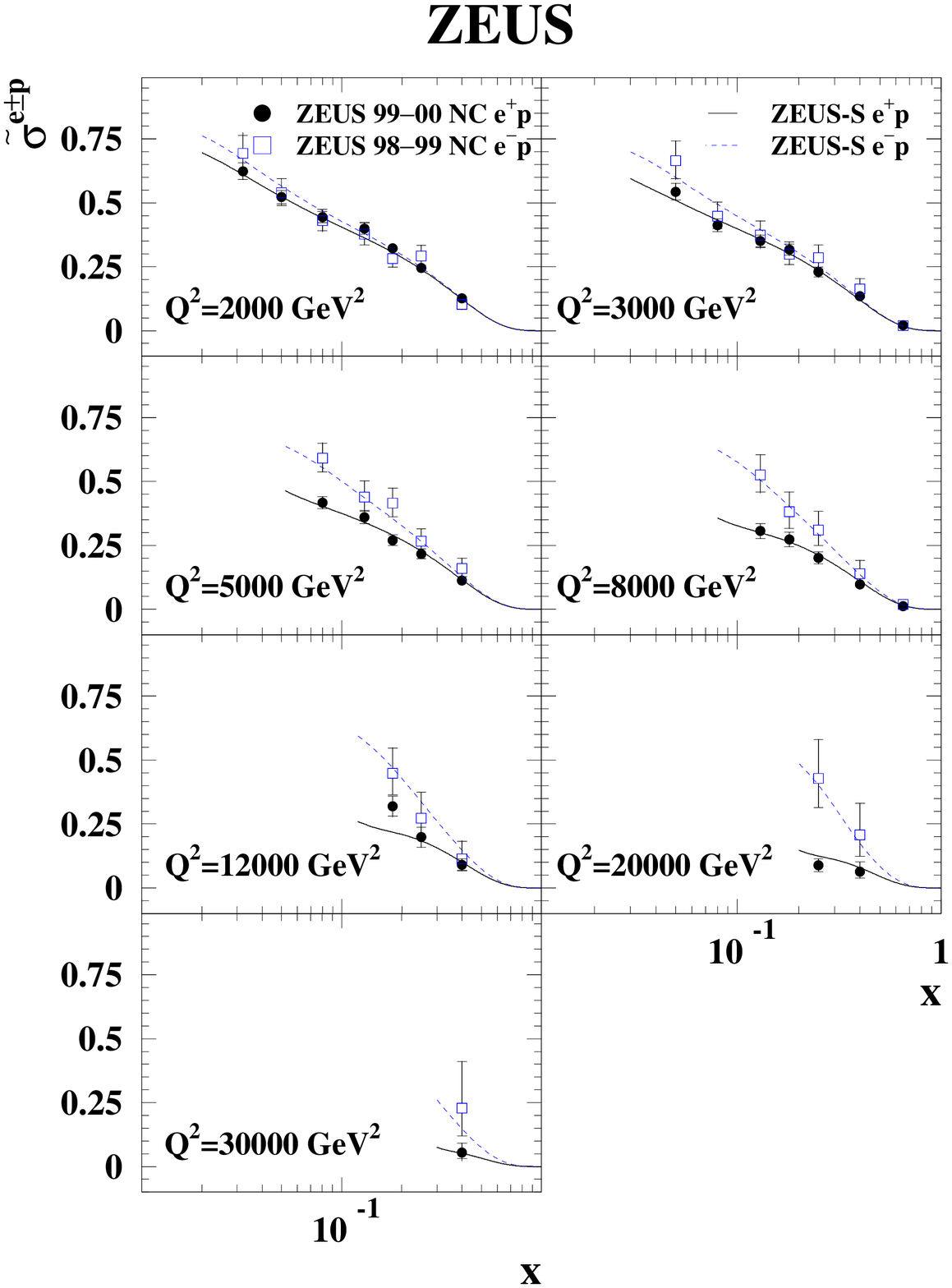}
  \end{center}
  \caption{
    The $e^+ p$ reduced cross section, $\tilde{\sigma}^{e^+ p}$, (solid
    points) plotted as a function of $x$ at fixed $Q^2$ between
    $2\,000 \gev^2$ and $30\,000 \gev^2$ compared  to
    $\tilde{\sigma}^{e^- p}$ (open squares).  
    The inner bars show the statistical uncertainty, while the
    outer ones show the statistical and systematic uncertainties added
    in quadrature.
    The Standard Model expectations, evaluated using the ZEUS-S PDFs,
    are shown as the solid ($e^+p$) and dashed ($e^-p$) lines.
  }
  \label{fig-RedCross2}
\end{figure}
%
%

%
%
\end{document}